\documentclass[journal,comsoc]{IEEEtran}
\usepackage{amsmath,amsfonts}
\usepackage{array}
\usepackage{textcomp}
\usepackage{stfloats}
\usepackage{url}
\usepackage{verbatim}
\usepackage{graphicx}
\usepackage{cite}
\usepackage{dsfont}
\usepackage{acronym}
\usepackage{xcolor}
\usepackage{mathrsfs}
\usepackage{mathtools}
\usepackage{pbalance}

\ifCLASSOPTIONcompsoc
    \usepackage[caption=false, font=normalsize, labelfont=sf, textfont=sf]{subfig}
\else
    \usepackage[caption=false, font=footnotesize]{subfig}
\fi

\makeatletter
\newcommand{\mathleft}{\@fleqntrue\@mathmargin0pt}
\newcommand{\mathcenter}{\@fleqnfalse}
\makeatother

\newacro{DoF}{degrees of freedom}
\newacro{LoS}{Line-Of-Sight}
\newacro{OFDM}{Orthogonal Frequency Division Multiplexing}
\newacro{WDM}{Wavenumber Division Multiplexing}
\newacro{CSI}{Channel State Information}
\newacro{SVD}{Singular Value Decomposition}
\newacro{MMSE}{Minimum Mean Square Error}
\newacro{PSWF}{Prolate Spheroidal Wave Function}
\newacro{MIMO}{Multiple Input Multiple Output}

\DeclareMathOperator{\rect}{rect}
\DeclareMathOperator{\sinc}{sinc}

\begin{document}

\title{On the Degrees of Freedom and Eigenfunctions of Line-of-Sight Holographic MIMO Communications}

\author{Juan Carlos Ruiz-Sicilia,~\IEEEmembership{Student Member,~IEEE,} Marco~Di~Renzo,~\IEEEmembership{Fellow,~IEEE,} Marco Donald Migliore,~\IEEEmembership{Senior Member,~IEEE,} Merouane Debbah,~\IEEEmembership{Fellow,~IEEE,} and H. Vincent Poor,~\IEEEmembership{Life Fellow,~IEEE} \vspace{-1cm}

\thanks{Manuscript received June 13, 2025. J. C. Ruiz-Sicilia and M. Di Renzo are with Universit\'e Paris-Saclay, CNRS, CentraleSup\'elec, Laboratoire des Signaux et Syst\'emes, 91192 Gif-sur-Yvette, France. (marco.di-renzo@universite-paris-saclay.fr). M. Di Renzo is also with King's College London, Centre for Telecommunications Research -- Department of Engineering, WC2R 2LS London, United Kingdom (marco.di\_renzo@kcl.ac.uk). M. D. Migliore is with the Dept. of Electrical and Information Engineering, Univ. of Cassino and Southern Lazio, 03043 Cassino, Italy. M. Debbah is with the Dept. of Electrical Engineering and Computer Science, Khalifa University, Abu Dhabi, United Arab Emirates. H. V. Poor is with the Dept. of Electrical and Computer Engineering, Princeton University, Princeton, NJ 08544, USA.}}



\maketitle

\begin{abstract}
\textcolor{black}{We consider a line-of-sight communication link between two holographic surfaces (HoloSs), and provide a closed-form expression for the effective degrees of freedom (eDoF), i.e., the number of orthogonal communication modes that can be established between them. The proposed framework can be applied to network deployments beyond the widely studied paraxial setting. This is obtained by partitioning the largest HoloS into sub-HoloSs, and proving that the supports of the Fourier transforms of the kernels of the obtained integral operators are limited and are almost disjoint in the wavenumber domain, provided that the sub-HoloSs are sufficiently small. Using the proposed approach, it is proved that (i) the eDoF correspond to an instance of Landau's second eigenvalue problem; (ii) the eigenvalues polarize asymptotically to multiple values; and (iii) the eDoF depend explicitly on the approximation accuracy according to Kolmogorov's $n$-width criterion. This result generalizes the analysis in the paraxial setting, in which it is known that the eigenvalues polarize asymptotically to two values. In addition, it is proved that the typical method of analysis utilized in the paraxial setting, which is based on a parabolic approximation of the wavefront in a local coordinates system, is equivalent to a quartic approximation of the wavefront in a general coordinates system. This facilitates the derivation of an explicit formula for the eDoF in terms of key system parameters, including the relative offset between the center-points of the HoloSs, and their relative rotation and tilt. We specialize the framework to canonical network deployments, and provide analytical expressions for the optimal, according to Kolmogorov's $n$-width criterion, basis functions (communication waveforms) for data encoding and decoding. With the aid of simulations, we validate the accuracy of the closed-form expressions for the eDoF and waveforms}.
\end{abstract}

\vspace{-0.2cm}

\vspace{-0.4cm}
\section{Introduction} \vspace{-0.2cm}
\IEEEPARstart{H}{olographic} multiple-input multiple-output (MIMO) is an emerging technology \cite{DBLP:journals/wc/HuangHAZYZRD20}, \cite{ChauCOMML}. The reference setup consists of two holographic surfaces (HoloSs) communicating with each other, with a HoloS being an electrically large antenna that is made of a virtually infinite number of radiating elements coupled with electronic circuits and a limited number of radio frequency chains \cite{MDR_EuCAP}. From a theoretical standpoint, the transmitting HoloS is capable of synthesizing any surface current density fulfilling Maxwell's equations, and, hence, any radiated electromagnetic field, and the receiving HoloS is capable of sensing any impinging electromagnetic field \cite{DardariD21}.

From a communication perspective, similar to conventional spatial multiplexing MIMO \cite{Telatar}, the theoretical characterization of the performance of a holographic MIMO link consists of identifying the so-called communication modes \cite{Miller:19}. A communication mode is defined as a spatial channel through which data can be transmitted in an interference-free manner. The number of interference-free (or orthogonal) spatial channels is referred to as the degrees of freedom (DoF). Generally speaking, the number of DoF may be infinite, but the finite size of the HoloSs and the finite spatial bandwidth of the transmission channel result in a finite number of strongly connected communication channels, where strongly connected refers to a spatial channel that carries a good portion of the total system (transmit) power. The number of strongly connected communication channels is referred to as the effective DoF (eDoF). A rigorous definition of eDoF originates from approximation theory in general multi-dimensional Hilbert spaces: The number of eDoF coincides with the minimum value of $n$ for which Kolmogorov's $n$-width is less than a predefined level of accuracy \cite{Kolmogorov}. In simple terms, the number of eDoF is the minimum number of optimal basis functions that is needed to represent any surface current density at the transmitting HoloS and any received electromagnetic field at the receiving HoloS within a certain maximum error \cite{FranceschettiBook}. Specifically, a set of $N$ basis functions is referred to as optimal if, among all the possible choices of $N$ orthonormal functions, it minimizes the approximation error for a given $N$. The optimal basis functions for the transmitting and receiving HoloSs are usually different but not independent. Interested readers are referred to \cite{Franceschetti2015}, \cite{Pizzo2022} and to the textbooks \cite{Kolmogorov} and \cite{FranceschettiBook} for a comprehensive overview and a historical perspective on the eDoF and Kolmogorov's $n$-width in approximation theory.

The theoretical characterization of the performance of a holographic MIMO link reduces, therefore, to the computation of the eDoF and the optimal basis functions to represent any surface current density and received electromagnetic field at the transmitting and receiving HoloSs, respectively \cite{MDM}. In the literature, broadly speaking, two analytical methods for the computation of the eDoF exist: The cut-set integral and the self-adjoint operator \cite{EuCAP_2025}. A comparison between these two approaches is available in \cite[Table I]{EuCAP_2025}. In the present paper, we focus our attention on the method based on the self-adjoint operator, as it enables the analysis for any approximation accuracy according to the definition of Kolmogorov's $n$-width, and the computation of the optimal basis functions. Also, we show that the proposed approach subsumes the cut-set integral method when the approximation accuracy is set to a sufficiently small value that is quantified analytically. 

To elaborate, the most related prior art includes \cite{Miller, Piestun:00, Dardari_2020, Decarli_2021, Pizzo2022, Lozano_2024, Niyato_2025, Hongliang_2025}. In \cite{Miller}, the author has introduced a general framework for estimating the eDoF and the optimal basis functions as solution of two eigenproblems. The approach is applicable to scalar electromagnetic fields in free-space, and leverages the theory of compact and self-adjoint operators over Hilbert spaces \cite{IntegralEq}. Under the assumption of a parabolic approximation for the wavefront of the electromagnetic waves and assuming a paraxial setting, the author shows that the optimal basis functions are related to prolate spheroidal wave functions (PSWFs), and the eDoF immediately follow from the energy concentration property of the PSWFs. The approach is generalized in \cite{Piestun:00} for application to vector electromagnetic fields and for propagation over general channels (beyond the free space scenario). The approach in \cite{Miller} and \cite{Piestun:00} is general, but, with the exception of the paraxial setting, it requires extensive numerical computations, and the eDoF and optimal basis functions can only be determined numerically. To overcome these limitations, the author of \cite{Dardari_2020} provides an approximate approach to compute the eDoF based on the calculation of an integral obtained from the wavevector between the transmitting and receiving HoloSs. Computationally, the approach is simpler than the solution of the eigenproblem in \cite{Miller} and \cite{Piestun:00}. Also, the author of \cite{Dardari_2020} provides a closed-form expression for the eDoF when one of the two HoloSs is sufficiently small compared with the distance, and the two HoloSs are either parallel or orthogonal to one another. In the general case, however, the approach is numerical and the optimal communication functions are not discussed. Also, it can be applied when the approximation accuracy according to Kolmogorov's $n$-width is sufficiently small. To fill these gaps, the authors of \cite{Decarli_2021} move from \cite{Dardari_2020} and introduce approximate basis functions based on the concept of focusing functions. The orthonormal functions are constructed iteratively, and, based on geometric considerations, an approximate expression for the eDoF is given when one of the HoloS is small enough compared to the distance between the HoloSs. The approach is applicable to lines, and leads to an approximate design. More recently, the authors of \cite{Pizzo2022} have established a connection between the eDoF of the eigenproblem formulated in \cite{Miller} for scalar electromagnetic fields and Landau's eigenvalue problem \cite{Landau1975}. Under the paraxial approximation, more precisely, the two problems are shown to be equivalent, and hence the authors compute the eDoF for some channel models. The approach is, however, applicable only under the paraxial approximation and no discussion about the optimal basis functions is given. The frameworks in \cite{Pizzo2022} and \cite{Miller} are based on a parabolic approximation for the wavefront of the electromagnetic waves in an appropriately chosen system of coordinates. \textcolor{black}{As recently remarked in \cite{ParabolicLozano}, the parabolic approximation is typically sufficiently accurate under the paraxial setting, but its accuracy in general deployments cannot be anticipated. The authors of \cite{9650519} offer a numerical study of the eDoF, and confirm the limitations of the paraxial approximation for application to general deployments. It was recently shown in \cite[Eq. 7]{Niyato_2025} that the approach proposed in \cite{9650519} is equivalent to that in \cite{Pizzo2022} and \cite{Miller}, since the eigenvalues of the holographic MIMO channel polarize to two values in the paraxial setup, as proved in \cite{Landau1975}. In \cite{Lozano_2024} and \cite{Hongliang_2025}, the authors generalize the approach in \cite{Pizzo2022} and \cite{Miller} in the presence of a reconfigurable intelligent surface. In \cite{Hongliang_2025}, the authors extend the approach in \cite{Dardari_2020} by  considering multiple receiving HoloSs that may not be located on the same plane, hence mutually blocking their fields of view with the transmitting HoloS. It is shown that the eDoF decrease in the presence of spatial blocking. Finally, recent results on the computation of the eDoF for holographic lines, which utilize the cut-set integral method, can be found in \cite{Strom_2022, Strom_2024, Kanatas_2025}}.

Motivated by these considerations, we introduce an analytical framework for estimating the eDoF of holographic MIMO beyond the paraxial setting, and provide closed-form expressions for the optimal communication waveforms for relevant wireless network deployments. To focus on the key aspects of the approach, the framework is elaborated for line-of-sight channels, which are receiving major attention from the research community especially in the context of (sub-)terahertz communications \cite{DoCPSLL21}, \cite{Bartoli2023}. The generalization in the presence of scattering objects is postponed to future research.

The contributions made by this paper are as follows:
\textcolor{black}{\begin{itemize}
\item We consider the eigenproblem in \cite{Miller}, which involves compact and self-adjoint operators, and provide a closed-form expression for the eDoF in non-paraxial settings. This is obtained by partitioning the largest HoloS into sub-HoloSs, and proving that the supports of the Fourier transforms of the kernels of the integral operators are finite and are almost disjoint in the wavenumber domain, provided that the sub-HoloSs are sufficiently small. 
\item The eDoF are formulated in a simple closed-form expression that depends on key system parameters, including the relative offset between the center-points of the HoloSs, and their relative rotation and tilt. This is obtained by proving that the typical method of analysis used in paraxial settings, which is based on a parabolic approximation of the wavefront in a local coordinate system, is equivalent to a quartic approximation of the wavefront in a general coordinate system.
\item By inspection of the obtained analytical formulation of the eDoF, we prove that (i) the eDoF correspond to an instance of Landau's second eigenvalue problem \cite{Landau1975}; (ii) the eigenvalues polarize asymptotically to multiple and distinct values, which are quantified analytically; and (iii) the eDoF depend explicitly on the approximation accuracy according to Kolmogorov's $n$-width criterion. This generalizes the analysis in the paraxial setting, where the eigenvalues polarize asymptotically to two values. 
\item We establish a simple relation between the cut-set integral and self-adjoint operator methods, by proving that the two approaches coincide when the approximation accuracy according to Kolmogorov's $n$-width criterion is arbitrarily small and is determined by the sub-HoloS whose eigenvalues polarize to the smallest non-zero value.
\item The proposed approach is applied to canonical network deployments, and analytical expressions for the optimal, according to Kolmogorov's $n$-width criterion, basis functions for data encoding and decoding, are computed. In the non-paraxial setting, it is proved that distinct communication waveforms are associated to different sub-HoloSs, which differ by a shift in the wavenumber domain and are spatially localized within each sub-HoloS.
\item The analytical findings are validated by numerically solving the exact formulation of the considered eigenproblem.
\end{itemize}
}

The remainder of the present paper is organized as follows. In Sec. II, we provide mathematical preliminaries on compact, self-adjoint operators and Kolmogorov's $n$-widths. In Sec. III, we introduce the system model and problem formulation. In Sec. IV, we summarize the proposed approach. In Sec. V, we provide the analytical framework to compute the eDoF. In Sec. VI, we analyze relevant network settings, and identify the optimal communication waveforms. In Sec. VII, we illustrate numerical results to validate the proposed approach in paraxial and non-paraxial settings. Conclusions are drawn in Sec. VIII.

\textit{Notation}: Bold lower and upper case letters represent vectors and matrices. Calligraphic letters denote sets. $(\cdot)$ is the scalar product. $(\cdot)^T$ and $(\cdot)^*$ are the transpose and conjugate transpose. $m(\mathcal{S})$ is the Lebesgue measure of $\mathcal{S}$. $j = \sqrt { - 1}$ is the imaginary unit. $\left\langle {f\left( {\bf{x}} \right),g\left( {\bf{x}} \right)} \right\rangle  = \int {f\left( {\bf{x}} \right){g^*}\left( {\bf{x}} \right)d{\bf{x}}}$ is the inner product and ${f\left( {\bf{x}} \right)} * {g\left( {\bf{x}} \right)}$ is the convolution of ${f\left( {\bf{x}} \right)}$ and ${g\left( {\bf{x}} \right)}$. $| {\cdot} |$ is the absolute value of scalars and functions. $\| {\cdot} \|$ is the ${\ell}^2$-norm of vectors. $\nabla= \left[\frac{\partial }{\partial x}, \frac{\partial }{\partial y}, \frac{\partial }{\partial z}\right]$ is the gradient. $\sinc(x)=\sin(\pi x)/(\pi x)$ is the sinc function. $\rect{(x)}$ is the boxcar function equal to one if $x \in (-1/2,1/2)$ and to zero elsewhere. $\det \mathbf{(\cdot)}$ is the determinant of matrices. $\mathds{1}_{\mathcal{P}}(\mathbf{x})$ is the indicator function, i.e., $\mathds{1}_{\mathcal{P}}(\mathbf{x}) = 1$ if $\mathbf{x} \in \mathcal{P}$ and zero elsewhere.

\vspace{-0.5cm}
\section{Mathematical Preliminaries}
\label{sec:preliminaries} \vspace{-0.25cm}
In this section, we summarize definitions and results from the mathematical literature, which are utilized in the rest of the paper. Specifically, this includes: (i) theory and results (notably the spectral theorem) pertaining to compact and self-adjoint operators over $n$-dimensional spaces; (ii) Kolmogorov's $n$-widths in approximation theory over $n$-dimensional spaces; and (iii) Landau's eigenvalue theorem  over $n$-dimensional spaces for a class of compact and self-adjoint operators.

\noindent \textbf{Definition 1.} Let $\mathbb{L}^2(\mathcal{S})$ be the Hilbert space of square-integrable complex-valued functions defined on the set $\mathcal{S} \subset	\mathbb{R}^n$, where $\mathbb{R}^n$ is the $n$-dimensional space of real numbers. An operator $K: \mathbb{L}^2(\mathcal{S}_i) \to \mathbb{L}^2(\mathcal{S}_o)$ with kernel ${k\left( {{\bf{x}},{\bf{y}}} \right)}$ for ${{\bf{x}}} \in \mathcal{S}_i$ and ${{\bf{y}}} \in \mathcal{S}_o$ applied to the function $\phi \left( {\bf{x}} \right) \in \mathbb{L}^2(\mathcal{S}_i)$ is defined as $\psi \left( {\bf{y}} \right) = \left( {K\phi } \right)\left( {\bf{y}} \right) = \int_{\mathcal{S}_i} {k\left( {{\bf{x}},{\bf{y}}} \right)\phi \left( {\bf{x}} \right)d{\bf{x}}} \in \mathbb{L}^2(\mathcal{S}_o)$ \cite{IntegralEq}.

\noindent \textbf{Definition 2.} Consider the operator $K$ in Def. 1. Assume that the kernel ${k\left( {{\bf{x}},{\bf{y}}} \right)}$ fulfills
$\int_{{{\mathcal{S}}_i}} {\int_{{{\mathcal{S}}_o}} {{{\left| {k\left( {{\bf{x}},{\bf{y}}} \right)} \right|}^2}d{\bf{x}}d{\bf{y}}} }  < \infty$, i.e., the kernel is a sufficiently well-behaved function. Then, the operator $K$ is bounded and compact \cite[Sec. 3.4]{IntegralEq}.

\noindent \textbf{Definition 3.} Consider the compact operator $K$ in Def. 2. A complex-valued scalar $\mu$ is an eigenvalue of $K$ if there exists a complex-valued function $\phi \left( {\bf{x}} \right) \in \mathbb{L}^2(\mathcal{S}_i)$ such that $\left( {K\phi } \right)\left( {\bf{x}} \right) = \mu \phi \left( {\bf{x}} \right)$. Also, $\phi \left( {\bf{x}} \right)$ is termed eigenfunction \cite[Def. 4.1]{IntegralEq}.

\noindent \textbf{Lemma 1.} Consider the compact operator $K$ in Def. 2. Let $\left\{ {{\mu _m}} \right\}$ be a (possibly infinite) sequence of distinct eigenvalues of $K$, as per Def. 3. Let $\left\{ {{\mu _m}} \right\}$ be ordered with non-increasing magnitude. Then, ${\mu _m} \to 0$ as $m \to \infty$ \cite[Th. 4.6]{IntegralEq}.

\noindent \textbf{Definition 4.} Consider the compact operator $K: \mathbb{L}^2(\mathcal{S}_i) \to \mathbb{L}^2(\mathcal{S}_o)$ in Def. 2.  The adjoint operator of $K$ is the compact operator $K_a: \mathbb{L}^2(\mathcal{S}_o) \to \mathbb{L}^2(\mathcal{S}_i)$ that fulfills the property $\left\langle {\left( {K\phi } \right)\left( {\bf{x}} \right),\psi \left( {\bf{x}} \right)} \right\rangle  = \left\langle {\phi \left( {\bf{x}} \right),\left( {{K_a}\psi } \right)\left( {\bf{x}} \right)} \right\rangle$ \cite[Def. 4.2]{IntegralEq}. By definition, the kernel, ${k_a\left( {{\bf{x}},{\bf{y}}} \right)}$, of $K_a$ is the complex conjugate of the kernel, ${k\left( {{\bf{x}},{\bf{y}}} \right)}$, of $K$, i.e., ${k_a\left( {{\bf{x}},{\bf{y}}} \right)} = {k^*\left( {{\bf{x}},{\bf{y}}} \right)}$.

\noindent \textbf{Definition 5.} Consider the compact operator $K: \mathbb{L}^2(\mathcal{S}) \to \mathbb{L}^2(\mathcal{S})$ and its adjoint $K_a \hspace{-0.055cm}: \mathbb{L}^2(\mathcal{S}) \hspace{-0.05cm}\to \hspace{-0.05cm} \mathbb{L}^2(\mathcal{S})$ in Def. 4. If $K\hspace{-0.05cm}=\hspace{-0.05cm}K_a$, $K$ is termed self-adjoint. Under the considered assumptions, $K$ is self-adjoint if and only if ${k\left( {{\bf{x}},{\bf{y}}} \right)}$ coincides with its complex conjugate, i.e., ${k\left( {{\bf{x}},{\bf{y}}} \right)} = {k^*\left( {{\bf{x}},{\bf{y}}} \right)} = {k_a\left( {{\bf{x}},{\bf{y}}} \right)}$ \cite[Sec. 4.4]{IntegralEq}. 

\noindent \textbf{Lemma 2.} Let $K\hspace{-0.075cm}: \hspace{-0.075cm} \mathbb{L}^2(\mathcal{S})  \hspace{-0.1cm} \to \hspace{-0.1cm} \mathbb{L}^2(\mathcal{S})$ be the self-adjoint operator in Def. 5. The eigenvalues of $K$ are real and the eigenfunctions of distinct eigenvalues are orthogonal \cite[Lemma 4.12]{IntegralEq}. 

\noindent \textbf{Lemma 3 (spectral theorem).} Let $K: \mathbb{L}^2(\mathcal{S}) \to \mathbb{L}^2(\mathcal{S})$ be a self-adjoint operator according to Def. 5. Then, there exists a, possibly finite, sequence $\left\{ {{\mu _m}} \right\}$ of real and non-zero eigenvalues of $K$ and a corresponding orthonormal sequence $\left\{ {{\phi _m}}\left( {\bf{x}} \right) \right\}$ of eigenfunctions, such that, for each $\phi \left( {\bf{x}} \right) \in \mathbb{L}^2(\mathcal{S})$, $\left( {K\phi } \right)\left( {\bf{x}} \right) = \sum\nolimits_{m = 1}^\infty  {{\mu _m}\left\langle {\phi \left( {\bf{x}} \right),{\phi _m}\left( {\bf{x}} \right)} \right\rangle {\phi _m}\left( {\bf{x}} \right)}$, with the sum being finite if the number of eigenvalues is finite. Consider the operator $\left( {K_N\phi } \right)\left( {\bf{x}} \right) = \sum\nolimits_{m = 1}^N {{\mu _n}\left\langle {\phi \left( {\bf{x}} \right),{\phi _m}\left( {\bf{x}} \right)} \right\rangle {\phi _m}\left( {\bf{x}} \right)}: \mathbb{L}^2(\mathcal{S}) \to \mathbb{L}^2(\mathcal{S})$. If the eigenvalues are infinitely many, then $\left\| {\left( {K\phi } \right)\left( {\bf{x}} \right) - \left( {{K_N}\phi } \right)\left( {\bf{x}} \right)} \right\| \to 0$ as $N \to \infty$ \cite[Th. 4.15]{IntegralEq}. 

\noindent \textbf{Definition 6.} Let $K: \mathbb{L}^2(\mathcal{S})  \to \mathbb{L}^2(\mathcal{S})$ be the self-adjoint operator in Def. 5. The eigenvalues $\mu_m$ of $K$ are bounded by the operator norm, i.e., $\mu_m \leq ||K||_{\mathrm{op}}$, which is defined by $||K||_{\mathrm{op}} = \sup \{||(K \phi) \left( {\bf{x}} \right)|| : ||\phi \left( {\bf{x}} \right)||\leq 1\}$ \cite[App. A]{IntegralEq}.

\noindent \textbf{Lemma 4.} Let $K: \mathbb{L}^2(\mathcal{S}) \to \mathbb{L}^2(\mathcal{S})$ and $\left\{ {{\phi _n}}\left( {\bf{x}} \right) \right\}$ be the operator and the eigenfunctions in Lemma 3. For any $\phi \left( {\bf{x}} \right) \in \mathbb{L}^2(\mathcal{S})$, there exists a $\phi_0 \left( {\bf{x}} \right) \in \mathbb{L}^2(\mathcal{S})$ so that $\phi \left( {\bf{x}} \right) = {\phi _0}\left( {\bf{x}} \right) +$ $\sum\nolimits_{m = 1}^\infty  {\left\langle {\phi \left( {\bf{x}} \right),{\phi _m}\left( {\bf{x}} \right)} \right\rangle {\phi _m}\left( {\bf{x}} \right)}$ and $\left( {K{\phi _0}} \right)\left( {\bf{x}} \right) \hspace{-0.05cm} = \hspace{-0.05cm} 0$ \cite[Cor. 4.16]{IntegralEq}. 

\noindent \textbf{Lemma 5.} Let $K: \mathbb{L}^2(\mathcal{S}) \to \mathbb{L}^2(\mathcal{S})$ and $\left\{ {{\phi _m}}\left( {\bf{x}} \right) \right\}$ be the operator and the eigenfunctions in Lemma 3. Then, $\left\{ {{\phi _m}}\left( {\bf{x}} \right) \right\}$ is a complete orthonormal basis in $\mathbb{L}^2(\mathcal{S})$ \cite[Cor. 4.17]{IntegralEq}.

Based on the definitions and lemmas summarized from \cite{IntegralEq}, we evince that the eigenfunctions of a compact and self-adjoint operator constitute a complete orthonormal basis. Thus, any function can be expressed as a (possibly infinite) linear combination of these eigenfunctions. Two important aspects remain, however, open: (i) the optimality of the eigenfunctions in terms of approximation accuracy provided by the truncated operator $K_N$ in Lemma 3 (given $N$); and (ii) the evaluation of the number, $N_{\rm{eDoF}}$, of effective eigenfunctions providing a non-negligible contribution. $N_{\rm{eDoF}}$ is referred to as the number of eDoF introduced in Sec. I \cite{Kolmogorov}. This is elaborated next.

\noindent \textbf{Definition 7.} Consider the compact operator $K: \mathbb{L}^2(\mathcal{S}_i)$ $\to \mathbb{L}^2(\mathcal{S}_o)$ in Def. 2 and denote $\text{X}_o = \mathbb{L}^2(\mathcal{S}_o)$. Let $\text{X}_k \subset \text{X}_o$ be the subspace of functions ${\psi_k \left( {\bf{y}} \right)}=\left( {K\phi } \right)\left( {\bf{y}} \right)$  determined by the kernel $K$ for ${\phi \left( {\bf{x}} \right)} \in \mathbb{L}^2(\mathcal{S}_i)$. The Kolmogorov $N$-width of $\text{X}_k$ in $\text{X}_o$ is defined as ${d_N}\left( {{{\rm{X}}_k};{{\rm{X}}_o}} \right) = {\inf _{{{\rm{X}}_{o,N}} \subset {{\rm{X}}_o}}}$ ${\sup _{_{{\psi _k}\left( {\bf{y}} \right) \in {{\rm{X}}_k}}}}{\inf _{{\psi _{o,N}}\left( {\bf{y}} \right) \in {{\rm{X}}_{o,N}}}}\left\| {{\psi _k}\left( {\bf{y}} \right) - {\psi _{o,N}}\left( {\bf{y}} \right)} \right\|$, where ${{{\rm{X}}_{o,N}}}$ is an $N$-dimensional subspace of ${{{\rm{X}}_{o}}}$ \cite[Ch. 2, Def. 1.1]{Kolmogorov}. Thus, ${D_{{{\rm{X}}_{o,N}}}}\left( {{{\rm{X}}_k}} \right) = {\sup _{_{{\psi _k}\left( {\bf{y}} \right) \in {{\rm{X}}_k}}}}{\inf _{{\psi _{o,N}}\left( {\bf{y}} \right) \in {{\rm{X}}_{o,N}}}}\left\| {{\psi _k}\left( {\bf{y}} \right) - {\psi _{o,N}}\left( {\bf{y}} \right)} \right\|$ is a measure of how well the worst function in ${{{\rm{X}}_k}}$ is approximated by ${{{\rm{X}}_{o,N}}}$. ${d_N}\left( {{{\rm{X}}_k};{{\rm{X}}_o}} \right)$ is the smallest ${D_{{{\rm{X}}_{o,N}}}}\left( {{{\rm{X}}_k}} \right)$ over all possible $N$-dimensional subspaces ${{{\rm{X}}_{o,N}}} \subset {{{\rm{X}}_{o}}}$.

\noindent \textbf{Lemma 6.} Consider the compact operator $K: \mathbb{L}^2(\mathcal{S}_i) \to \mathbb{L}^2(\mathcal{S}_o)$ in Def. 2 and its adjoint $K_a: \mathbb{L}^2(\mathcal{S}_o) \to \mathbb{L}^2(\mathcal{S}_i)$ in Def. 4. Define the operators $K_i=K K_a: \mathbb{L}^2(\mathcal{S}_i) \to \mathbb{L}^2(\mathcal{S}_i)$ and $K_o=K_a K: \mathbb{L}^2(\mathcal{S}_o) \to \mathbb{L}^2(\mathcal{S}_o)$ whose kernels are ${k_i}\left( {{{\bf{x}}_1},{{\bf{x}}_2}} \right) = \int\nolimits_{{{\mathcal{S}}_o}} {k\left( {{{\bf{x}}_1},{\bf{y}}} \right){k^*}\left( {{{\bf{x}}_2},{\bf{y}}} \right)d{\bf{y}}}$ and ${k_o}\left( {{{\bf{y}}_1},{{\bf{y}}_2}} \right) = \int\nolimits_{{{\mathcal{S}}_i}} {k\left( {{\bf{x}},{{\bf{y}}_1}} \right){k^*}\left( {{\bf{x}},{{\bf{y}}_2}} \right)d{\bf{x}} }$, respectively. The operators $K_i$ and $K_o$ are compact, self-adjoint, and non-negative, i.e., their eigenvalues are not negative \cite[Ch. 4, pg. 65]{Kolmogorov}.

\noindent \textbf{Definition 8.} Consider the compact operator $K$ and the compact, self-adjoint, non-negative operator $K_o$ with kernel $k_o$ defined in Lemma 6. The s-values of $K$ are defined as ${s_m}\left( K \right) = \sqrt {{\mu _m}\left( {{K_o}} \right)}$ for $m=1, 2, \ldots$, with $\mu _m$ being the eigenvalues of the operator $K_o$ with kernel $k_o$ \cite[pg. 65]{Kolmogorov}.

\noindent \textbf{Lemma 7.} Consider the compact operators $K$, its adjoint $K_a$, and the compact, self-adjoint, and non-negative operators $K_i$ and $K_o$ in Lemma 6. Let $\left\{ {{\mu _m}} \right\}$ and $\left\{ {{\phi _m}}\left( {\bf{x}} \right) \right\}$ be the sequence of non-zero and positive eigenvalues and orthonormal eigenfunctions of the operator $K_o$ with kernel $k_o$ defined in Lemma 6, respectively. Then, ${d_N}\left( {{{\rm{X}}_k};{{\rm{X}}_o}} \right) = {s_{N + 1}}\left( K \right)$, the functions $\left\{ {{\phi _m}}\left( {\bf{x}} \right) \right\}$ constitute a complete orthonormal basis in ${{\rm{X}}_{o,N}}$, and the subspace ${{\rm{X}}_{o,N}}$ constituted by all the possible linear combinations of $\left\{ {{\phi _m}}\left( {\bf{x}} \right) \right\}$, i.e., ${{\rm{X}}_{o,N}} = {\rm{span}}\left\{ {{\phi _1}\left( {\bf{x}} \right),{\phi _2}\left( {\bf{x}} \right), \ldots ,{\phi _N}\left( {\bf{x}} \right)} \right\}$ minimizes the Kolmogorov $N$-width ${d_N}\left( {{{\rm{X}}_k};{{\rm{X}}_o}} \right)$ \cite[Ch. 4, Th. 2.2]{Kolmogorov}. In simple terms, given a compact and self-adjoint operator $K$ from an input space to an output space and its adjoint $K_a$, the optimal basis functions, in terms of approximation accuracy according to Kolmogorov's definition, coincide with the eigenfunctions of the compact, self-adjoint, and non-negative operator given by $K_a K$ for the output space (and by $K K_a$ for the input space). Also, the approximation error by considering $N$ basis functions coincides with the square root of the $(N+1)$th eigenvalue of $K_a K$ for the output space (and of $K K_a$ for the input space).

\noindent \textbf{Definition 9.} Consider the compact operator $K$ and the compact, self-adjoint, non-negative operator $K_o$ defined in Lemma 6. The number of eDoF, ${{{N}}_{{\rm{eDoF}}}}$, corresponds to the minimum dimension of the approximating subspace such that the Kolmogorov $N$-width defined in Def. 7 is no greater than $\gamma$, i.e., the level of approximation accuracy is $\gamma$. By using the notation in Def. 7, it holds ${{{N}}_{{\rm{eDoF}}}}\left( \gamma  \right) = \min \left\{ {N:{d_N^2}\left( {{{\rm{X}}_k};{{\rm{X}}_o}} \right) \le \gamma } \right\}$ \cite{FranceschettiBook}.

\noindent \textbf{Definition 10.} Consider the compact operator $K$ and the compact, self-adjoint, and non-negative operator $K_o$ in Lemma 6 with $s_n$ being the s-values of $K$ defined in Def. 8. According to Lemma 7 and Def. 8, the eDoF, given the approximation accuracy $\gamma$, are ${{{N}}_{{\rm{eDoF}}}}\left( \gamma  \right) = \min \left\{ {N:{s_{N+1}^2}\left( K \right) = {\mu_{N+1}}\left( K_o \right) \le \gamma } \right\}$ \cite{FranceschettiBook}. From Def. 8, ${{{N}}_{{\rm{eDoF}}}}\left( \gamma  \right)$ can be interpreted as the number of eigenvalues of $K_o$ with a magnitude no smaller than $\gamma$.

\textcolor{black}{\noindent \textbf{Lemma 8.} Consider a function $f\left( {\bf{x}} \right)$ in the Hilbert space of square-integrable complex-valued functions, and its Fourier transform $F\left( {\boldsymbol{\nu}} \right) = \left( {Tf} \right)\left( {\boldsymbol{\nu}} \right)$, where $T$ is the Fourier operator. Consider the self-adjoint Hermitian operator $\left(A_r f\right)(\mathbf{x}) = \int \mathds{1}_{r\mathcal{Q}}(\mathbf{x}) \mathds{1}_{r\mathcal{Q}}(\mathbf{y}) h(\mathbf{x} - \mathbf{y}) f(\mathbf{y}) \, d\mathbf{y}$, with $\mathcal{Q}$ being a set in $\mathbb{R}^n$ and $P(\boldsymbol{\nu}) = \left(T h\right)(\boldsymbol{\nu})$ is real. Consider the set $\mathcal{S}_\gamma = \{\boldsymbol{\nu} : P(\boldsymbol{\nu})\geq \gamma\}$. From Def. 10, let ${N_{\mathrm{eDoF}}}\left( \gamma  \right)$ be the number of eigenvalues of $A_r$ no smaller than $\gamma$. Then, for any $\gamma > 0$, ${\lim _{r \to \infty }}{r^{ - n}}{{N_{\mathrm{eDoF}}}\left( \gamma  \right)} = {\left( {2\pi } \right)^{ - n}}m\left( {\mathcal{Q}} \right)m\left( {\mathcal{S}_\gamma} \right) $ \cite[Th. 2]{Landau1975}}.

\textcolor{black}{\noindent \textbf{Lemma 9.} Consider the Hermitian operator $A_r$ in Lemma 8. Assume that $\left( {T h} \right)\left( {\boldsymbol{\nu}} \right) = \mathds{1}_{\mathcal{P}} (\boldsymbol{\nu})$, i.e., $h(\mathbf{x})$ is an ideal pass-band filter with Fourier transform equal to $||A_r||_{\mathrm{op}}$ if $\boldsymbol{\nu} \in \mathcal{P}$ and to zero elsewhere. Then, for any $0 < \gamma \le ||A_r||_{\mathrm{op}}$, ${\lim _{r \to \infty }}{r^{ - n}}{{{N_{\mathrm{eDoF}}}\left( \gamma  \right)}} = {\left( {2\pi } \right)^{ - n}}m\left( {\mathcal{Q}} \right)m\left( {\mathcal{P}} \right) = {{N_{\mathrm{eDoF}}}}$ \cite[Th. 1]{Landau1975}. In simple terms, the eigenvalues of $A_r$ polarize asymptotically so that the number of leading, i.e., those nearly equal to the operator norm $||A_r||_{\mathrm{op}}$, eigenvalues is ${{N_{\mathrm{eDoF}}}}$ and the others are nearly equal to zero. In the asymptotic regime ${\lim _{r \to \infty }}$, in addition, the eDoF are independent of the approximation accuracy $\gamma$. In one-dimensional spaces ($n=1$), the width of the transition region between the leading (nearly $||A_r||_{\mathrm{op}}$) and the nearly zero eigenvalues is known \cite[Eq. (2.132)]{FranceschettiBook}, \cite{Landau1975}}.

\vspace{-0.075cm}
\section{Problem Formulation}
\label{sec:formulation}

We consider a transmitting and a receiving HoloS located at $\mathbf{r}_{Tx} = \left(x_{Tx}, y_{Tx}, z_{Tx}\right) \in \mathcal{S}_{Tx}$ and $\mathbf{r}_{Rx} = \left(x_{Rx}, y_{Rx}, z_{Rx}\right)\in \mathcal{S}_{Rx}$, with surface areas $A_{Tx} = m(\mathcal{S}_{Tx})$ and $A_{Rx} = m(\mathcal{S}_{Rx})$. The surface current density at $\mathbf{r}_{Tx} \in \mathcal{S}_T$ is denoted by $\mathbf{J}(\mathbf{r}_{Tx})$. We assume the transmission of a monochromatic electromagnetic wave at frequency $f_0$. By Maxwell's equations, the electric field observed at $\mathbf{r}_{Rx} \in \mathcal{S}_{Rx}$ is \cite[Eq. (3.3)]{Piestun:00} 
\begin{equation} 
\mathbf{E}(\mathbf{r}_{Rx})=\int_{\mathcal{S}_{Tx}} \bar{\bar{\mathbf{G}}}(\mathbf{r}_{Rx} - \mathbf{r}_{Tx}) \mathbf{J}(\mathbf{r}_{Tx}) d\mathbf{r}_{Tx} \label{Eq:Efield} 
\end{equation}
where $\bar{\bar{\mathbf{G}}}(\mathbf{r}_{Rx} - \mathbf{r}_{Tx})$ is the dyadic Green function. 

Equation \eqref{Eq:Efield} can be applied to any linear, in general shift-variant, system. Specifically, it can be applied for modeling the propagation of electromagnetic waves between transmitting and receiving domains in the presence of material bodies, e.g., blocking obstacles or scattering surfaces \cite{Piestun:00}. To highlight the key aspects of the proposed approach, we focus our attention on the free space scenario, which is receiving major renewed attention lately \cite{DoCPSLL21}, \cite{Bartoli2023}. In free space, the dyadic Green function $\bar{\bar{\mathbf{G}}}(\mathbf{r})$ can be expressed as \cite[Eq. 1.3.49]{chew1999waves} 
\begin{equation}
\label{eq:GDyadic}
\bar{\bar{\mathbf{G}}}(\mathbf{r}) = \frac{j\omega_0\mu}{4\pi}
\left(\bar{\bar{{\mathbf{I}}}} + \frac{1}{\kappa_0^2}\nabla\nabla \right) \frac{\exp(-j\kappa_0\|\mathbf{r}\|)}{\|\mathbf{r}\|} 
\end{equation}
where $\mathbf{r} = \mathbf{r}_{Rx} - \mathbf{r}_{Tx}$, $\kappa_0 = 2\pi / \lambda$ is the wavenumber, $\lambda = c/f_0$ is the wavelength, $\omega_0 = 2\pi f_0$ is the angular frequency, $\mu$ is the magnetic permeability, and $\bar{\bar{{\bf I}}} = \hat{\mathbf{u}}_x\hat{\mathbf{u}}_x^* + \hat{\mathbf{u}}_y\hat{\mathbf{u}}_y^* + \hat{\mathbf{u}}_z\hat{\mathbf{u}}_z^*$ is the identity dyadic tensor with $\hat{\mathbf{u}}_a$ denoting the unit-norm vector in the direction of the $a$-axis for $a \in \left\{ {x,y,z} \right\}$.

Usually, wireless communication systems do not operate in the reactive near-field, and we can assume $\|\mathbf{r}\| \gg \lambda$ \cite{MDR_EuCAP}. Under this assumption, $\bar{\bar{{\mathbf{G}}}}$ can be approximated as \cite[Eq. (3)]{Dardari_2020} 
\begin{equation}
\bar{\bar{{\mathbf{G}}}}(\mathbf{r}) \approx  G(\|\mathbf{r}\|)\left(\bar{\bar{{\mathbf{I}}}} - \hat{\mathbf{r}}\hat{\mathbf{r}}^* \right), \quad G(\|\mathbf{r}\|) = \frac{j\eta \exp(-j\kappa_0\|\mathbf{r}\|)}{2\lambda \|\mathbf{r}\|} 
\label{Eq:GreenRadiative}
\end{equation}
where $\hat{\mathbf{r}} = \mathbf{r}/\|\mathbf{r}\|$ and $\eta  = \sqrt {{\mu  \mathord{\left/
{\vphantom {\mu  \varepsilon }} \right. \kern-\nulldelimiterspace} \varepsilon }}$ with $\varepsilon$ the electric permittivity.

With no loss of generality, we assume that the considered communication system is probed by exciting one polarization of $\mathbf{J}(\mathbf{r}_{Tx})$ at a time. Considering the polarization along the $i$-axis, i.e., $\mathbf{J}(\mathbf{r}_{Tx}) = J_i(\mathbf{r}_{Tx}) \hat{\mathbf{u}}_i$ for $i \in \left\{ {x,y,z} \right\}$, and denoting $\mathbf{r} = x_{TR} \hat{\mathbf{u}}_x+ y_{TR} \hat{\mathbf{u}}_y + z_{TR} \hat{\mathbf{u}}_z$ with $a_{TR} = a_{Rx} - a_{Tx}$ for $a \in \left\{ {x,y,z} \right\}$, the received electric field reduces to 
\begin{align}
\mathbf{E}_i(\mathbf{r}_{Rx}) & \approx \int_{\mathcal{S}_{Tx}} G(\|\mathbf{r}\|)\left[ \bar{\bar{{\mathbf{I}}}} \hat{\mathbf{u}}_i  - \frac{\mathbf{r}\left( \mathbf{r}^* \hat{\mathbf{u}}_i \right)}{\|\mathbf{r}\|^2}\right] J_i(\mathbf{r}_{Tx}) d\mathbf{r}_{Tx} \\
&=\int_{\mathcal{S}_{Tx}} G(\|\mathbf{r}\|)\left(\hat{\mathbf{u}}_i - \frac{i_{TR}}{\|\mathbf{r}\|^2}{\mathbf{r}}\right) J_i(\mathbf{r}_{Tx}) d\mathbf{r}_{Tx} 
\end{align}
\textcolor{black}{where $i_{TR} = \mathbf{r}^* \hat{\mathbf{u}}_i$}.

Explicitly, the three Cartesian components of $\mathbf{E}_i(\mathbf{r}_{Rx}) = E_{i,x}(\mathbf{r}_{Rx}) \hat{\mathbf{u}}_x + E_{i,y}(\mathbf{r}_{Rx}) \hat{\mathbf{u}}_y + E_{i,z}(\mathbf{r}_{Rx}) \hat{\mathbf{u}}_z$ can be written as 
\begin{equation} \label{eq:E_x_gen}
E_{i,o}(\mathbf{r}_{Rx}) = \int_{\mathcal{S}_{Tx}} G(\|\mathbf{r}\|) {e_{i,o}}\left( {{{\bf{r}}_{Tx}},{{\bf{r}}_{Rx}}} \right) J_i(\mathbf{r}_{Tx}) d\mathbf{r}_{Tx} 
\end{equation}
where ${e_{i,o}}\left( {{{\bf{r}}_{Tx}},{{\bf{r}}_{Rx}}} \right) = {{{\bf{\hat u}}}_i} \cdot {{{\bf{\hat u}}}_o} - \left( {{i_{TR}}/{{\| {\bf{r}} \|}^2}} \right){\bf{r}} \cdot {{{\bf{\hat u}}}_o}$ accounts for the coupling between the $i$th component of the surface current density and the $o$th component of the received electric field.

From Def. 1, we evince that \eqref{eq:E_x_gen} is an operator $G_{i,o}:$ $\mathbb{L}^2(\mathcal{S}_{Tx}) \to \mathbb{L}^2(\mathcal{S}_{Rx})$, whose  kernel is $g_{i,o}\left( {{{\bf{r}}_{Tx}},{{\bf{r}}_{Rx}}} \right) = G\left( {\| {\bf{r}} \|} \right){e_{i,o}}\left( {{{\bf{r}}_{Tx}},{{\bf{r}}_{Rx}}} \right)$. Based on Def. 2, $G_{i,o}$ is compact, since $\int_{{{\mathcal{S}}_{Tx}}} {\int_{{{\mathcal{S}}_{Rx}}} {{{\left| {g_{i,o}\left( {{{\bf{r}}_{Tx}},{{\bf{r}}_{Rx}}} \right)} \right|}^2}d{{{\bf{r}}_{Tx}}}d{{{\bf{r}}_{Rx}}}} }  < \infty$ by virtue of the considered modeling assumptions. Based on Lemma 7, the eDoF and the optimal pair of communication waveforms at the transmitting (encoding) and receiving (decoding) HoloSs are solutions, respectively, of the following eigenproblems: 
\begin{equation}
\label{eq:eigenproblemIn}
\mu_{m}\phi_{m}(\mathbf{r}_{Tx}) = \int_{{\mathcal{S}}_{Tx}} G_{Tx}(\mathbf{r}_{Tx},\mathbf{r}_{Tx}') \phi_{m}(\mathbf{r}_{Tx}')d\mathbf{r}_{Tx}' 
\end{equation}
\begin{equation}
\label{eq:eigenproblemOut}
\mu_{m}\psi_{m}(\mathbf{r}_{Rx}) = \int_{{\mathcal{S}}_{Rx}} G_{Rx}(\mathbf{r}_{Rx},\mathbf{r}_{Rx}') \psi_{m}(\mathbf{r}_{Rx}')d\mathbf{r}_{Rx}' 
\end{equation}
where $\mu_{m}$ are the real-valued and positive eigenvalues for $m=1, 2, \ldots, {{{N}}_{{\rm{eDoF}}}}\left( \cdot  \right)$ according to Defs. 3 and 8, $\left\{ {{\phi _{m}}} \right\}$ and $\left\{ {{\psi _{m}}} \right\}$ are the corresponding pair of communication waveforms at the transmitting and receiving HoloSs, respectively, and the kernels $G_{Tx}(\mathbf{r}_{Tx},\mathbf{r}_{Tx}')$ and $G_{Rx}(\mathbf{r}_{Rx},\mathbf{r}_{Rx}')$ are defined as 
\begin{align}
\label{eq:selfAdjoint}
G_{Tx}(\mathbf{r}_{Tx},\mathbf{r}_{Tx}') & =\int_{{\mathcal{S}}_{Rx}} \hspace{-0.2cm} g_{i,o}^*(\mathbf{r}_{Tx},\mathbf{r}_{Rx}) g_{i,o}(\mathbf{r}_{Tx}',\mathbf{r}_{Rx})d\mathbf{r}_{Rx}\\
 \hspace{-0.2cm}G_{Rx}(\mathbf{r}_{Rx},\mathbf{r}_{Rx}') &=\int_{{\mathcal{S}}_{Tx}}  \hspace{-0.2cm} g_{i,o}^*(\mathbf{r}_{Tx},\mathbf{r}_{Rx})g_{i,o}(\mathbf{r}_{Tx},\mathbf{r}_{Rx}')d\mathbf{r}_{Tx} \,. 
\end{align}

Based on Lemma 6, $G_{Tx}(\mathbf{r}_{Tx},\mathbf{r}_{Tx}')$ and $G_{Rx}(\mathbf{r}_{Rx},\mathbf{r}_{Rx}')$ are compact, self-adjoint, and non-negative kernels. Our objective is to provide analytical expressions for ${{{N}}_{{\rm{eDoF}}}}\left( \cdot  \right)$, $\left\{ {{\mu _{m}}} \right\}$, $\left\{ {{\phi _{m}}} \right\}$, $\left\{ {{\psi _{m}}} \right\}$. With no loss of generality, we consider \eqref{eq:eigenproblemIn}.

\begin{figure}[!t]
    \centering \includegraphics[width=0.55\columnwidth]{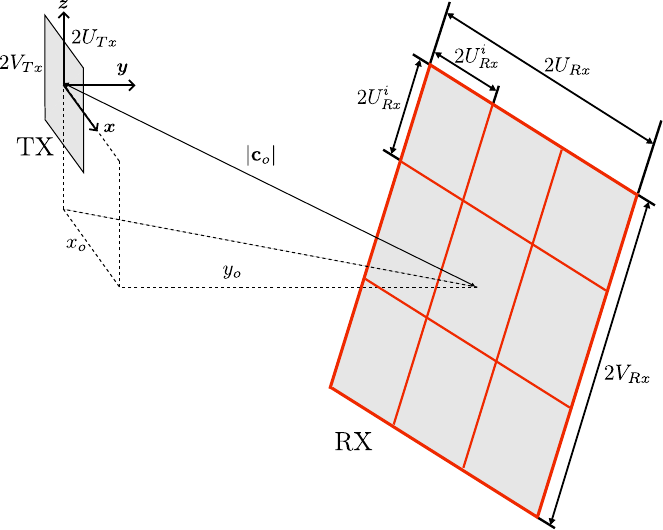}
    \vspace{-0.25cm}
    \caption{Network deployment and partitioning in a non-paraxial setting.}
    \label{fig:system_model} \vspace{-0.5cm}
\end{figure}

\vspace{-0.25cm}
\section{Proposed Approach}
\label{sec:approach} \vspace{-0.15cm}
The center-points of ${{\mathcal{S}}_{Tx}}$ and ${{\mathcal{S}}_{Rx}}$ are denoted by ${{\bf{c}}_{Tx}} = \left( {{x_{Tc}},{y_{Tc}},{z_{Tc}}} \right)$ and ${{\bf{c}}_{Rx}} = \left( {{x_{Rc}},{y_{Rc}},{z_{Rc}}} \right)$, and their distance is $\| {{{\bf{c}}_{Rx}} - {{\bf{c}}_{Tx}}} \|$. As shown in Fig. \ref{fig:system_model}, the proposed approach can be applied to general network deployments. For example, ${{\mathcal{S}}_{Tx}}$ and ${{\mathcal{S}}_{Rx}}$ are not necessarily parallel to one another, and an arbitrary tilt (denoted by the angle $\beta$) with respect to the $z$-axis, and an arbitrary rotation (denoted by the angle $\alpha$) on the $xy$-plane with respect to the $x$-axis are admissible. Notably, the proposed approach can be applied to non-paraxial settings, where the sides of the receiving HoloS are comparable with the transmission distance. The approach can, however, be applied under three assumptions: (i) by virtue of the approximation in \eqref{Eq:GreenRadiative}, ${{\mathcal{S}}_{Tx}}$ and ${{\mathcal{S}}_{Rx}}$ cannot be located in the reactive near-field of one another; (ii) the validity of the approximations applied to the amplitude and phase of the Green functions in $G_{Tx}(\mathbf{r}_{Tx},\mathbf{r}_{Tx}')$, as detailed next; and  (iii) one of the two HoloS (${{\mathcal{S}}_{Tx}}$ in Fig. \ref{fig:system_model}) is small compared to the other. \textcolor{black}{The motivation for this latter assumption is discussed in Appendix J. In a nutshell, Lemma 8 cannot be applied otherwise, since the resulting eigenproblem would involve a self-adjoint operator that cannot be formulated as a convolution-type integral}.

\vspace{-0.25cm}
\subsection{Non-paraxial Propagation Model}
\textcolor{black}{We commence by introducing the signal model in a general non-paraxial setting. In the next sub-section, the paraxial setting is discussed to better position the proposed approach with respect to the prior art overviewed in Sec. I}.

\textcolor{black}{Let us assume that the sides of ${{\mathcal{S}}_{Rx}}$ are comparable with the distance $\| {{{\bf{c}}_{Rx}} - {{\bf{c}}_{Tx}}} \|$ between the center-points ${{\bf{c}}_{Tx}}$ and ${{\bf{c}}_{Rx}}$. The proposed approach consists of partitioning ${{\mathcal{S}}_{Rx}}$ into smaller sub-surfaces (sub-HoloSs), so that the link between the transmitting HoloS and each receiving sub-HoloS fulfills the paraxial setting. To elaborate, ${{\mathcal{S}}_{Rx}}$ is partitioned into $N_r$ sub-HoloSs, the $n$th sub-HoloS is centered at $\mathbf{c}_{Rx}^n$, and it is denoted by $\mathcal{S}_{Rx}^n$. To ensure that each sub-HoloS fulfills the paraxial setting, the sides of ${{\mathcal{S}}_{Tx}}$ and $\mathcal{S}_{Rx}^n$ need to be much smaller than the distance $\| {{{\bf{c}}_{Rx}^n} - {{\bf{c}}_{Tx}}} \|$ between the center-points ${{\bf{c}}_{Tx}}$ and ${{\bf{c}}_{Rx}^n}$ (see \eqref{eq:validityModelParax} and \eqref{eq:validityModelNoParax} in Sec. V). The $N_r$ sub-HoloSs constitute a partition of ${{\mathcal{S}}_{Rx}}$, i.e., they are co-planar, disjoint, i.e., $\mathcal{S}_{Rx}^n \cap \mathcal{S}_{Rx}^m = \emptyset$ for any $n \not= m$, and $\bigcup_{n=1}^{N_r} \mathcal{S}_{Rx}^n = \mathcal{S}_{Rx}$}.

\textcolor{black}{By applying the partitioning, \eqref{eq:selfAdjoint} can be written as
\begin{equation}
\label{eq:partitioning}
G_{Tx}(\mathbf{r}_{Tx},\mathbf{r}_{Tx}')=\sum\nolimits_{n=1}^{N_r} G_{Tx}^n(\mathbf{r}_{Tx},\mathbf{r}_{Tx}')
\end{equation}
where
\begin{equation}
\label{eq:selfAdjoint_sub}
    G_{Tx}^n(\mathbf{r}_{Tx},\mathbf{r}_{Tx}') = \hspace{-0.1cm} \int_{{\mathcal{S}}_{Rx}^n} \hspace{-0.2cm} g_{i,o}^*(\mathbf{r}_{Tx},\mathbf{r}_{Rx}) g_{i,o}(\mathbf{r}_{Tx}',\mathbf{r}_{Rx}) d\mathbf{r}_{Rx} \, . 
\end{equation} }

\textcolor{black}{From \eqref{eq:partitioning}, the kernel of the considered operator can be formulated as the sum of the kernels of the sub-HoloSs. It is important to emphasize, however, that this does not imply either that the eDoF associated with the kernel $G_{Tx}^n(\mathbf{r}_{Tx},\mathbf{r}_{Tx}')$ are the sum of the eDoF associated with the kernels $G_{Tx}^n(\mathbf{r}_{Tx},\mathbf{r}_{Tx}')$ or that the communication waveforms can be optimized independently for each sub-HoloS. These aspects are, on the other hand, two contributions of this paper and are studied next}.

\textcolor{black}{The proposed approach of analysis is based on the modeling assumption that the sides of each sub-HoloS are small enough compared with the distance $\|\mathbf{c}_{Rx}^n - \mathbf{c}_{Tx}\|$. In mathematical terms, this condition is quantified in \eqref{eq:validityModelParax} and \eqref{eq:validityModelNoParax} (see Sec. V). In this case, ${{\mathcal{S}}_{Tx}}$ and ${{\mathcal{S}}_{Rx}^n}$ operate in the paraxial setting, and the kernel $G_{Tx}^n(\mathbf{r}_{Tx},\mathbf{r}_{Tx}')$  can be approximated accordingly}.

\textcolor{black}{In the literature, the typical approach utilized to approximate the self-adjoint kernel corresponding to two HoloSs in the paraxial setting consists of the following procedure \cite{Miller, ParabolicLozano, Pizzo2022, Fettweis}: (i) the coordinate system is chosen so that the center-points of both HoloSs are aligned along one of the coordinate axis; (ii) the HoloSs are projected onto planes perpendicular to the direction of the common coordinate axis; and (iii) a parabolic approximation of the wavefront is applied in the resulting coordinate system. If one applies this approach to each HoloS that constitutes the receiving HoloS, a different coordinate system and hence a different projection need to be applied  to each HoloS. The resulting approximations, therefore, cannot be directly combined together, as they are applied to different coordinate systems. To study the properties of the kernel $G_{Tx}^n(\mathbf{r}_{Tx},\mathbf{r}_{Tx}')$, it would be necessary to first undo the projections in a common coordinate system}.

\textcolor{black}{Next, we show that this conventional approach can be avoided by replacing a parabolic approximation of the wavefront with a quartic approximation, considering a single coordinate system and avoiding projections. The proposed approach offers a direct method of analysis and simple analytical expressions for the eDoF, as a function of relevant system parameters. In the paraxial setting, we prove that the conventional and proposed methods are equivalent, corroborating the correctness of the proposed method besides its simplicity}.

\vspace{-0.25cm}
\subsection{Paraxial Model in an Arbitrary Coordinate System}
Before analyzing the non-paraxial setting, we introduce a paraxial model for the link between ${{\mathcal{S}}_{Tx}}$ and ${{\mathcal{S}}_{Rx}^n}$ which is valid for any coordinate system. We aim to find an approximation for $g_{i,o}(\mathbf{r}_{Tx},\mathbf{r}_{Rx})$ for any $\mathbf{r}_{Tx} \in \mathcal{S}_{Tx}$ and $\mathbf{r}_{Rx} \in \mathcal{S}_{Rx}^n$.  For ease of writing, we use the notation $g_{i,o}^n(\mathbf{r}_{Tx},\mathbf{r}_{Rx})$ to indicate $\mathbf{r}_{Rx} \in \mathcal{S}_{Rx}^n$. To this end, $g_{i,o}(\mathbf{r}_{Tx},\mathbf{r}_{Rx})$ is rewritten as 
\begin{equation} \label{Eq:ExactKernel}
g_{i,o}(\mathbf{r}_{Tx},\mathbf{r}_{Rx}) = \bar{g}_{i,o}(\mathbf{r}_{Tx}, \mathbf{r}_{Rx}) \exp\left(-j\kappa_0\|\mathbf{r}_{Rx} - \mathbf{r}_{Tx}\| \right)
\end{equation}
where $\bar{g}_{i,o}(\mathbf{r}_{Tx}, \mathbf{r}_{Rx}) = ({j\eta} {e_{i,o}(\mathbf{r}_{Tx}, \mathbf{r}_{Rx})})/({2\lambda}{\|\mathbf{r}_{Rx} - \mathbf{r}_{Tx}\|})^{-1}$.

We consider two different approximations for the amplitude and phase of $g_{i,o} (\mathbf{r}_{Tx}, \mathbf{r}_{Rx})$.

\noindent {\bf{Amplitude}}: As for the term $\bar{g}_{i,o}(\mathbf{r}_{Tx}, \mathbf{r}_{Rx})$, we use the typical approximation $\bar{g}_{i,o}(\mathbf{r}_{Tx}, \mathbf{r}_{Rx}) \approx \bar{g}_{i,o}^n = \bar{g}_{i,o}(\mathbf{c}_{Tx}, \mathbf{c}_{Rx}^n)$. This implies that ${{\mathcal{S}}_{Tx}}$ and ${{\mathcal{S}}_{Rx}^n}$ are located in the radiative near-field (Fresnel region) of one another, but their sizes cannot be too large as compared with the distance $\| {{{\bf{c}}_{Rx}^n} - {{\bf{c}}_{Tx}}} \|$. 

\noindent {\bf{Phase}}: As for the phase term  $\exp \left( { - j{\kappa_0}\| \mathbf{r}_{Rx} - \mathbf{r}_{Tx} \|} \right)$, we use a quartic approximation that can be applied regardless of the coordinate system being considered. To elaborate, we define ${a_{Tx}} = {a_{Tc}} + \Delta {a_{Tx}}$ and $a_{Rx} = a_{Rc}^n + \Delta a_{Rx}^n$ for $a \in \left\{ {x,y,z}\right\}$, where $\Delta a_{Tx}$ and $\Delta a_{Rx}^n$ are the local coordinates with respect to the center-points ${a_{Tc}}$ and ${a_{Rc}}$. Also, we define $\mathbf{c}_o^n = (x_o^n, y_o^n, z_o^n) = \mathbf{c}_{Rc}^n - \mathbf{c}_{Tc}$ with $a_o^n = a_{Rc}^n - a_{Tc}$. Denoting $\sum\nolimits_a {}  = \sum\nolimits_{a \in \left\{ {x,y,z} \right\}} {}$, the distance ${\| \mathbf{r}_{Rx}^n - \mathbf{r}_{Tx} \|}$ simplifies to 
\begin{align} 
 \| \mathbf{r}_{Rx}^n - \mathbf{r}_{Tx} \| &= \sqrt {\sum\nolimits_{a} {{{\left( {{a_{Rx}^n} - {a_{Tx}}} \right)}^2}} } \label{Eq:WavefrontApprox_1} \\
& \hspace{-1.0cm}= \sqrt{\sum\nolimits_a \left[ \left( a_{Rc}^n - a_{Tc} \right) + \left( \Delta a_{Rx}^n - \Delta a_{Tx} \right) \right]^2} \nonumber \\
& \hspace{-1.0cm} =\| \mathbf{c}_{o}^n  \| \sqrt{1 + \frac{\rho^n(\mathbf{r}_{Tx}, \mathbf{r}_{Rx})}{\left| \mathbf{c}_{o}^n \right|^2}} \nonumber \\
& \hspace{-1.0cm} \mathop  \approx \limits^{\left( a \right)} \| \mathbf{c}_{o}^n \| \left[ 1 + \frac{ \rho^n(\mathbf{r}_{Tx}, \mathbf{r}_{Rx})}{2 \left| \mathbf{c}_{o}^n \right|^2} 
- \frac{ (\rho^n(\mathbf{r}_{Tx}, \mathbf{r}_{Rx}))^2}{8 \left| \mathbf{c}_{o}^n \right|^4} \right]  \label{Eq:WavefrontApprox_2} 
\end{align}
where $\rho^n \left( {{{\bf{r}}_{Tx}},{{\bf{r}}_{Rx}}} \right)$ is defined, for $\mathbf{r}_{Rx} \in \mathcal{S}_{Rx}^n$, as \vspace{-0.1cm} 
\begin{align} 
\rho^n(\mathbf{r}_{Tx}, \mathbf{r}_{Rx}) &= 2 \sum\nolimits_a a_{o}^n (\Delta a_{Rx}^n - \Delta a_{Tx}) \label{Eq:DistanceApprox_1} \\
& \hspace{0.3cm} + \sum\nolimits_a (\Delta a_{Rx}^n - \Delta a_{Tx})^2 \nonumber \\ & \mathop{\approx}^{(b)} 2 \sum\nolimits_a a_{o}^n(\Delta a_{Rx}^n - \Delta a_{Tx}) \label{Eq:DistanceApprox_2} 
\end{align}
and $(a)$ follows from Taylor's approximation $\sqrt {1 + t}  \approx 1 + t/2 - {t^2}/8$. As for the approximation of the amplitude, $(a)$ holds if ${{\mathcal{S}}_{Tx}}$ and ${{\mathcal{S}}_{Rx}^n}$ are not too large compared to the distance $\| \mathbf{c}_{o}^n \|$. The approximation in $(b)$ can be applied if   $\sum\nolimits_a {\left| {a_{o}^n}  \right|} \gg \sum\nolimits_a {\left| {\Delta {a_{Rx}^n} - \Delta {a_{Tx}}} \right|}$, i.e., when the misalignment between the center-points of ${{\mathcal{S}}_{Tx}}$ and ${{\mathcal{S}}_{Rx}^n}$ is larger than their sizes.

\textcolor{black}{Compared with the conventional parabolic approximation \cite{Miller, ParabolicLozano}, the approximation in \eqref{Eq:WavefrontApprox_2} includes the quadratic term ${(\rho^n(\mathbf{r}_{Tx}, \mathbf{r}_{Rx}))^2}$. This latter term is necessary to account for non-broadside deployments, i.e., the center-points $\mathbf{c}_{Rc}^n$ and $\mathbf{c}_{Tc}$ are not aligned along a coordinate axis. In the conventional approach, this term is not needed because one coordinate axis coincides with the segment connecting $\mathbf{c}_{Rc}^n$  and $\mathbf{c}_{Tc}$. As a result, the quadratic term in ${(\rho^n(\mathbf{r}_{Tx}, \mathbf{r}_{Rx}))^2}$ can be ignored. This latter term is hence essential for ensuring that the approximation is accurate in an arbitrary coordinate system.}

\textcolor{black}{Based on these considerations, the exact formulation in \eqref{Eq:DistanceApprox_1} is to be utilized to compute the first-order term ${\rho^n(\mathbf{r}_{Tx}, \mathbf{r}_{Rx})}$ in \eqref{Eq:WavefrontApprox_2}, while the approximation $(b)$ in \eqref{Eq:DistanceApprox_2} is sufficient to compute the second-order term ${(\rho^n(\mathbf{r}_{Tx}, \mathbf{r}_{Rx}))^2}$ in \eqref{Eq:WavefrontApprox_2}. This approximation is referred to as ``quartic approximation''}.

Accordingly, $g_{i,o}^n(\mathbf{r}_{Tx},\mathbf{r}_{Rx})$ can be expressed as
 \begin{equation}
 \label{eq:quarticApp}
\hspace{-0.05cm}     g_{i,o}^n(\mathbf{r}_{Tx},\mathbf{r}_{Rx}) = \bar{g}_{i,o}^n f_{{Rx}}^n (\mathbf{r}_{Rx}) p^n(\mathbf{r}_{Tx}, \mathbf{r}_{Rx}) \left[f_{{Tx}}^n (\mathbf{r}_{Tx}) \right]^*
 \end{equation}
where \vspace{-0.25cm}
\begin{align}
\label{eq:defFocFunction}
    f_{{Tx}}^n(\mathbf{r}_{Tx}) &= \exp \Bigg\{   j \frac{\kappa_0}{2 \|\mathbf{c}_o^n \|}\left[ \sum\nolimits_a (\Delta a_{Tx})^2 -2 \sum\nolimits_a a_{o}^n \Delta a_{Tx}  
    \vphantom{\frac{(z_{Tx})^2}{\|\mathbf{c}_d\|^2}}\right. \nonumber\\ 
    & \left. \qquad \qquad   - \frac{\left(\sum\nolimits_a a_{o}^n \Delta a_{Tx}\right)^2}{\|\mathbf{c}_o^n\|^2}\right]\Bigg\} 
\end{align} \vspace{-0.5cm}
\begin{align}
\label{eq:focFunctionRX}
    f_{{Rx}}^n(\mathbf{r}_{Rx}) = \exp \Bigg\{ & - j \frac{\kappa_0}{2\|\mathbf{c}_o^n\|}\left[  \sum\nolimits_a (\Delta a_{Rx}^n)^2 + 2\sum\nolimits_a a_{o}^n \Delta a_{Rx}^n  
    \vphantom{\frac{(z_{Tx})^2}{\|\mathbf{c}_d\|^2}}\right. \nonumber\\ 
    & \left. \qquad \qquad   - \frac{\left(\sum\nolimits_a a_{o}^n \Delta a_{Rx}^n\right)^2}{\|\mathbf{c}_o^n\|^2}\right]\Bigg\}
\end{align} \vspace{-0.5cm}
\begin{align}
\label{eq:P_Parax}
    p^n(\mathbf{r}_{Tx},\mathbf{r}_{Rx}) = \exp \Bigg\{& j \frac{k_0}{\|\mathbf{c}_o^n\|} \left[ \sum\nolimits_a  \Delta a_{Rx}^n \Delta a_{Tx}
     \vphantom{\frac{(z_{Tx})^2}{\|\mathbf{c}_o^n\|^2}} \right.  \nonumber\\
    & \hspace{-0.75cm}\left. \qquad  - \frac{(\sum\nolimits_a   a_{o}^n \Delta a_{Tx}) (\sum\nolimits_a a_{o}^n \Delta a_{Rx}^n ) }{\|\mathbf{c}_o^n\|^2}\right]\Bigg\} \, .
\end{align}

The equivalence between the quartic approximation in \eqref{Eq:WavefrontApprox_2} and the conventional parabolic approximation is proved in Appendix I. In the next section, we show that the quartic approximation leads to  expressions given in terms of key geometrical parameters that are easy to interpret.

\vspace{-0.25cm}
\subsection{Approximation of the Integral Kernel}
\label{subsec:ApproachDoF}
By inserting \eqref{eq:quarticApp} in \eqref{eq:selfAdjoint_sub}, the kernel of the self-adjoint operator corresponding to the $n$th HoloS can be expressed as
\begin{align}
\label{eq:selfAdjoint_subApprox}
    &G_{Tx}^n(\mathbf{r}_{Tx},\mathbf{r}_{Tx}') \approx f_{{Tx}}^n (\mathbf{r}_{Tx})  \left[f_{{Tx}}^n (\mathbf{r}_{Tx}') \right]^* \bar{G}_{Tx}^n(\mathbf{r}_{Tx},\mathbf{r}_{Tx}')
\end{align}
where
\begin{align}
\label{eq:genGBarTXn}
    \bar{G}_{Tx}^n(\mathbf{r}_{Tx},\mathbf{r}_{Tx}') & \nonumber\\
    &\hspace{-1.5cm}= |\bar{g}_{i,o}^n|^2 \int_{{\mathcal{S}}_{Rx}^n} \left[p^n(\mathbf{r}_{Tx}, \mathbf{r}_{Rx}) \right]^*  p^n(\mathbf{r}_{Tx}', \mathbf{r}_{Rx})  d\mathbf{r}_{Rx} \, .
\end{align}

The quartic approximation of the complete integral kernel is obtained by inserting \eqref{eq:selfAdjoint_subApprox} in \eqref{eq:partitioning}. The obtained expression is utilized in the next section to compute the eDoF in the paraxial and non-paraxial settings. The paraxial setting is analyzed to facilitate the comparison with prior art, and because some steps of the derivations are used to analyze the non-paraxial setting. The paraxial setting is obtained by letting $N_r =1$ in \eqref{eq:partitioning}.

\vspace{-0.25cm}
\section{Number of Effective Degrees of Freedom}
\label{sec:GeneralSetup} \vspace{-0.1cm}
To obtain simple and insightful expressions, we commence by introducing a convenient parametrization for ${{\mathcal{S}}_{Tx}}$ and ${{\mathcal{S}}_{Rx}^n}$, and utilize a simplified notation. Without loss of generality, we set $\mathbf{c}_{Tx} = \mathbf{c}_i=(0,0,0)$ and $\mathbf{c}_{Rx} = \mathbf{c}_o=(x_o,y_o,z_o)$. Thus, the distance between the center-points of ${{\mathcal{S}}_{Tx}}$ and ${{\mathcal{S}}_{Rx}}$ is $\|\mathbf{c}_o -\mathbf{c}_i\| = \|\mathbf{c}_o\|$. Also, we set $A_{s} = 4U_{s}V_{s}$ for $s = \left\{ {Tx,Rx} \right\}$, where $2U_{s}$ and $2V_{s}$ are the lengths of the sides of ${{\mathcal{S}}_{Tx}}$ and ${{\mathcal{S}}_{Rx}}$. 

Specifically, ${{\mathcal{S}}_{Tx}}$ is identified by the parametrization
%
\begin{equation}
\label{eq:transSurface}
\mathcal{S}_{Tx} = \{ (u_i, 0,v_i) + \mathbf{c}_i\, : \, |u_i| \leq U_{Tx}, \: |v_i| \leq V_{Tx}\} 
\end{equation}

As mentioned, ${{\mathcal{S}}_{Rx}}$ is partitioned into $N_r$ sub-HoloSs $\mathcal{S}_{Rx}^n$. Each sub-HoloS is centered in $\mathbf{c}_o^n = (x_o^n, y_o^n, z_o^n)$ and its area is $A_{Rx}^n = 4 U_{Tx}^n V_{Tx}^n$, where $2 U_{Tx}^n$ and $2 U_{Rx}^n$ are the lengths of its sides. Thus, $\mathcal{S}_{Rx}^n$ is identified by the parametrization
\begin{multline}
\label{eq:recSurfacen}
\mathcal{S}_{Rx}^n = \{  (u_o^n \cos \alpha - v_o^n \sin \beta \sin \alpha, v_o^n \sin \beta \cos \alpha \\ + u_o^n \sin \alpha, v_o^n \cos \beta) + \mathbf{c}_o^n\, : \, |u_o^n| \leq U_{Rx}^n, \: |v_o^n| \leq V_{Rx}^n\} 
\end{multline}
where $\alpha$ and $\beta$ are the angles defined in Sec. \ref{sec:approach}.

\noindent
\textbf{Lemma 10.} Consider the parametrizations in \eqref{eq:transSurface} and \eqref{eq:recSurfacen}. Then, ${ f}_{Tx}^n( \mathbf{r}_{Tx})$ in \eqref{eq:P_Parax} and ${\bar G}_{Tx}^n\left( {{{\bf{r}}_{Tx}},{\bf{r}}_{Tx}'} \right)$ in \eqref{eq:genGBarTXn} simplify to 
\begin{align}
\label{eq:focusingFunction}
{f}_{Tx}^n( \mathbf{r}_{Tx}) = \exp\bigg\{ j \frac{\kappa_0}{2\|\mathbf{c}_o^n\|}\left[u_i^2 + v_i^2 - 2x_o^n u_i -2 z_o^n v_i \right. \nonumber \\ \left.
- \frac{4}{\|\mathbf{c}_o^n\|^2}\left(x_o^n u_i + z_o^n v_i\right)^2 \right]\bigg\} 
\end{align}
\begin{align}
\label{eq:GBarTXn}
{\bar G}_{Tx}^n\left( {{{\bf{r}}_{Tx}},{\bf{r}}_{Tx}'} \right) &= |\bar{g}_{i,o}^n|^2 A_{Rx}^n \nonumber \\
 & \times \sinc \left[ U_o^n \left( \tau_{11}^n (u_i - u_i') + \tau_{12}^n (v_i - v_i') \right)\right] \nonumber \\
        & \times \sinc \left[ V_o^n \left( \tau_{21}^n (u_i - u_i') + \tau_{22}^n (v_i - v_i') \right)\right]  \vspace{-0.10cm}
\end{align}
where $U_o^n=\frac{2U_{Rx}^n}{\lambda\|\mathbf{c}_o^n\|}$, $V_o^n=\frac{2V_{Rx}^n}{\lambda\|\mathbf{c}_o^n\|}$, $\tau_{11}^n = \cos \alpha - x_o^n \tau_1^n$, $\tau_{12}^n =  - z_o^n \tau_1^n $, $\tau_{21}^n = - \sin \beta \sin \alpha - x_o^n \tau_2^n $, $\tau_{22}^n = \cos \beta - z_o^n \tau_2^n$, $\tau_1^n = \frac{x_o^n \cos \alpha + y_o^n \sin \alpha}{\|\mathbf{c}_o^n\|^2}$, and $\tau_2^n=\frac{-x_o^n \sin \beta \sin \alpha + y_o^n \sin \beta \cos \alpha + z_o^n \cos \beta}{\|\mathbf{c}_o^n\|^2}$.
\begin{IEEEproof} See Appendix A. \end{IEEEproof}

\begin{figure}[!t]
	\centering
	\includegraphics[width=0.75\columnwidth]{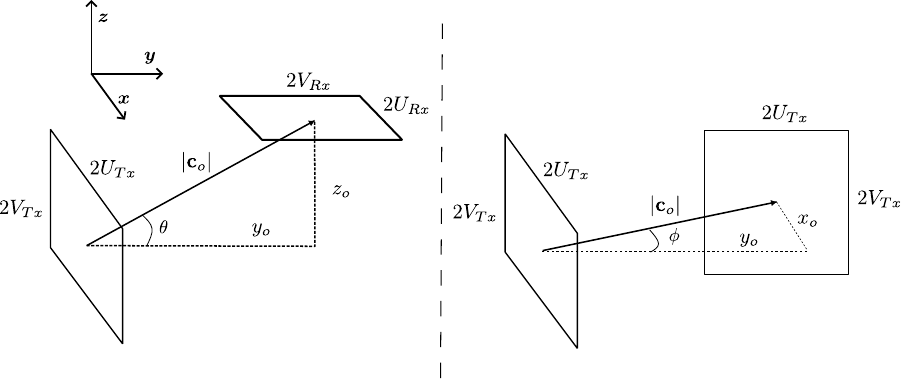}
    \vspace{-0.65cm}
	\caption{Examples of relevant network deployments for wireless applications. (left) Case study $x_o=0$, $\alpha=0$, $\beta=\pi/2$: ${{\mathcal{S}}_{Tx}}$ and ${{\mathcal{S}}_{Rx}}$ are deployed on a wall and on a ceiling, respectively. (right) Case study $z_o=0$, $\beta=0$, $\alpha=\pi/2$: ${{\mathcal{S}}_{Tx}}$ and ${{\mathcal{S}}_{Rx}}$ are deployed on two perpendicular walls.} \vspace{-0.5cm}
	\label{fig:particularCases}
\end{figure}
\textcolor{black}{The parametrizations in \eqref{eq:transSurface} and \eqref{eq:recSurfacen} can be applied to any network deployment. In wireless communications, several relevant network typologies are obtained by setting $\tau_{12} = \tau_{21} = 0$. Two illustrations are portrayed in Fig. \ref{fig:particularCases}, as examples. In Sec. VI-A, some analytical frameworks and results are specialized to the case studies $\tau_{12} = \tau_{21} = 0$, given their relevance for wireless communications and the engineering insights that we can learn from the obtained analytical frameworks}.

\subsection{Number of eDoF -- Paraxial Setting}
\label{sec:eDoFParaxial}
In this case, no partitioning for ${{\mathcal{S}}_{Rx}}$ is needed, and we consider $N_r =1$. By using \eqref{eq:selfAdjoint_subApprox}, the kernel in \eqref{eq:partitioning} becomes
\begin{equation}
    G_{Tx}(\mathbf{r}_{Tx},\mathbf{r}_{Tx}') \approx f_{{Tx}}^1 (\mathbf{r}_{Tx})  \left[f_{{Tx}}^1 (\mathbf{r}_{Tx}') \right]^* \bar{G}_{Tx}^1 (\mathbf{r}_{Tx},\mathbf{r}_{Tx}')
\end{equation}
with $\bar{G}_{Tx}^1 (\mathbf{r}_{Tx},\mathbf{r}_{Tx}')$ given in \eqref{eq:GBarTXn}. Next, for ease of notation, we omit the superscript that identifies the single sub-HoloS.

To compute the eDoF, the eigenproblem in \eqref{eq:eigenproblemIn} needs to be solved. To this end, without loss of generality, we look for eigenfunctions that can be expressed as $\phi_{m}(\mathbf{r}_{Tx}) = f_{{Tx}} (\mathbf{r}_{Tx}) \bar{\phi}_{m} (\mathbf{r}_{Tx})$. Accordingly, \eqref{eq:eigenproblemIn} simplifies as follows:
\begin{equation}
\label{eq:eigenproblemSimp}
\mu_{m} \bar{\phi}_{m} (\mathbf{r}_{Tx}) = \int_{{\mathcal{S}}_{Tx}}  \bar{G}_{Tx}(\mathbf{r}_{Tx},\mathbf{r}_{Tx}') \bar{\phi}_{m}(\mathbf{r}_{Tx}')d\mathbf{r}_{Tx}' \, . 
\end{equation}

By inspection of \eqref{eq:GBarTXn}, we see that $\bar{G}_{Tx}\left(\mathbf{r}_{Tx}, \mathbf{r}_{Tx}'\right)$ can be written as $\bar{G}_{Tx}\left(\mathbf{r}_{Tx}, \mathbf{r}_{Tx}'\right) = h_G^{\mathrm{parax}}\left(\mathbf{r}_{Tx} -  \mathbf{r}_{Tx}'\right)$. Then, in the asymptotic regime $\mathcal{S}_{Tx} = r \mathcal{S}_{Tx}'$ with $r \to \infty$, the eigenproblem in \eqref{eq:eigenproblemSimp} becomes a convolution integral. Thus, the kernel $\bar{G}_{Tx}\left(\mathbf{r}_{Tx}, \mathbf{r}_{Tx}'\right)$ plays the role, in the wavenumber domain, of a spatially invariant linear filter with impulse response $h_G^{\mathrm{parax}}\left(\mathbf{r}_{Tx}\right)$. \textcolor{black}{Next, we prove that the eigenproblem in \eqref{eq:eigenproblemSimp} fulfills the conditions in Lemma 9. This allows us to obtain, in an arbitrary coordinate system, an explicit expression for the eDoF as a function of key system parameters. To this end, we first analyze three important properties of $\bar{G}_{Tx}\left(\mathbf{r}_{Tx}, \mathbf{r}_{Tx}'\right)$ in the following three lemmas: (i) the operator norm; (ii) the relation between the eigenvalues in \eqref{eq:eigenproblemSimp} and the operator norm; and (iii) the Fourier transform in the wavenumber domain}.

\noindent
\textcolor{black}{\textbf{Lemma 11.} The operator norm of the self-adjoint operator $\bar G_{Tx}(\mathbf{r}_{Tx},\mathbf{r}_{Tx}')$ formulated in \eqref{eq:GBarTXn} is given by
\begin{equation} \label{Eq:OpNormExplicit}
    ||\bar{G}_{Tx}||_{\mathrm{op}} = |\bar{g}_{i,o}|^2 \frac{ \lambda^2 \|\mathbf{c}_o\|^ 2}{|\Upsilon(\mathbf{c}_o,\alpha,\beta)|}
\end{equation}
where $\Upsilon(\mathbf{c}_o,\alpha,\beta) = \tau_{11}\tau_{22} - \tau_{12}\tau_{21} = (y_o^2 \cos \alpha \cos \beta - y_o z_o \sin \beta  - x_o y_o \sin \alpha \cos \beta)\|\mathbf{c}_o\|^{-2}$}.

\begin{IEEEproof} \textcolor{black}{See Appendix B}. \end{IEEEproof}

\noindent
\textcolor{black}{\textbf{Lemma 12.} The largest eigenvalue of the eigenproblem in \eqref{eq:eigenproblemSimp} is upper-bounded by the operator norm $||\bar{G}_{Tx}||_{\mathrm{op}}$ in \eqref{Eq:OpNormExplicit}}.

\begin{IEEEproof} \textcolor{black}{Let us compute the supremum of the norm of both sides of \eqref{eq:eigenproblemSimp}: $\sup \left\| {{\mu_m}{{\bar \phi_m}}\left( {{{\bf{r}}_{Tx}}} \right)} \right\| = \sup \left\| {\left( {{G_{Tx}}{{\bar \phi }_m}} \right)\left( {{{\bf{r}}_{Tx}}} \right)} \right\|$. The eigenvalues in \eqref{eq:eigenproblemSimp} are real and positive, and $\left\| {{{\bar \phi_m}}} \right\|=1$ by definition, hence $\sup \left\| {{\mu_m}{{\bar \phi_m}}\left( {{{\bf{r}}_{Tx}}} \right)} \right\| = \sup({\mu_m})\sup \left\| {{{\bar \phi_m}}\left( {{{\bf{r}}_{Tx}}} \right)} \right\| = \sup({\mu _m})$. The proof follows by definition of operator norm (Def. 6): $||\bar{G}_{Tx}||_{\mathrm{op}}=\sup \left\| {\left( {{G_{Tx}}{{\bar \phi }_m}} \right)\left( {{{\bf{r}}_{Tx}}} \right)} \right\|$ for $\left\| {{{\bar \phi_m}}} \right\|=1$}. \end{IEEEproof}

\noindent
\textbf{Lemma 13.} The Fourier ($\mathcal{F}$) transform (in the wavenumber domain) of the impulse response $h_G^{\mathrm{parax}}\left(\mathbf{r}_{Tx}\right)$ is ${H_G^{\mathrm{parax}}}\left( {{{\boldsymbol {\kappa}}_{Tx}}} \right) = \int_{-\infty}^{+\infty} {\int_{-\infty}^{+\infty} {h_G^{\mathrm{parax}}\left(\mathbf{r}_{Tx}\right){e^{ - j \left( {u{\kappa_u} + v{\kappa_v}} \right)}}} dudv}$ and it is equal to 
%
\begin{align}
    \label{eq:fourierTransform}
    H_G^{\mathrm{parax}}(\boldsymbol{\kappa}_{Tx}) &= ||\bar{G}_{Tx}||_{\mathrm{op}} \mathds{1}_{\mathcal{H}_G}(\boldsymbol{\kappa}_{Tx})
\end{align}
where $||\bar{G}_{Tx}||_{\mathrm{op}}$ is the operator norm in Lemma 11 and
\begin{align}
    \mathcal{H}_G =& \biggl\{(\kappa_u, \kappa_v) \, : \, \frac{|\tau_{22} \kappa_u - \tau_{21} \kappa_v|}{|\tau_{11} \tau_{22} - \tau_{12}\tau_{21}|}\leq \frac{U_{Rx} \kappa_0}{\|\mathbf{c}_o\|}\, , \nonumber\\ 
    & \hspace{1.75cm}\frac{|- \tau_{12} \kappa_u + \tau_{11} \kappa_v|}{|\tau_{11} \tau_{22} - \tau_{12}\tau_{21}|}\leq \frac{V_{Rx} \kappa_0}{\|\mathbf{c}_o\|} \biggl \}
\end{align}
is the support of the Fourier transform of $h_G^{\mathrm{parax}}\left(\mathbf{r}_{Tx}\right)$. $H_G^{\mathrm{parax}}(\boldsymbol{\kappa}_{Tx})$ is the Fourier transform of a bidimensional ideal low-pass filter. The spatial bandwidth of $H_G^{\mathrm{parax}}(\boldsymbol{\kappa}_{Tx})$, i.e., the Lebesgue measure of $\mathcal{H}_G$ in the wavenumber domain, is
%
\begin{equation}
\label{eq:bandwidthFilter}
    m_G =4\pi^2 \frac{A_{Rx}}{\lambda^2 \|\mathbf{c}_o\|^2}|\Upsilon(\mathbf{c}_o,\alpha,\beta)| \,. 
\end{equation}
\begin{IEEEproof} See Appendix C. \end{IEEEproof}

Using Lemmas 11-13, the eDoF for an arbitrary deployment in the paraxial setting are given in the following proposition.

\noindent
\textbf{Proposition 1.} Consider the asymptotic regime of Lemma 9. Given the level of accuracy $0 < \gamma \leq  ||\bar{G}_{Tx}||_{\mathrm{op}}$, the eDoF are 
\begin{equation}
\label{eq:DoFparax}
        N_{{\rm{eDoF}}} (\gamma) =   N_{{\rm{eDoF}}} =  \max \left\{ {1,\frac{A_{Tx} A_{Rx}}{\lambda^2 \|\mathbf{c}_o\|^2} |\Upsilon(\mathbf{c}_o,\alpha,\beta)|} \right\}
\end{equation}
where $||\bar{G}_{Tx}||_{\mathrm{op}}$ is given in Lemma 11.
\begin{IEEEproof} 
The eigenproblem in \eqref{eq:eigenproblemSimp} is an instance of Landau's operator in Lemma 8, where $n=2$, $m(\mathcal{Q}) = m(\mathcal{S}_{Tx}) = A_{Tx}$, $m(\mathcal{P}) = m_G$ given in \eqref{eq:bandwidthFilter}. The $\max \left\{  \cdot  \right\}$ function ensures that $N_{{\rm{eDoF}}}$ is, as per its definition, no smaller than one.
\end{IEEEproof}

\textcolor{black}{Equation \eqref{eq:DoFparax} provides an accurate estimate for the eDoF, provided that the network deployment adheres to the asymptotic regime of Lemma 9. If the asymptotic regime is fulfilled, the eigenvalues polarize to two values: Either they are zero or they are equal to $||\bar{G}_{Tx}||_{\mathrm{op}}$ in \eqref{Eq:OpNormExplicit}, and the number of non-zero eigenvalues is equal to $N_{{\rm{eDoF}}}$ in \eqref{eq:DoFparax}. In Sec. VI, we discuss some network deployments in which \eqref{eq:DoFparax} has wider applicability than the asymptotic regime of Lemma 9}.

\textcolor{black}{In mathematical terms, Prop. 1 is sufficiently accurate if two conditions are fulfilled simultaneously \cite{Pizzo2022}: (i) the distance $\| {{\bf{c}}_{o}}\|$ between the center-points of ${{\mathcal{S}}_{Tx}}$ and ${{\mathcal{S}}_{Rx}}$ is much larger than the length of the sides of ${{\mathcal{S}}_{Tx}}$ and ${{\mathcal{S}}_{Rx}}$, i.e., ${\|\mathbf{c}_{o}\| }\gg \max \{{2U_{Tx}}, {2V_{Tx}}, {2U_{Rx}}, {2V_{Rx}}\} = U_{\max}$; (ii) the eDoF in \eqref{eq:DoFparax} are sufficiently many to ensure that the asymptotic regime of Lemma 9 is fulfilled, i.e., $N_{{\rm{eDoF}}} \gg 1$. Therefore, we obtain}
\begin{equation}
\label{eq:validityModelParax}
    \textcolor{black}{\sqrt{\frac{A_{Tx}}{\lambda^2}\frac{A_{Rx}}{\lambda^2}|\Upsilon(\mathbf{c}_o,\alpha,\beta)| } \gg  \frac{\|\mathbf{c}_{o}\| }{\lambda} \gg \frac{U_{\max}}{\lambda} }
\end{equation} 

\textcolor{black}{Equation \eqref{eq:validityModelParax} highlights that the accuracy of Proposition 1 depends on the network deployment through $|\Upsilon(\mathbf{c}_o,\alpha,\beta)|$. Thus, a misalignment between the center-points of ${{\mathcal{S}}_{Tx}}$ and ${{\mathcal{S}}_{Rx}}$, and their relative tilt and rotation, have an impact on the attainable eDoF and on the accuracy of the estimated eDoF. The following lemma provides $\alpha$ and $\beta$ that maximize \eqref{eq:DoFparax}}.

\noindent
\textbf{Lemma 14.} For a given network deployment, the values of $\alpha^\mathrm{opt}$ and $\beta^\mathrm{opt}$ that maximize $N_{{\rm{eDoF}}}$ in \eqref{eq:DoFparax} are \vspace{-0.10cm}
\begin{equation}
\label{eq:optimalAngles}
    \alpha^\mathrm{opt} = -{x_o}/{y_o} \qquad \beta^\mathrm{opt} = -{z_o}/{\sqrt{x_o^2 + y_o^2}} \;.  \vspace{-0.10cm}
\end{equation}
and the corresponding $N_{{\mathrm{eDoF}}}^\mathrm{opt}$ in \eqref{eq:DoFparax} is \vspace{-0.10cm}
\begin{equation}
\label{eq:optimalNDof}
N_{{\mathrm{eDoF}}}^\mathrm{opt} = \frac{A_{Tx} A_{Rx}}{\lambda^2 \|\mathbf{c}_o\|^2} \frac{\left|y_o\right|}{\|\mathbf{c}_o\|}  \;. \vspace{-0.10cm}
\end{equation}

\begin{IEEEproof} $|\Upsilon(\alpha,\beta,x_o,y_o,z_o)|$ is not differentiable, but it has the same critical points as $ \Upsilon(\alpha,\beta,x_o,y_o,z_o)$. Therefore, we compute the partial derivatives of $\Upsilon(\alpha,\beta,x_o,y_o,z_o)=f(\alpha,\beta) = y_o^2 \cos \alpha \cos \beta - y_o z_o \sin \beta  - x_o y_o \sin \alpha \cos \beta$. Computing ${{\partial f\left( {\alpha ,\beta } \right)} \mathord{\left/ {\vphantom {{\partial f\left( {\alpha ,\beta } \right)} {\partial \alpha }}} \right. \kern-\nulldelimiterspace} {\partial \alpha }}$ and ${{\partial f\left( {\alpha ,\beta } \right)} \mathord{\left/ {\vphantom {{\partial f\left( {\alpha ,\beta } \right)} {\partial \beta }}} \right. \kern-\nulldelimiterspace} {\partial \beta }}$, and setting them equal to zero, we obtain $\tan \alpha^\mathrm{opt}  = -{x_o}/{y_o}$ and
$\tan \beta^\mathrm{opt}  = {z_o}/{(x_o \sin \alpha^\mathrm{opt} - y_o \cos \alpha^\mathrm{opt})}$. Using the trigonometric identities $\sin x = {\tan x}/{\sqrt{1 + \tan^2 x}}$ and $\cos x = {1}/{\sqrt{1 + \tan^2 x}}$, we obtain $\alpha^\mathrm{opt}$ and $\beta^\mathrm{opt}$ in \eqref{eq:optimalAngles}. The eDoF follow from \eqref{eq:DoFparax}. \end{IEEEproof} 

To provide some engineering insights from Lemma 14, we employ a spherical coordinate system, by setting $x_o = \|\mathbf{c}_o\| \sin \phi_o \cos \theta_o$, $y_o = \|\mathbf{c}_o\| \cos \phi_o \cos \theta_o$ $z_o = \|\mathbf{c}_o\| \sin \theta_o$, where $\theta_o$ and $\phi_o$ denote the elevation and azimuth angles with respect to the origin (i.e., the center-point of ${{\mathcal{S}}_{Tx}}$). Converting \eqref{eq:optimalAngles} to spherical coordinates, we obtain $\alpha^\mathrm{opt} = - \phi_o$ and $\beta^\mathrm{opt} = - \theta_o$. This implies that the eDoF are maximized if ${{\mathcal{S}}_{Rx}}$ is oriented towards the center-point of ${{\mathcal{S}}_{Tx}}$. Also, let $N_{{\mathrm{eDoF}}}^\mathrm{broadside} = \frac{A_{Tx} A_{Rx}}{\lambda^2 \|\mathbf{c}_o\|^2}$ be the number of eDoF corresponding to the broadside deployment, i.e., $\alpha=0$, $\beta=0$ and the center-points of ${{\mathcal{S}}_{Tx}}$ and ${{\mathcal{S}}_{Rx}}$ are aligned along the same axis. Then,  \eqref{eq:optimalNDof} states that $N_{{\mathrm{eDoF}}}^\mathrm{opt} \le N_{{\mathrm{eDoF}}}^\mathrm{broadside}$, since $|y_0|/\|\mathbf{c}_o\| \le 1$ for any network deployment. This implies that, in the presence of a misalignment between the center-points of ${{\mathcal{S}}_{Tx}}$ and ${{\mathcal{S}}_{Rx}}$, an optimized tilt and rotation help increase the eDoF, but the best network deployment is always the broadside setting, i.e.,  $x_o=z_o=0$ with the considered parametrization in \eqref{eq:recSurfacen}.

\vspace{-0.25cm}
\subsection{Number of eDoF -- Non-Paraxial Setting}
\label{sec:eDoFNonParaxial}
\textcolor{black}{In this section, we analyze the non-paraxial setting in which the distance $\|\mathbf{c}_{o}\|$ is comparable with $U_{\max} = \max \{{2U_{Tx}}, {2V_{Tx}}, {2U_{Rx}}, {2V_{Rx}}\}$. To this end, we adopt the approach introduced in Section \ref{sec:approach}, by partitioning ${{\mathcal{S}}_{Rx}}$ into $N_r>1$ sub-HoloSs. By using \eqref{eq:partitioning}, \eqref{eq:selfAdjoint_sub} and \eqref{eq:quarticApp}-\eqref{eq:P_Parax}, the kernel of the self-adjoint operator can be approximated as
\begin{equation}
\label{eq:nonParaxialSA}
  \hspace{-0.25cm}  G_{Tx}(\mathbf{r}_{Tx},\mathbf{r}_{Tx}') \hspace{-0.05cm} \approx \hspace{-0.1cm}  \sum\limits_{n=1}^{N_r} f_{{Tx}}^n (\mathbf{r}_{Tx})  \left[f_{{Tx}}^n (\mathbf{r}_{Tx}') \right]^* \bar{G}_{Tx}^n(\mathbf{r}_{Tx},\mathbf{r}_{Tx}')
\end{equation}
with $f_{{Tx}}^n (\mathbf{r}_{Tx})$ and $\bar{G}_{Tx}^n(\mathbf{r}_{Tx},\mathbf{r}_{Tx}')$ given in \eqref{eq:focusingFunction} and \eqref{eq:GBarTXn}}. 

\textcolor{black}{The resulting eigenproblem in \eqref{eq:eigenproblemIn} cannot, however, be formulated in a form that fulfills the conditions of Lemma 8. This is because $G_{Tx}(\mathbf{r}_{Tx},\mathbf{r}_{Tx}')$ cannot be formulated as $G_{Tx}(\mathbf{r}_{Tx} - \mathbf{r}_{Tx}')$, due to the presence of quadratic terms in the function $f_{{Tx}}^n (\mathbf{r}_{Tx})$ in \eqref{eq:focusingFunction}. On the other hand, we note from \eqref{eq:GBarTXn} that $\bar{G}_{Tx}^n(\mathbf{r}_{Tx},\mathbf{r}_{Tx}') = \bar{G}_{Tx}^n(\mathbf{r}_{Tx}-\mathbf{r}_{Tx}')$. To overcome this difficulty, we propose to solve \eqref{eq:eigenproblemIn} by looking for solutions that can be formulated as $\phi_{m}(\mathbf{r}_{Tx}) = \bar{f}_{{Tx}} (\mathbf{r}_{Tx}) \bar{\phi}_{m} (\mathbf{r}_{Tx})$, where}
\begin{equation}
\label{eq:barFocFunction}
\textcolor{black}{\hspace{-0.12cm}\bar{f}_{Tx}( \mathbf{r}_{Tx}) = \exp\bigg\{ j \frac{\kappa_0}{2\|\mathbf{c}_o\|}\left[u_i^2 + v_i^2 
- \frac{4\left(x_o u_i + z_o v_i\right)^2}{\|\mathbf{c}_o\|^2} \right]\bigg\} }. 
\end{equation}

\textcolor{black}{Accordingly, the eigenproblem in \eqref{eq:eigenproblemIn} can be formulated as
\begin{align}
\label{eq:eigenproblemNewFocusingFunc}
&\mu_{m} \bar{\phi}_{m} (\mathbf{r}_{Tx}) \\ &= \sum\nolimits_{n=1}^{N_r}  \int\nolimits_{{\mathcal{S}}_{Tx}} \bar{F}_{Tx}^{n}(\mathbf{r}_{Tx},\mathbf{r}_{Tx}')  \bar{G}_{Tx}^n(\mathbf{r}_{Tx},\mathbf{r}_{Tx}') \bar{\phi}_{m}(\mathbf{r}_{Tx}')d\mathbf{r}_{Tx}' \, . \nonumber 
\end{align}
with $\bar{F}_{Tx}^{n}(\mathbf{r}_{Tx},\mathbf{r}_{Tx}') = \left[\bar{f}_{{Tx}}^{*} (\mathbf{r}_{Tx}) f_{{Tx}}^n (\mathbf{r}_{Tx})\right]  \left[\bar{f}_{{Tx}}^{*} (\mathbf{r}_{Tx}') f_{{Tx}}^n (\mathbf{r}_{Tx}')\right]^{*}$}.

\textcolor{black}{The next lemma provides an explicit approximation for the eigenproblem in \eqref{eq:eigenproblemNewFocusingFunc}, which is convenient for analysis}.

\textcolor{black}{\textbf{Lemma 15.} The eigenproblem in \eqref{eq:eigenproblemNewFocusingFunc} can be simplified as
\begin{equation}
\label{eq:eigenproblemSimpNP}
\mu_{m} \bar{\phi}_{m} (\mathbf{r}_{Tx}) = \int_{{\mathcal{S}}_{Tx}}  \bar{G}_{Tx}^{\mathrm{no-parax}} (\mathbf{r}_{Tx},\mathbf{r}_{Tx}') \bar{\phi}_{m}(\mathbf{r}_{Tx}')d\mathbf{r}_{Tx}' 
\end{equation}
where
\begin{align}
\label{eq:nonParaxialSA_app}
    \bar{G}_{Tx}^{\mathrm{no-parax}} (\mathbf{r}_{Tx},\mathbf{r}_{Tx}')  & \approx \sum\nolimits_{n=1}^{N_r} \bar{G}_{Tx}^n(\mathbf{r}_{Tx},\mathbf{r}_{Tx}')
   \\ & \times \exp \{- j \left( \Delta\kappa_u^{n} (u_i - u_i') + \Delta \kappa_v^{n} (v_i - v_i') \right)\} \nonumber
\end{align}
with $\Delta \kappa_u^{n} = \kappa_0 x_o^n/\|\mathbf{c}_o^n\|$ and $\Delta \kappa_v^{n} = \kappa_0 z_o^n/\|\mathbf{c}_o^n\|$.
\begin{IEEEproof} See Appendix D. \end{IEEEproof} 
}

\textcolor{black}{By inspection of \eqref{eq:nonParaxialSA_app}, we see that $\bar{G}_{Tx}^{\mathrm{no-parax}} (\mathbf{r}_{Tx},\mathbf{r}_{Tx}')$ can be written as $\bar{G}_{Tx}^{\mathrm{no-parax}} (\mathbf{r}_{Tx},\mathbf{r}_{Tx}') = h_G^{\mathrm{no-parax}}\left(\mathbf{r}_{Tx} -  \mathbf{r}_{Tx}'\right)$ with $h_G^{\mathrm{no-parax}}\left(\mathbf{r}_{Tx} -  \mathbf{r}_{Tx}'\right) = \sum\nolimits_{n=1}^{N_r} h_{G^n}^{\mathrm{no-parax}}\left(\mathbf{r}_{Tx} -  \mathbf{r}_{Tx}'\right)$. Similar to the paraxial setting, the eigenproblem in \eqref{eq:nonParaxialSA_app} becomes a convolution integral in the asymptotic regime $\mathcal{S}_{Tx} = r \mathcal{S}_{Tx}'$ with $r \to \infty$. In the non-paraxial setting as well, the kernel $\bar{G}_{Tx}^{\mathrm{no-parax}} (\mathbf{r}_{Tx},\mathbf{r}_{Tx}')$ plays hence the role, in the wavenumber domain, of a spatially invariant linear filter with impulse response $h_G^{\mathrm{no-parax}}\left(\mathbf{r}_{Tx}\right)$. Next, we prove that the eigenproblem in \eqref{eq:nonParaxialSA_app} fulfills the conditions of Lemma 9 instead of Lemma 8, as for the paraxial setting. In Section V-C, we discuss the main differences between the paraxial and non-paraxial settings}.

\textcolor{black}{To obtain an explicit expression for the eDoF as a function of key system parameters, we first analyze a few important properties of $\bar{G}_{Tx}^{\mathrm{no-parax}} (\mathbf{r}_{Tx},\mathbf{r}_{Tx}')$ in the following lemmas: (i) the Fourier transform of $h_{G}^{\mathrm{no-parax}}\left(\mathbf{r}_{Tx}\right)$ in the wavenumber domain; (ii) the overlap of the supports of the Fourier transforms of $h_{G^n}^{\mathrm{no-parax}}\left(\mathbf{r}_{Tx}\right)$ for any $n$; (iii) the operator norm; (iv) the range of admissible values for the eigenvalues and their relation with the operator norm; (v) the polarization of the eigenvalues; and (vi) the Lebesgue measure of the support of the Fourier transform of $h_{G}^{\mathrm{no-parax}}\left(\mathbf{r}_{Tx}\right)$ and $h_{G^n}^{\mathrm{no-parax}}\left(\mathbf{r}_{Tx}\right)$ as a function of a given approximation accuracy}.

\noindent
\textcolor{black}{\textbf{Lemma 16.} 
The Fourier ($\mathcal{F}$) transform (in the wavenumber domain) of $h_G^{\mathrm{no-parax}}\left(\mathbf{r}_{Tx}\right)$ is given by ${H_G^{\mathrm{no-parax}}}\left( {{{\boldsymbol {\kappa}}_{Tx}}} \right) = \int_{-\infty}^{+\infty} {\int_{-\infty}^{+\infty} {h_G^{\mathrm{no-parax}}\left(\mathbf{r}_{Tx}\right){e^{ - j \left( {u{\kappa_u} + v{\kappa_v}} \right)}}} dudv}$, and it is equal to}
\begin{align}
    \label{eq:fourierTransformNP}
    \textcolor{black}{H_G^{\mathrm{no-parax}}(\boldsymbol{\kappa}_{Tx}) = \sum\nolimits_{n=1}^{N_r} ||\bar{G}_{Tx}^n||_{\mathrm{op}}  \mathds{1}_{\mathcal{H}_G^n}(\boldsymbol{\kappa}_{Tx})}
\end{align}
\textcolor{black}{where $\Upsilon^n(\mathbf{c}_o^n,\alpha,\beta) = \tau_{11}^n \tau_{22}^n - \tau_{12}^n \tau_{21}^n$ and}
\begin{equation} \label{Eq:OpNormN}
\textcolor{black}{||\bar{G}_{Tx}^n||_{\mathrm{op}} = |\bar{g}_{i,o}^n|^2 \frac{ \lambda^2 \|\mathbf{c}_o^n\|^ 2}{|\Upsilon^n(\mathbf{c}_o^n,\alpha,\beta)|}}
\end{equation}
\begin{align} \label{Eq:SpectralBandNoParax}
    \mathcal{H}_G^n =& \biggl\{(\kappa_u, \kappa_v)  :  \frac{|\tau_{22}^n (\kappa_u - \Delta \kappa_u^{n}) - \tau_{21}^n (\kappa_v - \Delta \kappa_v^{n})|}{|\tau_{11}^n \tau_{22}^n - \tau_{12}^n \tau_{21}^n|}\leq \frac{U_{Rx}^n \kappa_0}{\|\mathbf{c}_o^n\|}\, , \nonumber\\ 
    & \hspace{0.1cm} \frac{|- \tau_{12}^n (\kappa_u - \Delta \kappa_u^{n}) + \tau_{11}^n (\kappa_v - \Delta \kappa_v^{n})|}{|\tau_{11}^n \tau_{22}^n - \tau_{12}^n \tau_{21}^n|}\leq \frac{V_{Rx}^n \kappa_0}{\|\mathbf{c}_o^n\|} \biggl \} \, .
\end{align}

\begin{IEEEproof} \textcolor{black}{The proof follows along the lines of the proof of Lemma 13, by using the linearity of the Fourier transform and by noting that the exponential term in the spatial domain in \eqref{eq:nonParaxialSA_app} corresponds to a shift in the wavevumber domain}. \end{IEEEproof}

\textcolor{black}{From Lemma 16, we evince that the supports of the Fourier transforms of $h_{G^n}^{\mathrm{no-parax}}\left(\mathbf{r}_{Tx}\right)$ are finite in the wavenumber domain. According to \eqref{eq:bandwidthFilter}, we obtain $m_{G^n} =4\pi^2 {A_{Rx}^n}|\Upsilon^n(\mathbf{c}_o^n,\alpha,\beta)|/{(\lambda^2 \|\mathbf{c}_o^n\|^2)} $. In contrast to the paraxial setting, however, the Fourier transform in \eqref{eq:fourierTransformNP} is not an ideal pass-band filter. The bandwidth of $h_{G}^{\mathrm{no-parax}}\left(\mathbf{r}_{Tx}\right)$ in the wavenumber domain is further analyzed in Lemma 21}.

\noindent
\textcolor{black}{\textbf{Lemma 17.} If ${U_{Rx}^n}/{\|\mathbf{c}_o^n\|}$ and ${V_{Rx}^n}/{\|\mathbf{c}_o^n\|}$ are sufficiently small, the supports of the Fourier transforms of $h_{G^n}^{\mathrm{no-parax}}\left(\mathbf{r}_{Tx}\right)$ are almost disjoint in the wavenumber domain. If $\tau_{12} = \tau_{21}= 0$ (see Fig. \ref{fig:particularCases}), the overlap in the wavenumber domain is upper-bounded by functions that scale with ${U_{Rx}^n}/{\|\mathbf{c}_o^n\|}$ and ${V_{Rx}^n}/{\|\mathbf{c}_o^n\|}$.
\begin{IEEEproof} See Appendix E. \end{IEEEproof} 
}

\textcolor{black}{From Lemma 17, we reveal that the overlap of the supports of the Fourier transforms of $h_{G^n}^{\mathrm{no-parax}}\left(\mathbf{r}_{Tx}\right)$ for any $n$ can be made sufficiently small, in the wavenumber domain, if each sub-HoloS fulfills the paraxial setting, i.e., it is small enough}.

\noindent
\textcolor{black}{\textbf{Lemma 18.} 
The operator norm of the self-adjoint operator $\bar{G}_{Tx}^{\mathrm{no-parax}} (\mathbf{r}_{Tx},\mathbf{r}_{Tx}')$ formulated in \eqref{eq:nonParaxialSA_app} is given by} \vspace{-0.1cm}
\begin{equation}
\label{Eq:OpNomNP}
\textcolor{black}{||\bar{G}_{Tx}^{\mathrm{no-parax}}||_{\mathrm{op}} = {\max \nolimits_n}\left\{ { ||\bar{G}_{Tx}^n||_{\mathrm{op}}} \right\}} . \vspace{-0.1cm}
\end{equation}

\begin{IEEEproof} \textcolor{black}{See Appendix F.} \end{IEEEproof}

\textcolor{black}{The proof of Lemma 18 unveils important properties of the supports (in the spatial and wavenumber domains) of the eigenfunctions of \eqref{eq:eigenproblemSimpNP}. This is further detailed in Sec. VI}.

\noindent
\textcolor{black}{\textbf{Lemma 19.} 
The largest eigenvalue of the eigenproblem in \eqref{eq:eigenproblemSimpNP} is upper-bounded by $\mu_m \le {\max \nolimits_n}\left\{ { ||\bar{G}_{Tx}^n||_{\mathrm{op}}} \right\}$ and the smallest non-zero eigenvalue of the eigenproblem in \eqref{eq:eigenproblemSimpNP} is lower-bounded by $\mu_m \ge {\min \nolimits_n}\left\{ { ||\bar{G}_{Tx}^n||_{\mathrm{op}}} \right\}$}.

\begin{IEEEproof} \textcolor{black}{The upper-bound follows directly from the proof of Lemma 18 and by definition of operator norm (see Def. 6). The lower-bound follows by computing the infimum instead of the supremum, and by using similar analytical steps}. \end{IEEEproof}

\begin{figure}[!t]
	\centering
	\includegraphics[width=0.95\columnwidth]{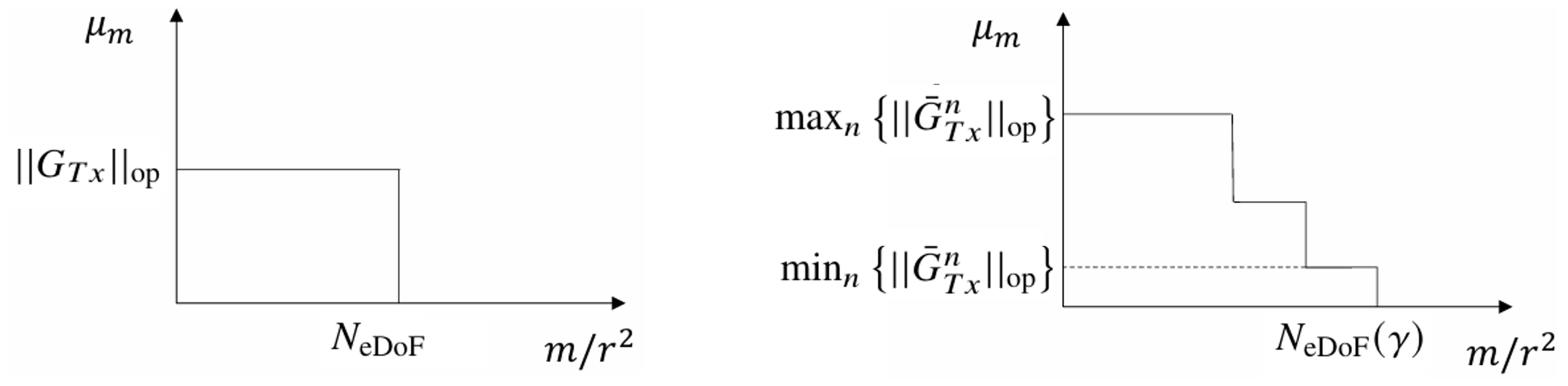}
	\vspace{-0.35cm} \caption{Distribution of the eigenvalues (asymptotic regime) for the operator kernels $\bar{G}_{Tx}$ (left) and ${\bar G}_{Tx}^{\mathrm{no-parax}}$ (right).}
	\label{fig:asymptoticRegime} \vspace{-0.55cm}
\end{figure}
\noindent
\textcolor{black}{\textbf{Lemma 20.} 
Consider the asymptotic regime of Lemma 8, i.e., $\mathcal{S}_{Tx} = r \mathcal{S}_{Tx}'$ with $r \to \infty$. The non-zero eigenvalues of the eigenproblem in \eqref{eq:eigenproblemSimpNP} polarize to $N_r$ different values. Each sub-HoloS contributes independently with a number of eigenvalues \vspace{-0.1cm}
\begin{equation} \label{Eq:NDoFn}
    N_{{\rm{eDoF}}}^n = \frac{A_{Tx} A_{Rx}^n}{\lambda^2 \|\mathbf{c}_o^n\|^2} |\Upsilon(\mathbf{c}_o^n,\alpha,\beta)| \vspace{-0.1cm}
\end{equation}
and the intensity of these eigenvalues is $||\bar{G}_{Tx}^n||_{\mathrm{op}}$ in \eqref{Eq:OpNormN}}.

\textcolor{black}{\begin{IEEEproof} See Appendix G. \end{IEEEproof} }

\textcolor{black}{Lemma 20 highlights a fundamental difference between the operator kernels $\bar{G}_{Tx}$ (paraxial) and ${\bar G}_{Tx}^{\mathrm{no-parax}}$ (non-paraxial): The eigenvalues of ${\bar G}_{Tx}^{\mathrm{no-parax}}$ polarize to more than two values, as shown in Fig. \ref{fig:asymptoticRegime}. This is further elaborated in Sec. V-C}.

\noindent
\textcolor{black}{\textbf{Lemma 21.} 
Given the approximation accuracy $\gamma > 0$, the Lebesgue measure of the support of the Fourier transform in the wavenumber domain of $h_{G}^{\mathrm{no-parax}}\left(\mathbf{r}_{Tx}\right)$ is
\begin{equation} \label{Eq:BandNoP}
    m_G(\gamma) = 4\pi^2 \sum\nolimits_{n \in \mathcal{N}_\gamma}  \frac{A_{Rx}^n}{\lambda^2 \|\mathbf{c}_o^n\|^2}|\Upsilon(\mathbf{c}_o^n,\alpha,\beta)|
\end{equation}
where $\mathcal{N}_{\gamma} = \{n : ||\bar{G}_{Tx}^n||_{\mathrm{op}} \geq \gamma\}$ with $||\bar{G}_{Tx}^n||_{\mathrm{op}}$ given in \eqref{Eq:OpNormN}}.

\begin{IEEEproof} \textcolor{black}{It follows from Lemmas 16 and 17, since the supports of the Fourier transforms of $h_{G^n}^{\mathrm{no-parax}}\left(\mathbf{r}_{Tx}\right)$ are finite and do not overlap in the wavenumber domain, as well as the additive property of the Lesbegue measure for disjoint sets}. \end{IEEEproof}

\noindent
\textcolor{black}{\textbf{Proposition 2.} 
Consider the asymptotic regime of Lemma 8 and the accuracy level $0 < \gamma \leq  ||\bar{G}_{Tx}^{\mathrm{no-parax}}||_{\mathrm{op}}$, the eDoF are 
\begin{equation}
\label{eq:DoFNP}
        N_{{\rm{eDoF}}} (\gamma) =  \max \left\{ {1,\sum\nolimits_{n \in \mathcal{N}_\gamma} N_{{\rm{eDoF}}}^n } \right\}
\end{equation}
where $N_{{\rm{eDoF}}}^n$ is given in \eqref{Eq:NDoFn} and $\mathcal{N}_\gamma$ is defined in \eqref{Eq:BandNoP}}.

\begin{IEEEproof} \textcolor{black}{It follows directly from Lemma 8, considering the Lesbegue measure computed in Lemma 21}. \end{IEEEproof}

\textcolor{black}{Assuming the asymptotic regime of Lemma 8, Prop. 2 showcases a major difference between the paraxial and non-paraxial settings: In the former setting, the eDoF are independent of the approximation accuracy $\gamma$, while in the later setting the eDoF are determined by the approximation accuracy $\gamma$. This is due to the fact that the Fourier transform in \eqref{eq:fourierTransformNP} does not correspond to an ideal low-pass filter in the wavenumber domain}. 

\textcolor{black}{Also, Prop. 2 is consistent with \cite[Prop. 1]{Hongliang_2025} obtained with the cut-set integral method. As elaborated in Sec. V-C, it is a more general result, as it can be applied to any value of Kolmogorov's approximation accuracy $\gamma$, while the cut-set integral method can be applied only if $\gamma$ is arbitrarily small}.

\textcolor{black}{Similar to Proposition 1, \eqref{eq:DoFNP} in Proposition 2 provides an accurate estimate for the eDoF, provided that the network deployment adheres to the asymptotic regime of Lemma 8. Thus, Proposition 2 is sufficiently accurate if the same conditions as for Proposition 1 are fulfilled for each sub-HoloS. Based on \eqref{eq:validityModelNoParax}, this leads to the following inequalities}: \vspace{-0.15cm}
\begin{equation}
\label{eq:validityModelNoParax}
    \textcolor{black}{\sqrt{\frac{A_{Tx}}{\lambda^2}\frac{A_{Rx}^n}{\lambda^2}|\Upsilon^n(\mathbf{c}_o^n,\alpha,\beta)| } \gg  \frac{\|\mathbf{c}_{o}^n\| }{\lambda} \gg \frac{U_{\max}^n}{\lambda}} \vspace{-0.15cm}
\end{equation}
\textcolor{black}{where $U_{\max}^n= \max \{{2U_{Tx}^n}, {2V_{Tx}^n}, {2U_{Rx}^n}, {2V_{Rx}^n}\}$. In the non-paraxial setting, however, an additional condition needs to be fulfilled for Proposition 2 to be accurate: The supports of the Fourier transforms of $h_{G^n}^{\mathrm{no-parax}}\left(\mathbf{r}_{Tx}\right)$ need to be almost disjoint in the wavenumber domain, according to Lemma 17. To elaborate, $U_{\max}^n / \|\mathbf{c}_{o}^n\|$ needs to be sufficiently small, which is typically the case if each sub-HoloS complies with the paraxial setting. Also, this condition is compatible with \eqref{eq:validityModelNoParax}}.

\vspace{-0.4cm}
\subsection{Number of eDoF -- Discussion and Comparison} \vspace{-0.1cm}
\textcolor{black}{In this section, we elaborate on important properties and differences between the paraxial and non-paraxial settings, and we compare our approach with others available in prior works}.

\subsubsection{Paraxial Setting}
\textcolor{black}{In the paraxial setting, an important point to prove is the equivalence between the proposed quartic approximation and the conventional parabolic approximation. This is proved in Appendix I. In a nutshell, the two approximations are equivalent, when they are applied in two different coordinate systems (as mentioned in Sec. IV-B). As detailed in Sec. IV-A, the quartic approximation facilitates the analysis of the non-paraxial setting, as in this latter case all the sub-HoloSs need to be formulated in the same coordinate system to be able to analyze the eDoF consistently and rigorously}.

\subsubsection{Non-paraxial Setting}
\textcolor{black}{In Sec. IV-A, we mentioned that our approach can be applied if one of the two HoloSs (${{\mathcal{S}}_{Tx}}$ in this paper) is small compared to the other. This is proved in App. J. The generalization to network deployments in which both ${{\mathcal{S}}_{Tx}}$ and ${{\mathcal{S}}_{Rx}}$ have a large size is left to a future research work, as Lemma 8 cannot be applied and hence a different approach is needed. The cut-set integral method in \cite{Dardari_2020, Hongliang_2025} can be applied when both ${{\mathcal{S}}_{Tx}}$ and ${{\mathcal{S}}_{Rx}}$ have a large size, but the eDoF are obtained numerically, and the approximation accuracy needs to be arbitrarily small, as detailed next}.

\textcolor{black}{Next, we compare the proposed approach with the numerical method in \cite{9650519}, and the cut-set integral method in \cite{Dardari_2020, Hongliang_2025}}. 

\paragraph{Comparison with \cite{9650519}}
\textcolor{black}{As shown in \cite{Niyato_2025}, the approach in \cite{9650519} is equivalent to ours and to prior art in the paraxial setting, since \cite[Eq. 12]{9650519} coincides with \eqref{eq:DoFparax} in the asymptotic regime where the eigenvalues polarize to two values. In the non-paraxial setting, nothing can be said about the accuracy of \cite{9650519}, since \cite[Eq. 12]{9650519} is independent of the approximation accuracy $\gamma$, as defined by Kolmogorov. In the non-paraxial setting, in addition, Lemma 20 shows that the non-zero eigenvalues polarize to more than two values}.

\paragraph{Comparison with \cite{Dardari_2020, Hongliang_2025}} \textcolor{black}{The approach introduced in \cite{Dardari_2020} and generalized in \cite{Hongliang_2025} can be viewed as a special case of Proposition 2 when Kolmogorov's approximation accuracy $\gamma$ is arbitrarily small. This is proved in Appendix H, by considering the case study $\alpha = \beta = 0$ analyzed in \cite{Dardari_2020} and \cite{Hongliang_2025}. The reason why $\gamma$ needs to be small is apparent by comparing \eqref{Eq:AppH_1} against \eqref{eq:DoFNP}: In \eqref{Eq:AppH_1}, all the $N_r$ sub-HoloSs are summed up regardless of the desired level of accuracy, while only the set of sub-HoloSs whose operator norm is greater than the specified level of accuracy is considered in \eqref{eq:DoFNP}. This makes the proposed approach more general. In addition, Proposition 2 coincides with \cite{Dardari_2020} and \cite{Hongliang_2025} in the limiting regime that the sub-HoloSs are infinitesimally small and hence the sum in \eqref{eq:DoFNP} tends to an integral. Thus, the proposed partitioning avoids the computation of integrals as well. Finally, the methods in \cite{Dardari_2020, Hongliang_2025, 9650519} cannot be applied to compute the eigenfunctions of \eqref{eq:eigenproblemSimp} and \eqref{eq:eigenproblemSimpNP}. On the other hand, we analyze the eigenfunctions in the next section}.

\begin{table}[t]
\centering
\caption{Paraxial vs. Non-Paraxial Settings (asymptotic regime)}
\label{tab:TableI} \vspace{-0.15cm}
\begin{tabular}{c||c|c}
\hline 
\textbf{eDoF} & \textbf{Paraxial Setting}                                                          & \textbf{Non-Paraxial Setting}                        \\ \hline \hline
Landau's Theorem \cite{Landau1975} & Theorem 1 & Theorem 2 \\ \hline
Polarization & Two values & Multiple values \\ \hline
Approximation accuracy & Independent & Dependent \\ \hline
\end{tabular} \vspace{-0.65cm}
\end{table}
\subsubsection{Paraxial vs. Non-paraxial Settings}
\textcolor{black}{By comparing the results in Secs. V-A and V-B, we conclude that major differences between the paraxial and non-paraxial settings exist. To facilitate the comparison, a summary is given in Table \ref{tab:TableI}}.

\vspace{-0.6cm}
\section{Optimal Communication Waveforms} \vspace{-0.25cm}
In this section, we compute the eigenfunctions of the eigenproblem in \eqref{eq:eigenproblemIn}, which correspond to the optimal communication waveforms in wireless communications. To the best of our knowledge, closed-form expressions for the eigenfunctions of \eqref{eq:eigenproblemIn} are known only for one-dimensional problems, and are known to be PSWFs \cite{Slepian1961}. In two or higher dimensional spaces, on the other hand, the eigenfunctions of \eqref{eq:eigenproblemIn} are usually computed numerically, e.g., by discretizing the problem at hand by using the Garlekin \cite{Piestun:00} or singular value decomposition \cite{Dardari_2020} methods. In \cite{Miller}, the author has shown that the product of two PSWFs is optimal in two-dimensional spaces, under the assumption $\alpha = \beta = 0$ and by considering the paraxial setting. 

Thanks to the analytical formulation of $\bar{G}_{Tx}^{\mathrm{no-parax}} (\mathbf{r}_{Tx},\mathbf{r}_{Tx}')$ in \eqref{eq:nonParaxialSA_app}, we derive analytical expressions for the eigenfunctions of \eqref{eq:eigenproblemIn} in the non-paraxial setting. We prove that the eigenfunctions in the non-paraxial setting can be formulated in terms of the eigenfunctions of the operator kernels corresponding to the sub-HoloSs in the paraxial setting. If these latter eigenfunctions are known in closed form, the eigenfunctions in the non-paraxial setting are formulated in closed form too.

\vspace{-0.35cm}
\subsection{Paraxial Setting} \vspace{-0.15cm}
Thus, we first generalize \cite{Miller} for application to the paraxial setting, considering some non-broadside settings. Motivated by Fig. \ref{fig:particularCases}, we analyze network configurations fulfilling the conditions $\tau_{12} = \tau_{21} = 0$. In this latter case, we also show that a more accurate analysis of the eDoF can be done, which is not restricted to the asymptotic regime assumed in Lemma 9.

\noindent
\textbf{Proposition 3.} Assume $\tau_{12} = \tau_{21} = 0$. Let $P_m\left( {\xi ;p} \right)$ denote the $m$th PSWF with parameter $p$ that is solution of the equation in \cite[Eq. (11)]{Slepian1961}. The optimal eigenfunctions (encoding waveforms) of the eigenproblem in \eqref{eq:eigenproblemSimp} are given by
\vspace{-0.1cm}
\begin{equation}
\label{eq:genEigenfunctions}
{{\bar \phi }_m}\left(\mathbf{r}_{Tx}\right) = {{\bar \phi }_m}\left( {{u_i},{v_i}} \right) = P_{m_u}\left( {{u_i};{p_u}} \right)P_{m_v}\left( {{v_i};{p_v}} \right)
    \vspace{-0.1cm}
\end{equation}
where 
\vspace{-0.1cm}
\begin{equation}
{p_u} = \frac{{2\pi }}{\lambda }\frac{{{U_{Tx}}{U_{Rx}}}}{{\| {{{\bf{c}}_o}}\|}}{\tau _{11}},\quad {p_v} = \frac{{2\pi }}{\lambda }\frac{{{V_{Tx}}{V_{Rx}}}}{{\| {{{\bf{c}}_o}} \|}}{\tau _{22}} \,. \vspace{-0.1cm}
\end{equation}
\begin{IEEEproof} See Appendix K. \end{IEEEproof} 

From Prop. 3, we evince that the eigenfunctions are not PSWFs but are the product of two PSWFs whose parameters depend on the network deployment, e.g., $\alpha$ and $\beta$ through the coefficients $\tau_{11}$ and $\tau_{22}$. As for the computation of the PSWFs in Prop. 3,  many algorithms can be utilized. In Sec. VII, we utilize the numerical algorithms reported in \cite{PSWFgen}.

\noindent
\textbf{Eigenvalues and eDoF in the Non-asymptotic Regime}:
In Sec. V, we have analyzed the eDoF in the asymptotic regime defined in Lemma 9. Specifically, Prop. 1 can be applied to any network deployment in the paraxial setting (including the cases $\tau_{12} \ne 0$  and/or $\tau_{21} \ne 0$). If $\tau_{12} = \tau_{21} = 0$, the eigenvalues and hence the eDoF can be analyzed beyond the asymptotic regime thanks to the properties of the PSWFs. To elaborate, from the proof in App. K, the eigenvalues ${{{\mu _m}} }$ are given by
\vspace{-0.1cm}
\begin{equation} \label{Eq:EigenValuesPSWF}
{\mu _m} = \frac{{{\lambda ^2}{{\| {{{\bf{c}}_o}} \|}^2}}}{{{\tau _{11}}{\tau _{22}}}}{\left| {{{\bar g}_{i,o}}\left( {{{\bf{c}}_{Tx}};{{\bf{c}}_{Rx}}} \right)} \right|^2}{{\tilde \mu }_{m,i,o,u}}{{\tilde \mu }_{m,i,o,v}} \vspace{-0.1cm}
\end{equation}
where ${{{{\tilde \mu }_{m,i,o,u}}} }$ is the eigenvalue obtained by solving \cite[Eq. (11)]{Slepian1961} with ${T_u} = 2{U_{Tx}}$ and ${\Omega _u} = 2{p_u}/{T_u}$, and ${{{{\tilde \mu }_{m,i,o,v}}}}$ is the eigenvalue obtained by solving \cite[Eq. (11)]{Slepian1961} with ${T_v} = 2{V_{Tx}}$ and ${\Omega _v} = 2{p_v}/{T_v}$. The eigenvalues ${{{{\tilde \mu }_{m,i,o,u}}} }$ and ${{{{\tilde \mu }_{m,i,o,v}}} }$ have the property to be nearly equal to the operator norm of the associated one-dimensional eigenproblem for $m \le N_u = 2p_u/\pi$ and $m \le N_v = 2p_v/\pi$, and to fall off to zero very rapidly for $m > N_u$ and $m > N_v$, respectively \cite{Slepian1961}. Thus, the number of eigenvalues  in Prop. 3 that are nearly equal to the operator norm $||\bar{G}_{Tx}||_{\mathrm{op}}$ given in Lemma 11 is equal to
\vspace{-0.1cm}
\begin{equation} \label{Eq:NeffPSWF}
{N_{{\rm{eDoF}}}^{\rm{PSWF}}} \approx {N_u}{N_v} = \frac{{2{p_u}}}{\pi }\frac{{2{p_v}}}{\pi } = \frac{{{A_{Tx}}{A_{Rx}}}}{{{\lambda ^2}{{\| {{{\bf{c}}_o}} \|}^2}}}\left|{\tau _{11}}{\tau _{22}} \right|\vspace{-0.1cm}
\end{equation}
which coincides with \eqref{eq:DoFparax} if $\tau_{12} = \tau_{21} = 0$. 

Prop. 1 and Prop. 3 provide hence consistent results, with Prop. 1, based on Lemma 9 and Prop. 3, following from the properties of PSWFs. Prop. 3 allows us to better characterize the distribution of the eigenvalues compared with Prop. 1. It is known, in fact, that the transition region between nearly equal to $||\bar{G}_{Tx}||_{\mathrm{op}}$ and nearly equal to zero eigenvalues is known only in one-dimensional spaces \cite{Landau1975}, \cite[Eq. (2.132), Fig. 2.13]{FranceschettiBook}. In simple terms, this is the reason why the number of eDoF in Prop. 1 is independent of the approximation accuracy $\gamma$. In the paraxial setting and if $\tau_{12} = \tau_{21} = 0$, Prop. 3 allows us to gain more information about the relationship between the eDoF and the level of approximation accuracy $\gamma$. 

Based on \cite[Eq. (2.132)]{FranceschettiBook}, $N_u = 2p_u/\pi$ and $N_v = 2p_v/\pi$ correspond (approximately) to the number of eigenvalues greater than half of the operator norm of the associated one-dimensional eigenproblems. If $\gamma = 0.5$, in fact, \cite[Eq. (2.132)]{FranceschettiBook} tends to $N_u = 2p_u/\pi$ and $N_v = 2p_v/\pi$. Accordingly, $N_{{\rm{eDoF}}}$ in \eqref{eq:DoFparax} and ${N_{{\rm{eDoF}}}^{\rm{PSWF}}}$ in \eqref{Eq:NeffPSWF} provide, if $\tau_{12} = \tau_{21} = 0$, an estimate of the number of eigenvalues that are no smaller than half of $||\bar{G}_{Tx}||_{\mathrm{op}}$ given in Lemma 11. \textcolor{black}{In mathematical terms}
\vspace{-0.1cm}
\begin{equation} \label{Eq:PSWFproduct}
\textcolor{black}{N_{{\rm{eDoF}}} = N_{{\rm{eDoF}}}^{{\rm{PSWF}}} \approx \max \left\{ {m:{||\bar{G}_{Tx}||_{\mathrm{op}}}/2 \le {\mu _m} \le {||\bar{G}_{Tx}||_{\mathrm{op}}}} \right\} \,.} \nonumber
\vspace{-0.1cm}
\end{equation}

\textcolor{black}{This remark provides a justification about the reason why some authors, e.g., \cite{DardariD21}, \cite{Hongliang_2025}, have assumed $\gamma=0.5$ to validate their analytical frameworks for the eDoF against numerical methods: The considered network deployments are consistent with the assumption $\tau_{12} = \tau_{21} = 0$ and the paraxial setting}.

\vspace{-0.35cm}
\subsection{Non-paraxial Setting}
\textcolor{black}{The following proposition provides a general result about the eigenfunctions of the eigenproblem in \eqref{eq:eigenproblemSimpNP}, as well their relationship with the eigenfunctions in the paraxial setting}.

\textcolor{black}{\textbf{Proposition 4.} Consider the eigenproblem in \eqref{eq:eigenproblemSimpNP}, by setting $\mathcal{S}_{Tx} = r \mathcal{S}_{Tx}'$ with $\mathcal{S}_{Tx}'$ being a fixed set. In the asymptotic regime $r \rightarrow \infty$, the eigenfunctions of \eqref{eq:eigenproblemSimpNP} are given by} \vspace{-0.1cm}
\begin{equation}
\label{eq:NP_eigenfunctions}
    \textcolor{black}{\bar{\phi}_m(\mathbf{r}_{Tx}) = \exp\biggl\{- j \left( \Delta k_u^{n} u_i + \Delta k_v^{n} v_i \right)\biggl\} {\bar{\phi}}_{m}^{n}(\mathbf{r}_{Tx})} \vspace{-0.1cm}
\end{equation}
\textcolor{black}{where ${\bar{\phi}}_{m}^{n}(\mathbf{r}_{Tx})$ are the eigenfunctions of the eigenproblem in \eqref{eq:eigenproblemSimp} with the kernel defined in \eqref{eq:GBarTXn} for $n=1, 2, \ldots, N_r$}.
\begin{IEEEproof} \textcolor{black}{See Appendix L}. \end{IEEEproof} 

\textcolor{black}{Based on Prop. 4, three main observations about the eigenfunctions in the non-paraxial setting can be made: (i) Eq. \eqref{eq:NP_eigenfunctions} is general and can be applied even if $\tau_{12} \ne 0$ or $\tau_{21} \ne 0$; (ii) if the eigenfunctions in the paraxial setting are formulated in closed-form (e.g., based on Prop. 3), then those in the non-paraxial regime are obtained by applying a translation in the wavenumber domain (as apparent from the exponential function in \eqref{eq:NP_eigenfunctions}); and (iii) each sub-HoloS contributes with independent eigenfunctions whose supports in the wavenumber domain do not overlap by virtue of Lemma 17. If $\tau_{12} = 0$ or $\tau_{21} = 0$, in addition, we note that Prop. 3 and Prop. 4 combined together provide closed-form expressions for the eigenfunctions in the non-paraxial setting, which can be formulated in terns of products of PSWFs. The numerical results illustrated in the next section will show that the eigenfunctions are spatially localized within the sub-HoloS, thanks to the properties of the PSWFs}.

\vspace{-0.35cm}
\section{Numerical Results}
\label{sec:numericalResults} \vspace{-0.15cm}
We show numerical results to validate the main findings of the analysis. We assume $f_c = 28$ GHz (i.e., $\lambda = 1.07$ cm), and, similar to the notation in Lemma 14, we define $x_o = \|\mathbf{c}_o\| \sin \phi_o \cos \theta_o$, $y_o = \|\mathbf{c}_o\| \cos \phi_o \cos \theta_o$, $z_o = \|\mathbf{c}_o\| \sin \theta_o$. The paraxial and non-paraxial settings are both analyzed.

\vspace{-0.20cm}
\subsection{Paraxial Setting}
To fulfill the paraxial setting, we set $2U_{Tx} = 2V_{Tx} = 2U_{Rx} = 2V_{Rx} = 32 \lambda = 34.24$ cm, $\|\mathbf{c}_o\| = 256 \lambda =2.74$ m. Also, we consider $\alpha = 0$ and $\beta = \pi/2$ as an example. 

\begin{figure*}
\begin{minipage}[t]{0.33\linewidth} \hspace{-1.0cm}
\centering
\includegraphics[scale=.31]{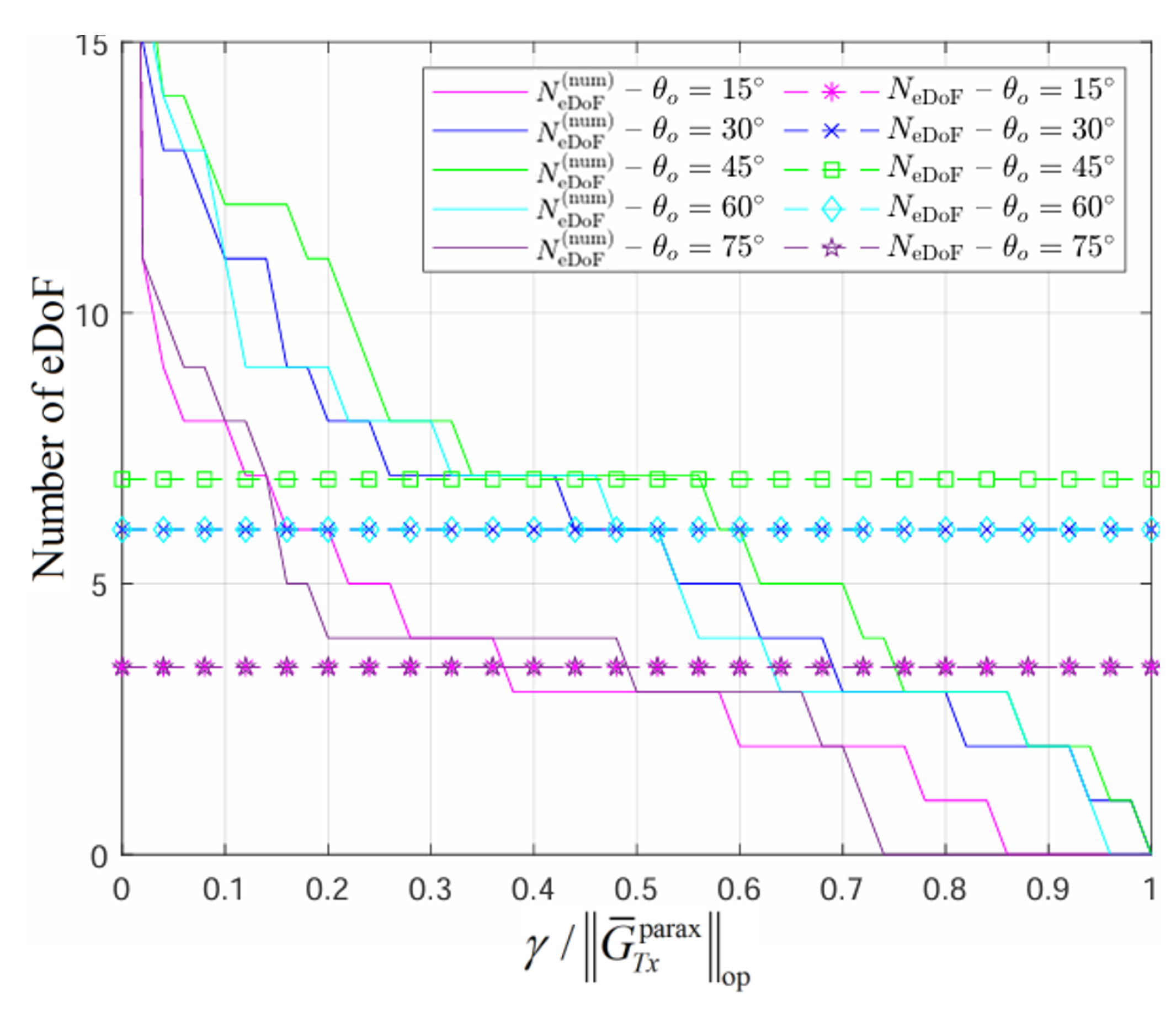}\\
\vspace{-0.40cm}
\caption{eDoF (paraxial setting, $\phi_o = \pi/6$) - Solid lines: Numerical solution. Markers: $N_{\mathrm{eDoF}}$ in \eqref{eq:DoFparax}.}
\label{fig:P_DoF}
\end{minipage}
\begin{minipage}[t]{0.02\linewidth}
\end{minipage}
\begin{minipage}[t]{0.33\linewidth} \hspace{-1.0cm}
\centering
\includegraphics[scale=.31]{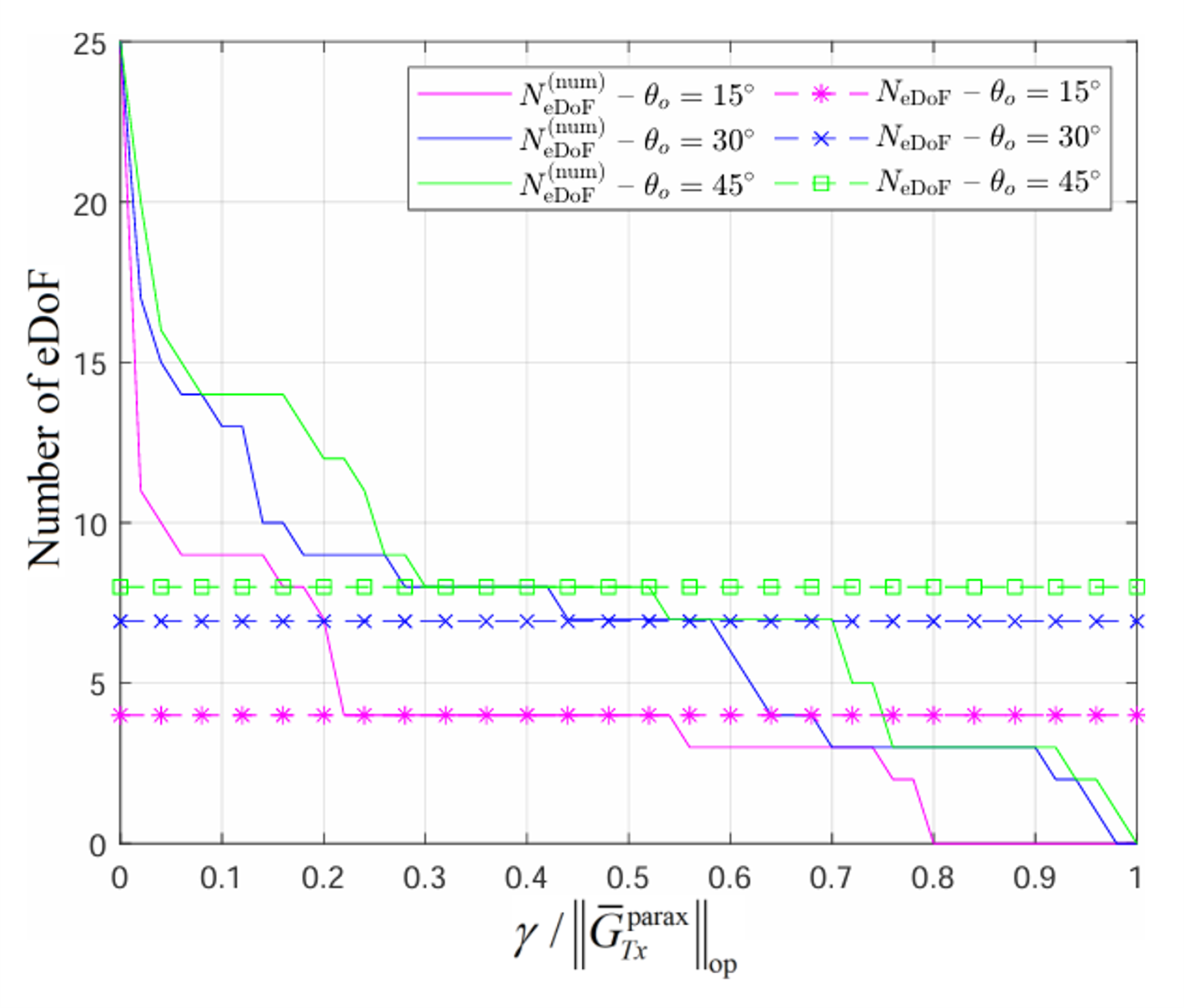}\\
\vspace{-0.40cm}
\caption{Eigenvalues (paraxial setting, $\phi_o = 0$) - Solid lines: Numerical solution. Markers: PSWF in \eqref{Eq:EigenValuesPSWF}.}
\label{fig:P_DoFPSWF}
\end{minipage}
\begin{minipage}[t]{0.02\linewidth}
\end{minipage}
\begin{minipage}[t]{0.33\linewidth} \hspace{-1.0cm}
\centering
\includegraphics[scale=.32]{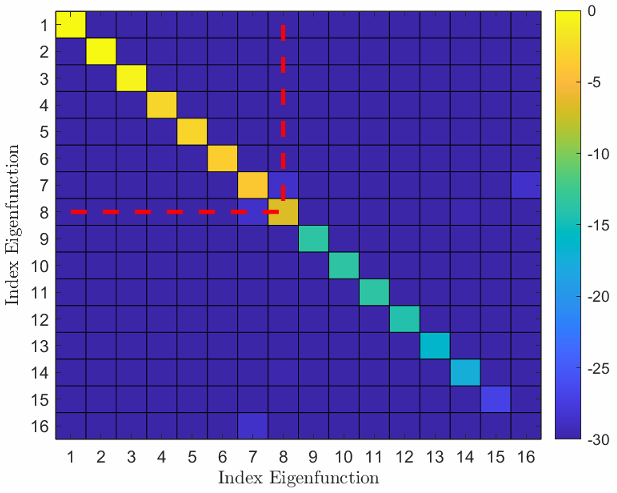}\\
\vspace{-0.40cm}
\caption{Correlation matrix (in dB) in the paraxial setting. The red lines indicate $N_{\mathrm{eDoF}}$ in \eqref{eq:DoFparax}.}
\label{fig:P_eigenfunctions}
\end{minipage}
\vspace{-0.50cm}
\end{figure*}

In Fig. \ref{fig:P_DoF}, we portray the eDoF as a function of the approximation accuracy $\gamma$ (normalized by the operator norm ${||\bar{G}_{Tx}||_{\mathrm{op}}}$) and $\theta_o$. We compare the analytical framework in \eqref{eq:DoFparax} against the empirical distribution of the eigenvalues, which is obtained numerically by using the same approach as in \cite{Dardari_2020} and \cite{Hongliang_2025}, i.e., the singular value decomposition. According to  Sec. VI-A, \eqref{eq:DoFparax} provides a good estimate for the number of eigenvalues no less than half of the operator norm. In Fig. \ref{fig:P_DoFPSWF}, a similar study is shown by depicting the formula in \eqref{Eq:EigenValuesPSWF}. Similar conclusions can be drawn, as expected. 

In Fig. \ref{fig:P_eigenfunctions}, we analyze the orthonormality of the eigenfunctions in \eqref{eq:genEigenfunctions} when they are observed at the receiver. As an example, we consider $\theta_o = \pi/4$ and $\phi_o = 0$. The PSWFs in \eqref{eq:genEigenfunctions} are computed by using \cite{PSWFgen}, and the numerical waveforms are obtained by applying the singular value decomposition to the exact eigenproblem in \eqref{eq:eigenproblemIn}. Each eigenfunction obtained from \eqref{eq:genEigenfunctions} is inserted in \eqref{Eq:Efield}, and the electric field at the receiving HoloS is computed numerically. The resulting electric field is multiplied by each eigenfunction, and it is integrated (cross-correlation) across the receiving HoloS. We see that the orthonormal waveforms at the transmitting HoloS are still orthonormal at the receiving HoloS, even though they are obtained by applying the proposed approximation. We see that only $N_{\mathrm{eDoF}}$ eigenfunctions, with $N_{\mathrm{eDoF}}$ given in \eqref{eq:DoFparax}, are strongly coupled. This confirms the analytical derivation.

\vspace{-0.15cm}
\subsection{Non-paraxial Setting}
\textcolor{black}{To fulfill the non-paraxial setting, we set $2U_{Tx} = 2V_{Tx} = 8 \lambda = 8.56$ cm, $\|\mathbf{c}_o\| = 32\lambda = 34.24$ cm. Also, we consider $\theta_o = 0$, $\phi_o = 0$, $\alpha = 0$, and $\beta = 0$, as an example}. 

\begin{figure*}
\begin{minipage}[t]{0.33\linewidth} \hspace{-1.0cm}
\centering
\includegraphics[scale=.31]{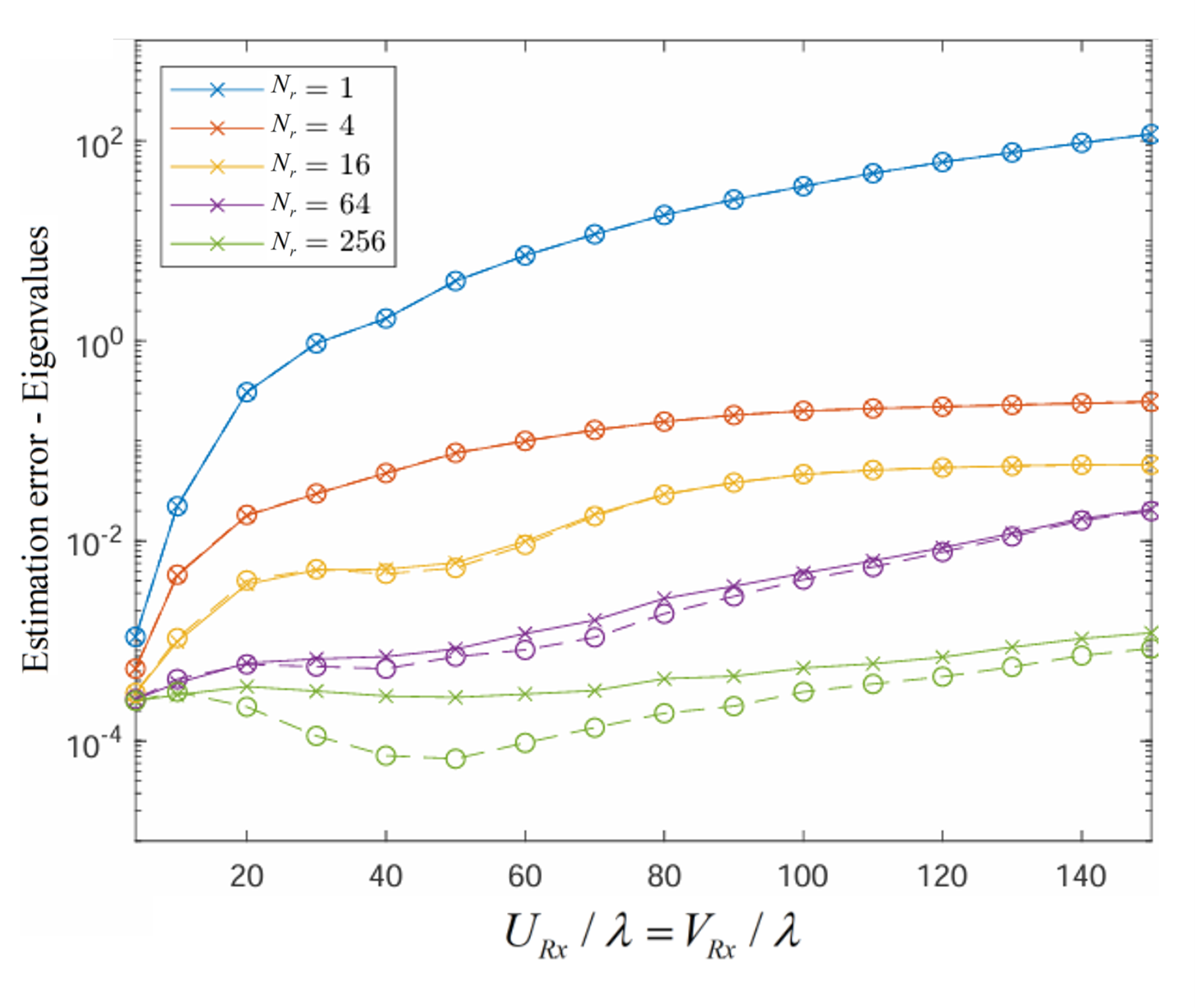}\\
\vspace{-0.30cm}
\caption{Error in estimating the eigenvalues (non-paraxial) - Solid and dashed lines from \eqref{eq:nonParaxialSA} and \eqref{eq:nonParaxialSA_app}.}
\label{fig:NP_kernel}
\end{minipage}
\begin{minipage}[t]{0.02\linewidth}
\end{minipage}
\begin{minipage}[t]{0.33\linewidth} \hspace{-1.0cm}
\centering
\includegraphics[scale=.31]{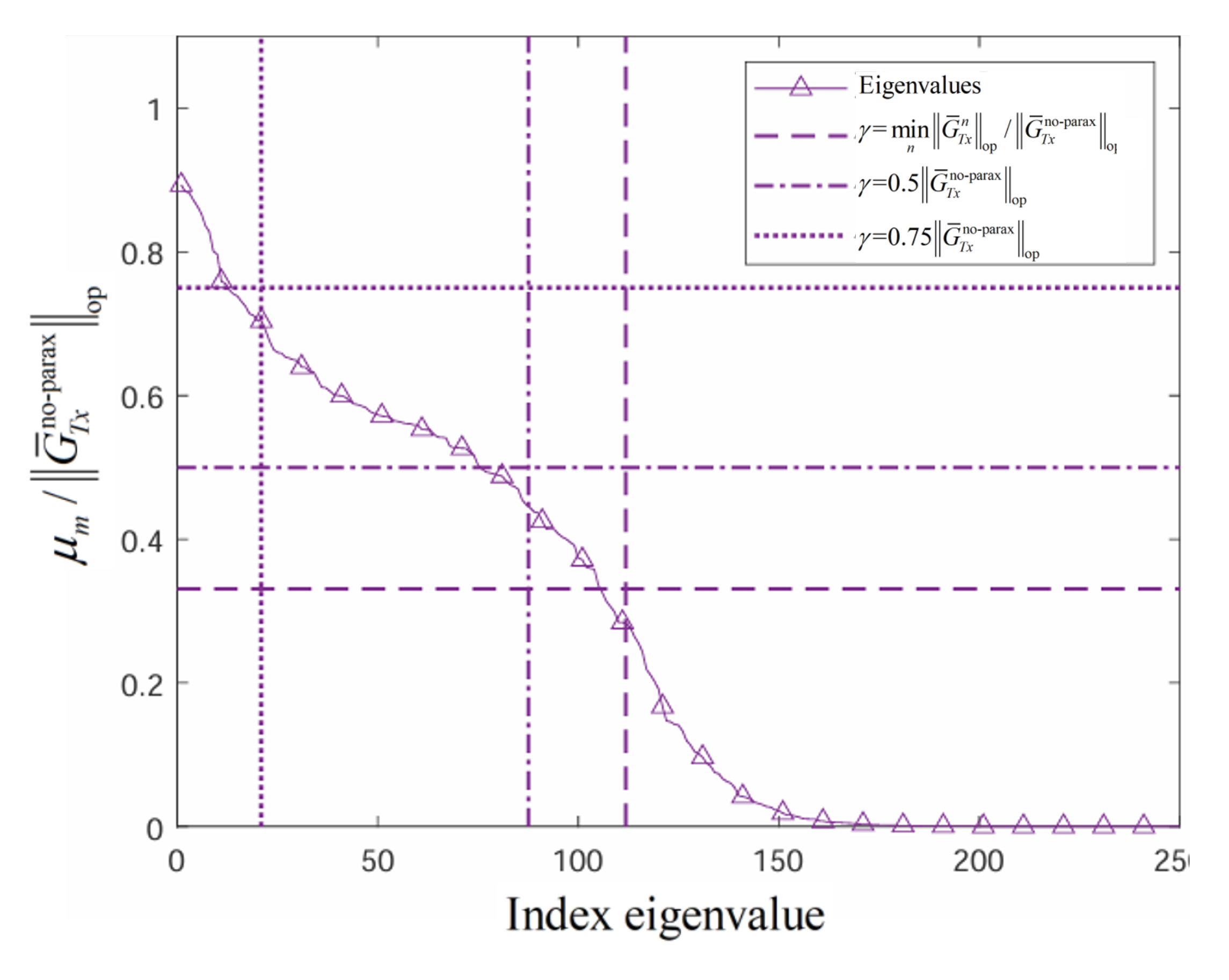}\\
\vspace{-0.30cm}
\caption{$N_{\mathrm{eDoF}}(\gamma)$ (non-paraxial) - Horizontal lines: Values of $\gamma$. Vertical lines: $N_{\mathrm{eDoF}}(\gamma)$ in \eqref{eq:DoFNP}.}
\label{fig:NP_eigenvalues}
\end{minipage}
\begin{minipage}[t]{0.02\linewidth}
\end{minipage}
\begin{minipage}[t]{0.32\linewidth} \hspace{-1.0cm}
\centering
\includegraphics[scale=.32]{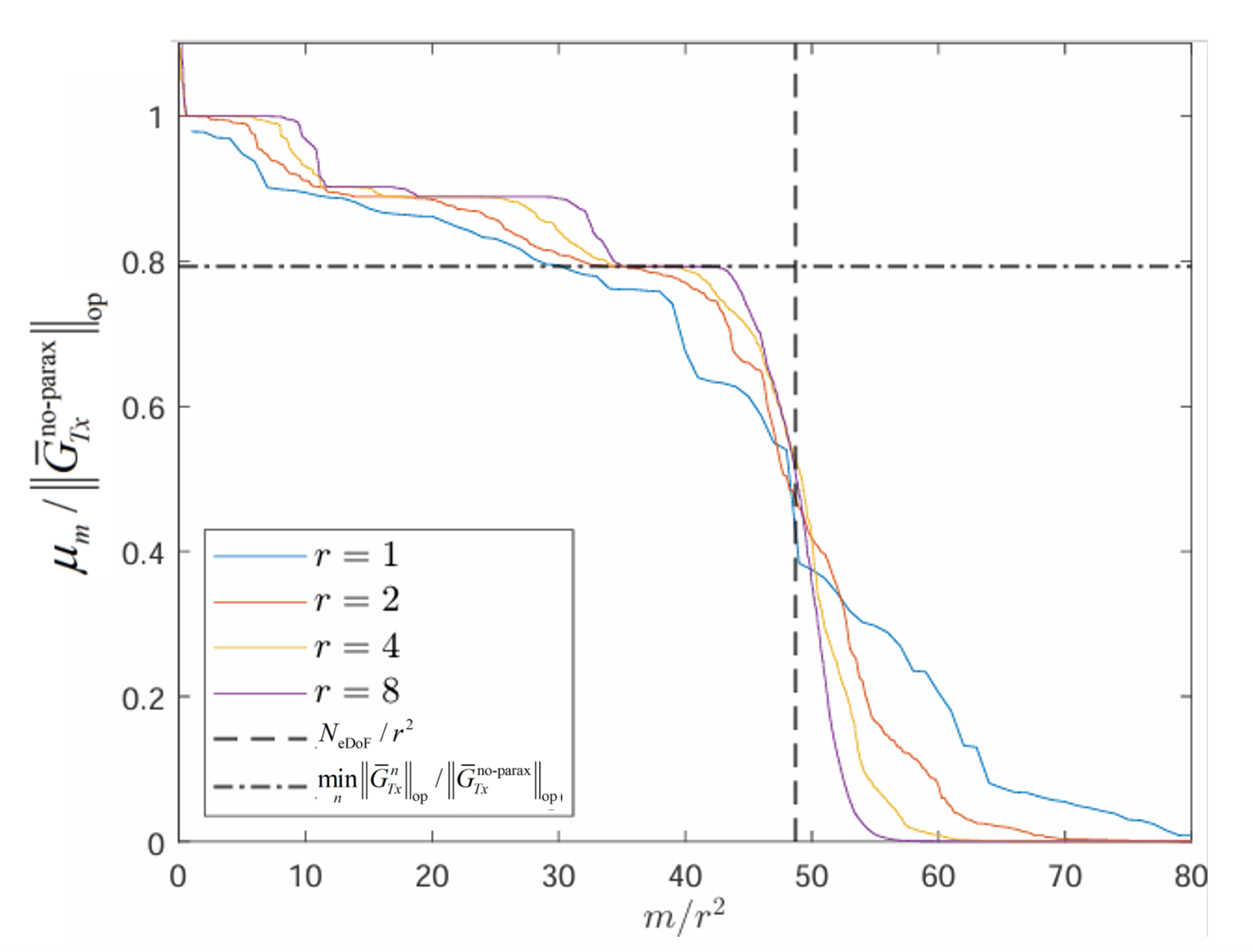}\\
\vspace{-0.30cm}
\caption{Asymptotic behavior of the eigenvalues in the non-paraxial setting ($2U_{Rx} = 2V_{Rx} = 32\lambda$).}
\label{fig:NP_asymptotic}
\end{minipage}
\vspace{-0.55cm}
\end{figure*}

\textcolor{black}{In Fig. \ref{fig:NP_kernel}, we analyze the accuracy of the proposed approximations against numerical estimates of the eigenvalues, which are obtained as in the paraxial case. The figure portrays the error function ${\sum\left|\mu_{e,m} - \mu_{e,a}\right|^2}/    {\sum\left|\mu_{e,m}\right|^2}$, with $\mu_{e,m}$ and $\mu_{a,m}$ the eigenvalues obtained numerically without and with the proposed approximations, respectively. The approximations in \eqref{eq:nonParaxialSA} and \eqref{eq:nonParaxialSA_app} are compared. We see that no major inaccuracy is introduced by the approximation in Lemma 15. As expected, the accuracy improves as the number of sub-HoloSs increases. Next, we set $2U_{Rx}^n = 2V_{Rx}^n = 8 \lambda$, i.e., $N_r = 64$ sub-HoloSs, as the estimation error is sufficiently small for the setup at hand}. 

\textcolor{black}{In Fig. \ref{fig:NP_eigenvalues}, we depict $N_{\mathrm{eDoF}}(\gamma)$ in \eqref{eq:DoFNP} for three values of the approximation accuracy $\gamma$ (normalized to the operator norm $||\bar{G}_{Tx}^{\mathrm{no-parax}}||_{\mathrm{op}}$ in \eqref{Eq:OpNomNP}). The comparison against numerical simulations shows a good agreement with the analytical framework in \eqref{eq:DoFNP}. Also, the dependence on $\gamma$ is sufficiently accurate. The distribution of the eigenvalues obtained with the numerical methods does not manifest, however, a clear step-like behavior (as in Fig. \ref{fig:asymptoticRegime}). This is because the asymptotic regime of Lemma 8 is fulfilled only approximately. To illustrate this  aspect, Fig. \ref{fig:NP_asymptotic} shows the empirical distribution of the eigenvalues for different values of $r$, i.e., setting $\mathcal{S}_{Tx} = r \mathcal{S}_{Tx}'$ with $\mathcal{S}_{Tx}' = 4U_{Tx} V_{Tx}$ and $2U_{Tx} = 2V_{Tx} = 8 \lambda$. It is apparent that the larger $r$, the more well-defined the step-like behavior}. 

\begin{figure}
\begin{minipage}[t]{0.45\textwidth} 
\centering
\includegraphics[scale=.31]{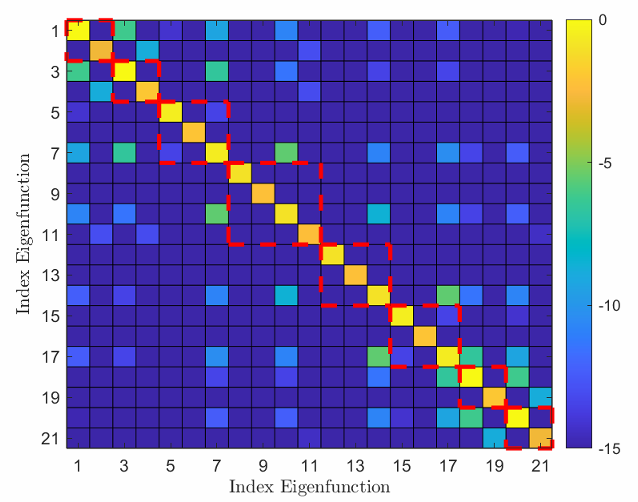}\\
\vspace{-0.35cm}
\caption{Correlation matrix (in dB) for the non-paraxial setting. The red squares group the eigenfunctions that belong to the same sub-HoloS.}
\label{fig:NP_orthogonality} 
\end{minipage} 
\begin{minipage}[t]{0.45\textwidth} \vspace{0.1cm}
\centering
\includegraphics[scale=.33]{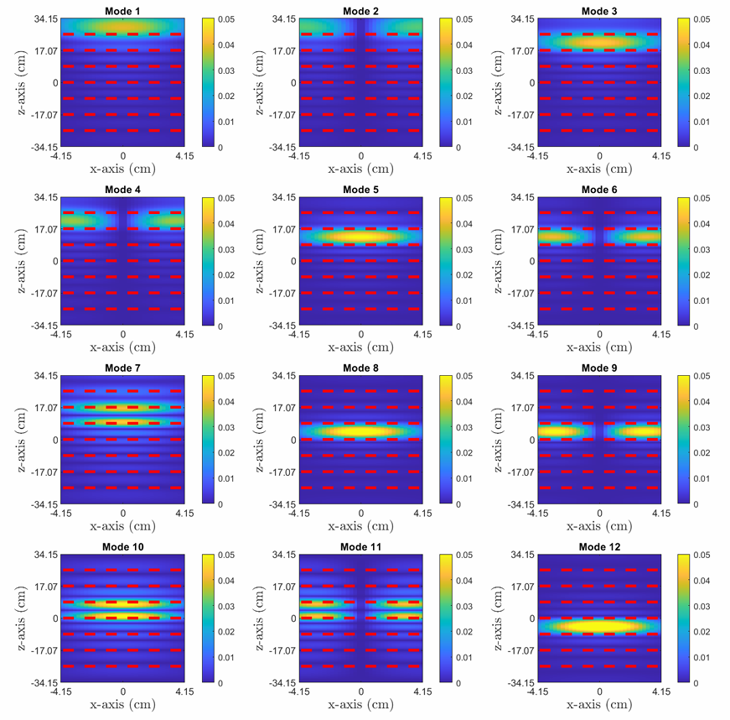}\\
\vspace{-0.4cm}
\caption{Received signal (non-paraxial) when transmitting the first 12 eigenfunctions. The red lines indicate the boundary of the sub-HoloSs.}
\label{fig:NP_eigenfunctions}
\end{minipage}
\vspace{-0.40cm}
\end{figure}

\textcolor{black}{In Fig. \ref{fig:NP_orthogonality}, we analyze the correctness of the eigenfunctions in \eqref{eq:NP_eigenfunctions}. To ease the numerical evaluation, we set $2U_{Rx} = 64 \lambda$, $2 V_{Rx} = 8 \lambda$. Since $2U_{Tx} = 2V_{Tx} = 8 \lambda$, this implies that the receiving HoloS is a strip. Thus, the eigenfunctions of each HoloS can be expressed in terms of PSWFs, as discussed in Sec. VI-B. Fig. \ref{fig:NP_orthogonality} is obtained as Fig. \ref{fig:P_eigenfunctions}. We see an excellent orthogonality among the eigenfunctions from each sub-HoloS (those in the red boxes). The small cross-correlation inaccuracies among the eigefunctions not contained within the same red box are obtained because the asymptotic regime $r \to \infty$ is fulfilled only approximately, as shown in Fig. \ref{fig:NP_asymptotic}}. 

\textcolor{black}{In Fig. \ref{fig:NP_eigenfunctions}, we portray the intensity of the electric field at the receiving HoloS, when the eigenfuctions in \eqref{eq:NP_eigenfunctions} are emitted by the transmitting HoloS. The numerical results are obtained by using \eqref{eq:E_x_gen}, setting the surface current density equal to the eigenfuctions in \eqref{eq:NP_eigenfunctions}. The figure confirms that the eigenfunctions are almost confined within the corresponding sub-HoloSs. The small leakage of energy is due to the asymptotic approximations made to derive the eigenfunctions in \eqref{eq:NP_eigenfunctions}}.

\vspace{-0.25cm}
\section{Conclusion} \vspace{-0.15cm}
\textcolor{black}{We have introduced an analytical framework to estimate the eDoF of holographic MIMO systems. The proposed approach can be applied to general antenna deployments, in paraxial and non-paraxial settings, provided that a HoloS is small compared to the other. The obtained analytical framework has unveiled fundamental differences between paraxial and non-paraxial deployment scenarios. In the non-paraxial setting, the eigenvalues polarize asymptotically to multiple non-zero eigenvalues and the eDoF depend on the approximation accuracy. From a system perspective, a communication link in the paraxial and non-paraxial settings behaves as an ideal band-pass filer and as an ideal multi-band band-pass filter with adjacent bands. Also, we have proved that, in some network deployments, the optimal communication waveforms for non-paraxial settings are shifted versions, in the wavenumber domain, of the optimal communication waveforms for non-paraxial settings. Finally, we have discussed the relationship between the cut-set integral and self-adjoint operator methods. Possible generalizations of this paper include the analysis of holographic MIMO where both HoloSs are large compared to the transmission distance}.

\vspace{-0.25cm}
\section*{Appendix A -- Proof of Lemma 10} \vspace{-0.1cm}
Equation \eqref{eq:focusingFunction} follows from \eqref{eq:defFocFunction}, by using the parametrizations in \eqref{eq:transSurface} and \eqref{eq:recSurfacen}. By using \eqref{eq:transSurface} and \eqref{eq:recSurfacen} to compute $p^n(\mathbf{r}_{Tx},\mathbf{r}_{Rx})$ in \eqref{eq:P_Parax}, the operator kernel in \eqref{eq:genGBarTXn} simplifies to 
\begin{multline}
    \bar{G}_{Tx}^n\left( \mathbf{r}_{Tx},\mathbf{r}_{Tx}' \right) \approx |\bar{g}_{i,o}^n|^2 \\ \times \int_{-U_{Rx}^n}^{U_{Rx}^n} \int_{-V_{Rx}^n}^{V_{Rx}^n} 
    \exp\bigg\{-j \frac{\kappa_0}{\|\mathbf{c}_o^n\|} \left[(\tau_{11}^n(u_i-u_i^{\prime}) +\tau_{12}^n(v_i-v_i^{\prime}))u_o^n \right.\\ \hspace{0.75cm}\left.
+ (\tau_{21}^n(u_i-u_i^{\prime})  +\tau_{22}^n(v_i-v_i^{\prime}))v_o^n\right]\bigg\} d v_o^n d u_o^n 
\end{multline}

Since $\int_{-a}^{a} \exp(-jbx)dx = 2 a  \sinc (ba/\pi)$, the expression in \eqref{eq:GBarTXn} follows noting that $\kappa_0=2\pi/\lambda$ and $A_{Rx}^n=(2 U_{Rx}^n)(2 V_{Rx}^n)$.

\vspace{-0.25cm}
\section*{Appendix B -- Proof of Lemma 11} \vspace{-0.1cm}
\textcolor{black}{According to Definition 6, the operator norm for the eigenproblem in \eqref{eq:eigenproblemSimp} can be formulated as follows}: \vspace{-0.1cm}
\begin{align} 
    ||\bar{G}_{Tx}||_{\mathrm{op}} &= \mathop{\sup}\nolimits_{||\bar{\phi}|| = 1}  \Biggl|\Biggl|\int_{\mathcal{S}_{Tx}} h_G^{\mathrm{parax}}(\mathbf{r}_{Tx} - \mathbf{r}_{Tx}') \bar{\phi}(\mathbf{r}_{Tx}') \, d\mathbf{r}_{Tx}'\Biggl|\Biggl| \nonumber \\
    &= \mathop{\sup}\nolimits_{||\bar{\phi}|| = 1} ||h_G^{\mathrm{parax}}(\mathbf{r}_{Tx}) * \left[ \mathds{1}_{\mathcal{S}_{Tx}}(\mathbf{r}_{Tx}) \bar{\phi}(\mathbf{r}_{Tx})\right] || \,. \nonumber \vspace{-0.1cm}
\end{align}

\textcolor{black}{Applying Parseval's theorem \cite[Prop. 2.12]{dubois2019multidimensional} and using the Fourier transform in \eqref{eq:fourierTransform}, the operator norm simplifies to} \vspace{-0.1cm}
\begin{align} 
||\bar{G}_{Tx}||_{\mathrm{op}} &= \mathop{\sup}\nolimits_{||\bar{\phi}|| = 1} || a_G \mathds{1}_{\mathcal{H}_G}(\mathbf{k}_{Tx})\mathcal{F}\left[ \mathds{1}_{\mathcal{S}_{Tx}}(\mathbf{r}_{Tx}) \bar{\phi} (\mathbf{r}_{Tx})\right] || \nonumber \vspace{-0.1cm}
\end{align}
\textcolor{black}{where $a_G = |\bar{g}_{i,o}|^2 { \lambda^2 \|\mathbf{c}_o\|^ 2}/{|\tau_{11} \tau_{22} - \tau_{12}\tau_{21}|}$. The supremum of $||\bar{G}_{Tx}||_{\mathrm{op}}$ is attained by choosing $\bar{\phi} (\mathbf{r}_{Tx})$  so that its support (in the spatial domain) is contained in $\mathcal{S}_{Tx}$ and the support of its Fourier transform (in the wavenumber domain) is contained in $\mathcal{H}_G$. These conditions can be achieved only asymptotically, as a non-zero function cannot have finite supports in both domains simultaneously. Under these conditions, $||\bar{G}_{Tx}||_{\mathrm{op}} = a_G$}.

\vspace{-0.25cm}
\section*{Appendix C -- Proof of Lemma 13}
\label{app:Theo1} \vspace{-0.1cm}
To calculate the Fourier transform of ${h_G^{\mathrm{parax}}}\left(\mathbf{r}_{Tx}\right)$ in the wavenumber domain, we first compute the Fourier transform of $h_0(\mathbf{q}) = |\bar{g}_{i,o}|^2 A_{Rx} \sinc(s) \sinc(t)$, and we then apply the linear transformation $h_G^{\mathrm{parax}}(\mathbf{r}_{Tx}) = h_0(\mathbf{q} = \mathbf{A} \mathbf{r}_{Tx})$, where 
\vspace{-0.1cm}
\begin{equation}
    \mathbf{A}=
    \begin{bmatrix} U_o \tau_{11} & U_o \tau_{12}\\ V_o \tau_{21} & V_o \tau_{22} \end{bmatrix} \,. \vspace{-0.1cm}
\end{equation}

Specifically, the Fourier transform of $h_0(\mathbf{q})$ is  \cite[Tab. 2.2]{dubois2019multidimensional} \vspace{-0.1cm}
\begin{equation} \label{Eq:H0}
    H_0(\boldsymbol{\kappa}_q) = |\bar{g}_{i,o}|^2 A_{Rx} \mathds{1}_{\mathcal{H}_0}(\boldsymbol{\kappa}_q)  \vspace{-0.1cm}
\end{equation}
with ${{\mathcal{H}}_0} = \left\{ {\left( {{\kappa_s},{\kappa_t}} \right):\left| {{\kappa_s}} \right| \le \pi ,\left| {{\kappa_t}} \right| \le \pi } \right\}$.

Using \cite[Prop. 2.6]{dubois2019multidimensional}, the Fourier transform of $h_G^{\mathrm{parax}}(\mathbf{r}_{Tx}) = h_0(\mathbf{q} = \mathbf{A} \mathbf{r}_{Tx})$ is $H_G^{\mathrm{parax}}(\boldsymbol{\kappa}_{Tx})= H_0(\boldsymbol{\kappa}_q=\mathbf{A}^{-T}\boldsymbol{\kappa}_{Tx})$ $/|\det{\mathbf{A}}|$, with ${{\bf{A}}^{ - T}} = {\left( {{{\bf{A}}^{ - 1}}} \right)^T}$. $H_G^{\mathrm{parax}}(\boldsymbol{\kappa}_{Tx})$ is given in \eqref{eq:fourierTransform}.

We see that $H_0(\boldsymbol{\kappa}_q)$ is an ideal low-pass filter that is non-zero in the set ${{\mathcal{H}}_0}$, whose Lebesgue measure is ${m_0} = m\left( {{{\mathcal{H}}_0}} \right) = 4{\pi ^2}$. ${m_0}$ is often called spatial bandwidth of $H_0(\boldsymbol{\kappa}_q)$. The spatial bandwidth of $H_G^{\mathrm{parax}}(\boldsymbol{\kappa}_{Tx})$ can be obtained from the transformation $\boldsymbol{\kappa}_{q} = \mathbf{A}^{-T}\boldsymbol{\kappa}_{Tx}$ applied to \eqref{Eq:H0} in the wavenumber domain. By applying, therefore, the change of variable $\boldsymbol{\kappa}_{Tx} = \mathbf{A}^{T}\boldsymbol{\kappa}_{q}$, the spatial bandwidth of $H_G^{\mathrm{parax}}(\boldsymbol{\kappa}_{Tx})$ is 
\begin{align}
    m_G & = m(\mathcal{H}_G) =\int_{-\pi}^{\pi} \int_{-\pi}^{\pi} |\det{\mathbf{A}^T}| dk_s dk_t \\ & = 4 \pi^2  |\det{\mathbf{A}^T}| = 4 \pi^2 \frac{A_{Rx}}{\lambda^2 \|\mathbf{c}_o|^2} \|\tau_{11} \tau_{22} - \tau_{12}\tau_{21}|
\end{align}
since $A_{Rx} = (2 U_{Rx})(2 V_{Rx})$.

\vspace{-0.15cm}
\section*{Appendix D -- Proof of Lemma 15}
\textcolor{black}{The self-adjoint kernel in \eqref{eq:eigenproblemNewFocusingFunc} can be written explicitly as}
\begin{align}
    \bar{G}_{Tx}^{\mathrm{no-parax}} (\mathbf{r}_{Tx},\mathbf{r}_{Tx}') &=   \sum\nolimits_{n=1}^{N_r} \left[f_{{Tx}}^n (\mathbf{r}_{Tx})  \left(\bar{f}_{Tx}( \mathbf{r}_{Tx})\right)^* \right] \nonumber \\
    & \hspace{-1.25cm} \times \left[f_{{Tx}}^n (\mathbf{r}_{Tx}') \left(\bar{f}_{Tx}( \mathbf{r}_{Tx}')\right)^*\right]^* \bar{G}_{Tx}^n(\mathbf{r}_{Tx},\mathbf{r}_{Tx}') \, .
\end{align}

\textcolor{black}{Let us consider the definitions of $f_{{Tx}}^n (\mathbf{r}_{Tx})$ and $\bar{f}_{Tx}( \mathbf{r}_{Tx})$ in \eqref{eq:defFocFunction} and \eqref{eq:barFocFunction}, respectively, and let us analyze the product $f_{{Tx}}^n (\mathbf{r}_{Tx})  \left(\bar{f}_{Tx}( \mathbf{r}_{Tx})\right)^* = \exp\{j \kappa_0 \Phi_{{Tx}}(\mathbf{r}_{Tx})\}$, where}
\begin{equation} \label{Eq:Phase}
   \Phi_{{Tx}}(\mathbf{r}_{Tx}) = \Psi_{{Tx}}^n(\mathbf{r}_{Tx}) + \Gamma_{{Tx}}^n(\mathbf{r}_{Tx}) - \Gamma_{{Tx}}^o(\mathbf{r}_{Tx})
\end{equation}
\begin{align}
    \Psi_{{Tx}}^n(\mathbf{r}_{Tx}) &= -\frac{1}{\|\mathbf{c}_o^n\|}\left[x_o^n u_i + z_o^n v_i\right] \label{Eq:Phase1} \\
    \Gamma_{{Tx}}^n(\mathbf{r}_{Tx}) &= \frac{1}{2\|\mathbf{c}_o^n\|}\left[u_i^2 + v_i^2 - \frac{4}{\|\mathbf{c}_o^n\|^2}\left(x_o^n u_i + z_o^n v_i\right)^2\right] \label{Eq:Phase2} \\
    \Gamma_{{Tx}}^o(\mathbf{r}_{Tx}) &= \frac{1}{2\|\mathbf{c}_o\|}\left[u_i^2 + v_i^2 - \frac{4}{\|\mathbf{c}_o\|^2}\left(x_o u_i + z_o v_i\right)^2\right] \, . \label{Eq:Phase3}
\end{align}

\textcolor{black}{By inspection of \eqref{Eq:Phase1}-\eqref{Eq:Phase3}, we evince that $\Phi_{{Tx}}(\mathbf{r}_{Tx}) \approx \Psi_{{Tx}}^n(\mathbf{r}_{Tx})$. The reason is the following: (i) if the sub-HoloSs are located nearby the sub-HoloS centered in $(x_o, y_o, z_o)$, then $x_o^n \approx x_o$, $z_o^n \approx z_o$, and $\|\mathbf{c}_o^n\| \approx \|\mathbf{c}_o\|$. Therefore, $\Gamma_{{Tx}}^n(\mathbf{r}_{Tx}) \approx \Gamma_{{Tx}}^o(\mathbf{r}_{Tx})$ and their difference almost cancel out in \eqref{Eq:Phase}; (ii) if the sub-HoloSs are not located nearby the sub-HoloS centered in $(x_o, y_o, z_o)$, then $\Psi_{{Tx}}^n(\mathbf{r}_{Tx})$ scales with $(x_o^n/\|\mathbf{c}_o^n\|)u_i + (z_o^n/\|\mathbf{c}_o^n\|)v_i$, while $\Gamma_{{Tx}}^n(\mathbf{r}_{Tx})$ and $\Gamma_{{Tx}}^o(\mathbf{r}_{Tx})$ scale with $(u_i/\|\mathbf{c}_o^n\|)u_i + (v_i/\|\mathbf{c}_o^n\|)v_i$ and $(u_i/\|\mathbf{c}_o\|)u_i + (v_i/\|\mathbf{c}_o\|)v_i$, respectively. Therefore, the dominant term is still $\Psi_{{Tx}}^n(\mathbf{r}_{Tx})$ since $u_i/\|\mathbf{c}_o^n\| \ll 1$, $u_i/\|\mathbf{c}_o^n\| \ll 1$, and $u_i/\|\mathbf{c}_o\| \ll 1$, $u_i/\|\mathbf{c}_o\| \ll 1$ by virtue of the partitioning, which is based on the condition $\max\{2U_{Tx},2V_{Tx}\} \ll \|\mathbf{c}_o^n\|$ and $\max\{2U_{Tx},2V_{Tx}\} \ll \|\mathbf{c}_o\|$ (paraxial setting between ${{\mathcal{S}}_{Tx}}$ and ${{\mathcal{S}}_{Rx}^n}$, but non-paraxial setting between ${{\mathcal{S}}_{Tx}}$ and ${{\mathcal{S}}_{Rx}}$). Eq. \eqref{eq:nonParaxialSA_app} follows by using the approximation $\Phi_{{Tx}}(\mathbf{r}_{Tx}) \approx \Psi_{{Tx}}^n(\mathbf{r}_{Tx})$}.

\vspace{-0.15cm}
\section*{Appendix E -- Proof of Lemma 17} 
\textcolor{black}{Let us introduce the notation ${\Delta ^n} = \kappa_0 \left| {\tau _{11}^n\tau _{22}^n - \tau _{12}^n\tau _{21}^n} \right|/{\left\| {{\bf{c}}_o^n} \right\|}$, $\Delta _1^n =  - \tau _{12}^n\Delta \kappa _u^n + \tau _{11}^n\Delta \kappa _v^n$, $\Delta _2^n = \tau _{22}^n\Delta \kappa _u^n - \tau _{21}^n\Delta \kappa _v^n$, and ${\Delta ^n} = \kappa_0 \left| {\tau _{11}^n\tau _{22}^n - \tau _{12}^n\tau _{21}^n} \right|/{\left\| {{\bf{c}}_o^n} \right\|}$. Then,  \eqref{Eq:SpectralBandNoParax} simplifies to}
\begin{align} \label{Eq:AppE_1}
\left\{ \begin{array}{l}
\Delta _2^n - U_{Rx}^n{\Delta ^n} \le \tau _{22}^n{\kappa _u} - \tau _{21}^n{\kappa _v} \le \Delta _2^n + U_{Rx}^n{\Delta ^n}\\
\Delta _1^n - V_{Rx}^n{\Delta ^n} \le  - \tau _{12}^n{\kappa _u} + \tau _{11}^n{\kappa _v} \le \Delta _1^n + V_{Rx}^n{\Delta ^n} \, .
\end{array} \right.
\end{align}

\textcolor{black}{In the asymptotic regime where $U_{Rx}^n$ and $V_{Rx}^n$ are sufficiently small, i.e., $U_{Rx}^n \to 0$ and $V_{Rx}^n \to 0$, we see that the spatial bandwidth reduces to two points $(\bar \kappa _u, \bar \kappa _v)$ fulfilling the conditions $\tau _{22}^n{\bar \kappa _u} - \tau _{21}^n{\bar \kappa _v} = \Delta _2^n$ and $- \tau _{12}^n{\bar \kappa _u} + \tau _{11}^n{\bar \kappa _v} = \Delta _1^n$. As a result, the supports of the Fourier transforms of $h_{G^n}^{\mathrm{no-parax}}\left(\mathbf{r}_{Tx}\right)$ are, asymptotically, almost disjoint for different values of $n$}.

\textcolor{black}{To gain further insights, consider deployments fulfilling the conditions $\tau_{12}^n = \tau_{21}^n= 0$ (see Fig. \ref{fig:particularCases}). For ease of exposition, we consider $\tau_{11}^n > 0$, $\tau_{22}^n > 0$ and $U_{Rx}^n = V_{Rx}^n = L_{Rx}$ for all the sub-HoloSs. In this case, the two expressions in \eqref{Eq:AppE_1} are decoupled in the $(\kappa_u,\kappa_v)$-domain. The supports of two sub-HoloSs in the wavenumber domain can be disjoint (not overlapping) in the (i) $\kappa_u$-domain; (ii) $\kappa_v$-domain; (iii) both in the $\kappa_u$-domain and $\kappa_v$-domain. The analysis of these three cases is similar, so we consider the first case as an example}. 

\textcolor{black}{To this end, we study the first expression in \eqref{Eq:AppE_1} and consider two generic HoloSs denoted by the superscripts $a$ and $b$. We use the notation $\tilde \Delta _2^n = \Delta _2^n / \tau_{22}^n$ and $\tilde \Delta^n = \Delta^n / \tau_{22}^n$. Also, we note that $\tilde \Delta^n > 0$ and $\tilde \Delta _2^n$ can be positive or negative. For the two sub-HoloSs $a$ and $b$, we then have the following}:
\begin{align} \label{Eq:AppE_2}
& \tilde \Delta _2^a - {L_{RX}}{{\tilde \Delta }^a} \le {\kappa _u} \le \tilde \Delta _2^a + {L_{RX}}{{\tilde \Delta }^a}\\
& \tilde \Delta _2^b - {L_{RX}}{{\tilde \Delta }^b} \le {\kappa _u} \le \tilde \Delta _2^b + {L_{RX}}{{\tilde \Delta }^b} \, .
\end{align}

\textcolor{black}{To illustrate the approach, we consider the case study when the sub-HoloS $a$ is located to the left of sub-HoloS $b$, i.e.}, 
\begin{equation} \label{Eq:AppE_3}
\tilde \Delta _2^a - \tilde U_{Rx}^a{{\tilde \Delta }^a} \le \tilde \Delta _2^b + \tilde U_{Rx}^a{{\tilde \Delta }^b} \,.
\end{equation}

\textcolor{black}{The other case study can be analyzed analogously. Three scenarios need to be considered}:\\ 
\begin{equation} \label{Eq:AppE_4}
{\rm{S1}}: \; \tilde \Delta _2^b - {L_{RX}}{{\tilde \Delta }^b} \le \tilde \Delta _2^a + {L_{RX}}{{\tilde \Delta }^a} \le \tilde \Delta _2^b + {L_{RX}}{{\tilde \Delta }^b}
\end{equation}
\begin{align} \label{Eq:AppE_5}
{\rm{S2}}: \; \left\{ \begin{array}{l}
\tilde \Delta _2^a + {L_{RX}}{{\tilde \Delta }^a} \ge \tilde \Delta _2^b + {L_{RX}}{{\tilde \Delta }^b}\\
\tilde \Delta _2^a - {L_{RX}}{{\tilde \Delta }^a} \ge \tilde \Delta _2^b - {L_{RX}}{{\tilde \Delta }^b}
\end{array} \right.
\end{align}
\begin{equation} \label{Eq:AppE_6}
{\rm{S3}}: \; \left\{ \begin{array}{l}
\tilde \Delta _2^a + {L_{RX}}{{\tilde \Delta }^a} \ge \tilde \Delta _2^b + {L_{RX}}{{\tilde \Delta }^b}\\
\tilde \Delta _2^a - {L_{RX}}{{\tilde \Delta }^a} \le \tilde \Delta _2^b - {L_{RX}}{{\tilde \Delta }^b}
\end{array} \right. \, .
\end{equation}

\textcolor{black}{As for S3, the supports of sub-HoloSs $a$ and $b$ are completely overlapped in the $\kappa_u$-domain. Then, this scenario needs to be analyzed by considering the second expression in  \eqref{Eq:AppE_1}. We then focus on the first two scenarios. The objective is to upper-bound the overlap of the supports of sub-HoloSs $a$ and $b$ in the wavenumber domain, showing that it can be made arbitrary small if ${L_{RX}}$ is sufficiently small compared with the distance between the center-points of the sub-HoloSs}.

\textcolor{black}{As far as S1 is concerned, the overlap in the $\kappa_u$-domain is $\Delta O = \left( {\tilde \Delta _2^a - \tilde \Delta _2^b} \right) + {L_{RX}}\left( {{{\tilde \Delta }^a} + {{\tilde \Delta }^b}} \right)$. By combining the inequalities in \eqref{Eq:AppE_3} and \eqref{Eq:AppE_4}, we obtain $0 \le \Delta O \le 2{L_{RX}}{{\tilde \Delta }^b}$}.

\textcolor{black}{As far as S2 is concerned, the overlap in the $\kappa_u$-domain is $\Delta O = \left( {\tilde \Delta _2^b - \tilde \Delta _2^a} \right) + {L_{RX}}\left( {{{\tilde \Delta }^a} + {{\tilde \Delta }^b}} \right)$. By combining the inequalities in \eqref{Eq:AppE_3} and \eqref{Eq:AppE_5}, we obtain $0 \le \Delta O \le 2{L_{RX}}{{\tilde \Delta }^b}$}.

\textcolor{black}{Thus, the overlap of the supports in the wavenumber domain can be made smaller than $\epsilon > 0$, if $2{L_{RX}}{{\tilde \Delta }^b} \le \epsilon$. Since ${{\tilde \Delta }^b}$ scales with the reciprocal of ${\left\| {{\bf{c}}_o^n} \right\|}$, this concludes the proof}.

\vspace{-0.15cm}
\section*{Appendix F -- Proof of Lemma 18}
\textcolor{black}{According to Definition 6, the operator norm for the eigenproblem in \eqref{eq:eigenproblemSimpNP} can be formulated as follows (np=no-parax)}:
\begin{align} 
    ||\bar{G}_{Tx}^{\mathrm{np}}||_{\mathrm{op}} &= \mathop{\sup}\nolimits_{||\bar{\phi}|| = 1}  \Biggl|\Biggl|\int_{\mathcal{S}_{Tx}} h_G^{\mathrm{np}}(\mathbf{r}_{Tx} - \mathbf{r}_{Tx}') \bar{\phi}(\mathbf{r}_{Tx}') \, d\mathbf{r}_{Tx}'\Biggl|\Biggl| \nonumber \\
    & \hspace{-0.5cm}= \mathop{\sup}\nolimits_{||\bar{\phi}|| = 1} ||h_G^{\mathrm{np}}(\mathbf{r}_{Tx}) * \left[ \mathds{1}_{\mathcal{S}_{Tx}}(\mathbf{r}_{Tx}) \bar{\phi}(\mathbf{r}_{Tx})\right] || \nonumber \, .
\end{align}

\textcolor{black}{Applying Parseval's theorem \cite[Prop. 2.12]{dubois2019multidimensional} and using the Fourier transform in \eqref{eq:fourierTransformNP}, the operator norm simplifies to}
\begin{align} 
& ||\bar{G}_{Tx}^{\mathrm{np}}||_{\mathrm{op}} = \mathop{\sup}\nolimits_{||\bar{\phi}|| = 1} || H_G^{\mathrm{np}}(\boldsymbol{\kappa}_{Tx})\mathcal{F}\left[ \mathds{1}_{\mathcal{S}_{Tx}}(\mathbf{r}_{Tx}) \bar{\phi} (\mathbf{r}_{Tx})\right] || \nonumber \\
&  = \mathop{\sup}\limits_{||\bar{\phi}|| = 1} || \sum\nolimits_{n=1}^{N_r} ||\bar{G}_{Tx}^n||_{\mathrm{op}}  \mathds{1}_{\mathcal{H}_G^n}(\boldsymbol{\kappa}_{Tx})     \mathcal{F}\left[ \mathds{1}_{\mathcal{S}_{Tx}}(\mathbf{r}_{Tx}) \bar{\phi} (\mathbf{r}_{Tx})\right] || \nonumber  \\  
& \mathop  \approx \limits^{\left( a \right)} \mathop{\sup}\limits_{||\bar{\phi}|| = 1}  \sum\nolimits_{n=1}^{N_r} ||\bar{G}_{Tx}^n||_{\mathrm{op}}  ||\mathds{1}_{\mathcal{H}_G^n}(\boldsymbol{\kappa}_{Tx})     \mathcal{F}\left[ \mathds{1}_{\mathcal{S}_{Tx}}(\mathbf{r}_{Tx}) \bar{\phi} (\mathbf{r}_{Tx})\right] ||     \nonumber
\end{align}
\textcolor{black}{where $(a)$ follows from Lemma 17, i.e., the supports ${\mathcal{H}_G^n}(\boldsymbol{\kappa}_{Tx})$ do not overlap, asymptotically, in the wavenumber domain}. 

\textcolor{black}{Let us define $E^{n}= ||\mathds{1}_{\mathcal{H}_G^n}(\boldsymbol{\kappa}_{Tx})     \mathcal{F}\left[ \mathds{1}_{\mathcal{S}_{Tx}}(\mathbf{r}_{Tx}) \bar{\phi} (\mathbf{r}_{Tx})\right] ||$ and $||\bar{G}_{Tx}^{\max}||_{\mathrm{op}} = {\max _n}\left\{ { ||\bar{G}_{Tx}^n||_{\mathrm{op}}} \right\}$. Also, we note that $\sum\nolimits_{n=1}^{N_r} E^n = 1$ if $\bar{\phi} (\mathbf{r}_{Tx})$ is chosen so that its support (in the spatial domain) is contained in $\mathcal{S}_{Tx}$ and the support of its Fourier transform (in the wavenumber domain) is contained in the support of the Fourier transform $H_G^{\mathrm{np}}(\boldsymbol{\kappa}_{Tx})$, i.e., there is no energy loss. Assuming $\sum\nolimits_{n=1}^{N_r} E^n = 1$, the supremum of $||\bar{G}_{Tx}^{\mathrm{np}}||_{\mathrm{op}}$ is attained by choosing $\bar{\phi} (\mathbf{r}_{Tx})$ so that the support of its Fourier transform (in the wavenumber domain) is contained in $\mathcal{H}_G^{\max}$, i.e., the sub-HoloS corresponding to $||\bar{G}_{Tx}^{\max}||_{\mathrm{op}}$, which provides the greatest weight in the summation of $||\bar{G}_{Tx}^{\mathrm{np}}||_{\mathrm{op}}$. These conditions can be achieved only asymptotically, as a non-zero function cannot have finite supports in both domains simultaneously. Under these conditions, we obtain $||\bar{G}_{Tx}^{\mathrm{np}}||_{\mathrm{op}} = ||\bar{G}_{Tx}^{\max}||_{\mathrm{op}}$}.

\vspace{-0.15cm}
\section*{Appendix G -- Proof of Lemma 20}
\textcolor{black}{In \cite{Landau1975}, the authors proved the result for $N_r=1$, i.e., the non-zero eigenvalues of the eigenproblem in \eqref{eq:eigenproblemSimp} (paraxial setting) polarize to the single value $||\bar{G}_{Tx}||_{\mathrm{op}}$ in \eqref{Eq:OpNormExplicit}. We generalize the proof in \cite{Landau1975} considering the eigenproblem in \eqref{eq:eigenproblemSimpNP} (non-paraxial setting). Similar to \cite{Landau1975}, we study the sum and the sum of the square of the eigenvalues of \eqref{eq:eigenproblemSimpNP}}. 

\textcolor{black}{By virtue of Mercer's theorem \cite[Th. 4.24, Cor. 4.26]{IntegralEq}}
\begin{equation} \label{Eq:AppG_1}
{r^{ - 2}}\sum {{\mu _m}}  = {r^{ - 2}}\int\nolimits_{r{{\mathcal{S}}_{Tx}}} {\bar G_{Tx}^{{\rm{no - parax}}}\left( {{{\bf{r}}_{Tx}},{{\bf{r}}_{Tx}}} \right)d{{\bf{r}}_{Tx}}} \, .
\end{equation}

\textcolor{black}{Since ${\bar G_{Tx}^{{\rm{no - parax}}}\left( {{{\bf{r}}_{Tx}},{{\bf{r}}_{Tx}}} \right)} = {h_G^{{\rm{no - parax}}}\left( {\bf{0}} \right)}$ and $m\left( {r{{\mathcal{S}}_{Tx}}} \right) = \int_{r{{\mathcal{S}}_{Tx}}} {d{{\bf{r}}_{Tx}}}$, we obtain $\int_{r{{\mathcal{S}}_{Tx}}} {\bar G_{Tx}^{{\rm{no - parax}}}\left( {{{\bf{r}}_{Tx}},{{\bf{r}}_{Tx}}} \right)d{{\bf{r}}_{Tx}}}  = h_G^{{\rm{no - parax}}}\left( {\bf{0}} \right)m\left( {r{{\mathcal{S}}_{Tx}}} \right)$. Then, $h_G^{{\rm{no - parax}}}\left( {\bf{0}} \right)$ can be expressed in terms of inverse Fourier transform, obtaining}
\begin{align} 
{r^{ - 2}}\sum {{\mu _m}}  &= \frac{{{r^{ - 2}}m\left( {r{{\mathcal{S}}_{Tx}}} \right)}}{{{{\left( {2\pi } \right)}^2}}}\int_{{{\mathbb{R}}^2}} {H_G^{{\rm{no - parax}}}\left( {{{\boldsymbol{\kappa }}_{Tx}}} \right)d{{\boldsymbol{\kappa }}_{Tx}}} \\
&\hspace{-1cm}\mathop  = \limits^{\left( a \right)} {r^{ - 2}}\sum\nolimits_{n = 1}^{{N_r}} {{{\left\| {\bar G_{Tx}^n} \right\|}_{{\rm{op}}}}  \left[{{m\left( {r{{\mathcal{S}}_{Tx}}} \right)}{m\left( {{\mathcal{H}}_G^n} \right)}}/{{{{\left( {2\pi } \right)}^2}}} \right]  }  \label{Eq:AppG_2}
\end{align}
\textcolor{black}{where $(a)$ follows from \eqref{eq:fourierTransformNP} with  $m\left( {{\mathcal{H}}_G^n} \right) = \int_{{\mathcal{H}}_G^n} {d{{\boldsymbol{\kappa }}_{Tx}}}$. From Prop. 1, we note that the term insides the square brackets is the eDoF of sub-HoloS $n$, considered individually, as $r \to \infty$}.

\textcolor{black}{Using \cite[Th. 4.19]{IntegralEq}, we obtain}
\begin{equation} \label{Eq:AppG_3}
{r^{ - 2}}\sum {\mu _m^2}  = {r^{ - 2}}\int_{r{{\mathcal{S}}_{Tx}}} {\int_{r{{\mathcal{S}}_{Tx}}} {{{\left| {\bar G_{Tx}^{{\rm{no - parax}}}\left( {{{\bf{r}}_{Tx}},{\bf{r}}_{Tx}'} \right)} \right|}^2}d{{\bf{r}}_{Tx}}d{\bf{r}}_{Tx}'} } \nonumber \, .
\end{equation}

\textcolor{black}{Also, $\bar G_{Tx}^{{\rm{no - parax}}}\left( {{{\bf{r}}_{Tx}},{\bf{r}}_{Tx}'} \right) = h_G^{{\rm{no - parax}}}\left( {{{\bf{r}}_{Tx}} - {\bf{r}}_{Tx}'} \right)$. Using the change of variables ${\bf{r}} = {{\bf{r}}_{Tx}} - {\bf{r}}_{Tx}'$ and ${\bf{s}} = {\bf{r}}_{Tx}'$, and letting $r \to \infty$, we obtain the following}:
\begin{align} \label{Eq:AppG_4}
& {r^{ - 2}}\sum {\mu _m^2} = m\left( {{{\mathcal{S}}_{Tx}}} \right)\int_{{{\mathbb{R}}^2}} {{{\left| {h_G^{{\rm{no - parax}}}\left( {\bf{r}} \right)} \right|}^2}d{\bf{r}}} \\
&\mathop  = \limits^{\left( a \right)} {\left( {2\pi } \right)^{ - 2}} m\left( {{{\mathcal{S}}_{Tx}}} \right)\int_{{{\mathbb{R}}^2}} {{{\left| {H_G^{{\rm{no - parax}}}\left( {\boldsymbol{\kappa }} \right)} \right|}^2}d{\boldsymbol{\kappa }}}  \\
& \mathop  = \limits^{\left( b \right)} {\left( {2\pi } \right)^{ - 2}}m\left( {{{\mathcal{S}}_{Tx}}} \right)\sum\nolimits_{n = 1}^{{N_r}} {\sum\nolimits_{l = 1}^{{N_r}} {{{\left\| {\bar G_{Tx}^n} \right\|}_{{\rm{op}}}}{{\left\| {\bar G_{Tx}^l} \right\|}_{{\rm{op}}}}m\left( {{\mathcal{H}}_G^{n,l}} \right)} } \nonumber
\end{align}
\textcolor{black}{where $(a)$ follows from Parseval's theorem \cite[Prop. 2.12]{dubois2019multidimensional}, and $(b)$ by using ${H_G^{{\rm{no - parax}}}\left( {\boldsymbol{\kappa }} \right)}$ in \eqref{eq:fourierTransformNP}. Also, $m\left( {{\mathcal{H}}_G^n \cap {\mathcal{H}}_G^l} \right) = \int_{{{\mathbb{R}}^2}} {{{\mathds{1}}_{{\mathcal{H}}_G^n}}\left( {\boldsymbol{\kappa }} \right){{\mathds{1}}_{{\mathcal{H}}_G^l}}\left( {\boldsymbol{\kappa }} \right)d{\boldsymbol{\kappa }}}$ represents the area where ${{\mathcal{H}}_G^n}$ and ${{\mathcal{H}}_G^l}$ in \eqref{Eq:SpectralBandNoParax} overlap. By virtue of Lemma 17, $m\left( {{\mathcal{H}}_G^n \cap {\mathcal{H}}_G^l} \right) \approx 0$ is $n \ne l$. Therefore, we obtain the following}:
\begin{equation} \label{Eq:AppG_5}
\hspace{-0.20cm}{r^{ - 2}}\sum {\mu _m^2}  \approx \sum\nolimits_{n = 1}^{{N_r}} {{{\left\| {\bar G_{Tx}^n} \right\|}_{{\rm{op}}}^2}\left[ {m\left( {{{\mathcal{S}}_{Tx}}} \right)m\left( {{\mathcal{H}}_G^n} \right)/{{\left( {2\pi } \right)}^2}} \right]}  \, .
\end{equation}

\textcolor{black}{From Prop. 1, similar to \eqref{Eq:AppG_2}, the term insides the square brackets is the eDoF of sub-HoloS $n$, considered individually. By comparing \eqref{Eq:AppG_2} and \eqref{Eq:AppG_5}, the proof follows}.

\vspace{-0.15cm}
\section*{Appendix H -- Relationship Between Proposition 2 and The Method in \cite{Dardari_2020} and \cite{Hongliang_2025}} 
\textcolor{black}{For consistency with \cite{Dardari_2020} and \cite{Hongliang_2025}, we consider the case study $\alpha = \beta = 0$. From \cite[Eqs. (13), (21)]{EuCAP_2025}, we obtain}
\begin{align}  
{N_o}&\mathop  = {\left( {2\pi } \right)^{ - 2}} \int_{{{\mathcal{S}}_{Rx}}} { { {{\mathcal{J}}\left( {{u_o},{v_o}} \right)} } d{u_o}d{v_o}} \label{Eq:AppH_1} \\
&\mathop  = \limits^{\left( a \right)} {\left( {2\pi } \right)^{ - 2}}  \sum\nolimits_{n = 1}^{{N_r}} {\int_{{\mathcal{S}}_{Rx}^n} {\left| { {{\mathcal{J}}\left( {u_o^n,v_o^n} \right)}} \right|du_o^ndv_o^n} } \label{Eq:AppH_2}
\end{align}
\textcolor{black}{where ${\mathcal{J}}\left( {{u_o},{v_o}} \right) = \int_{{{\mathcal{S}}_{Tx}}} {\left| {\det \left( {{\bf{J}}\left( {{u_i},{v_i},{u_o},{v_o}} \right)} \right)} \right|d{u_i}d{v_i}}$ and $(a)$ applies the partioning in Sec. III to $\mathcal{S}_{Rx}$. For the considered setting, ${\kappa _a}\left( {u_i, v_i, u_o^n,v_o^n} \right) ={\kappa _0}\left( {{a_{Rx}^n}\left( {u_o^n,v_o^n} \right) - {a_{Tx}}\left( {u_i, v_i} \right)} \right)/$ $\left\| {{{\bf{r}}_{Rx}}\left( {u_o^n,v_o^n} \right) - {{\bf{r}}_{Tx}}\left( {u_i, v_i} \right)} \right\|$ and $\det \left( {{\bf{J}}\left( {u_i, v_i, u_o^n,v_o^n} \right)} \right) = \left( {{{\partial {\kappa _x}} \mathord{\left/
 {\vphantom {{\partial {\kappa _x}} {\partial {u_i}}}} \right.
 \kern-\nulldelimiterspace} {\partial {u_i}}}} \right)\left( {{{\partial {\kappa _z}} \mathord{\left/
 {\vphantom {{\partial {\kappa _z}} {\partial {v_i}}}} \right.
 \kern-\nulldelimiterspace} {\partial {v_i}}}} \right) - \left( {{{\partial {\kappa _x}} \mathord{\left/
 {\vphantom {{\partial {\kappa _x}} {\partial {v_i}}}} \right.
 \kern-\nulldelimiterspace} {\partial {v_i}}}} \right)\left( {{{\partial {\kappa _z}} \mathord{\left/
 {\vphantom {{\partial {\kappa _z}} {\partial {u_i}}}} \right.
 \kern-\nulldelimiterspace} {\partial {u_i}}}} \right)$. Thanks to the partitioning, ${{\mathcal{S}}_{Rx}^n}$ is, similar to ${{\mathcal{S}}_{Tx}}$, small. Then, the integral in \eqref{Eq:AppH_2} is equal to $N_o^{n}=\kappa _0^2A_{Tx}A_{Rx}^n{\left\| {{\bf{c}}_o^n} \right\|^{ - 2}}\Upsilon \left( {{\bf{c}}_o^n,0,0} \right)$ with $\Upsilon(\cdot)$ given in \eqref{Eq:OpNormExplicit}. Thus, $N_o$ in \eqref{Eq:AppH_1} coincides with $N_{{\rm{eDoF}}} (\gamma)$ in \eqref{eq:DoFNP}, under the assumption that $\gamma$ is arbitrarily small}.

\vspace{-0.15cm}
\section*{Appendix I -- Equivalence Between the Quartic and Conventional Parabolic Approximations} 
\textcolor{black}{We introduce a new coordinate system identified by the versors $\left( {{{{\bf{\hat e}}}_x},{{{\bf{\hat e}}}_y},{{{\bf{\hat e}}}_z}} \right)$, where ${{{\bf{\hat e}}}_y} = {{\bf{c}}_o}/\left\| {{{\bf{c}}_o}} \right\|$ identifies the direction connecting the two center-points of $\mathcal{S}_{Tx}$ and $\mathcal{S}_{Rx}$, and the pair $\left( {{{{\bf{\hat e}}}_x},{{{\bf{\hat e}}}_z}} \right)$ is chosen to fulfill the equality ${{{\bf{\hat e}}}_x}{\bf{\hat e}}_x^T + {{{\bf{\hat e}}}_y}{\bf{\hat e}}_y^T + {{{\bf{\hat e}}}_z}{\bf{\hat e}}_z^T = {\bf{I}}$. In the new coordinate system, $\mathcal{S}_{Tx}$ and $\mathcal{S}_{Rx}$ are identified by the points ${\bf{r}}_{Tx}^p$ and ${\bf{r}}_{Rx}^p$, respectively. Then, we obtain}
\begin{align}  \label{Eq:AppI_1}
\left\| {{\bf{r}}_{Rx}^p - {\bf{r}}_{Tx}^p} \right\| &= \sqrt {\sum\nolimits_a {{{\left( {a_{Rx}^p - a_{Tx}^p} \right)}^2}} } \\
& = \left\| {{\bf{c}}_o^p} \right\|\sqrt {1 + {{\left\| {{\bf{c}}_o^p} \right\|}^{ - 2}}\sum\nolimits_a {{{\left( {\delta {a^p}} \right)}^2} + 2\delta {a^p}\delta a_c^p} } \nonumber
\end{align}
\textcolor{black}{where $\delta {a^p} = \Delta a_{Rx}^p - \Delta a_{Tx}^p$, $\delta a_c^p = \Delta a_{Rc}^p - \Delta a_{Tc}^p$, and ${\left\| {{\bf{c}}_o^p} \right\|^2} = \sum\nolimits_a {{{\left( {\delta a_c^p} \right)}^2}}$. In the new coordinate system, it is legitimate to apply the parabolic approximation $\sqrt {1 + x}  \approx 1 + x/2$ and to simplify the resulting expression by taking into account that $\delta {y^p} \ll \delta y_c^p$, since $\mathcal{S}_{Tx}$ and $\mathcal{S}_{Rx}$ are concentrated around $\bf{0}$ and ${\bf{c}}_o$, respectively. The expression obtained can be formulated in terms of the original coordinate system by applying the transformations $a_{Tx}^p = {\bf{\hat e}}_a^T{{\bf{r}}_{Tx}}$ and $a_{Rx}^p = {\bf{\hat e}}_a^T{{\bf{r}}_{Rx}}$. The resulting expression coincides with \eqref{Eq:WavefrontApprox_1}-\eqref{Eq:DistanceApprox_2} when the approximations $(a)$ and $(b)$ are both applied. This concludes the proof}.

\vspace{-0.15cm}
\section*{Appendix J -- Non-Paraxial Setting with Large-Size Transmitting and Receiving HoloSs} 
\textcolor{black}{Due to space limitations, we report a sketch of the proof. We consider the setting in which both HoloSs $\mathcal{S}_{Tx}$ and $\mathcal{S}_{Rx}$ have a large size. Similar to $\mathcal{S}_{Rx}$, we partition $\mathcal{S}_{Tx}$ into $N_t$ sub-HoloSs that are identified by the superscript $m$. With a notation similar to Sec. IV-B, we write $a_{Tx} = a_{Tc}^m + \Delta a_{Tx}^m$. The integral in \eqref{eq:eigenproblemIn} can hence be formulated in terms of the summation of the $N_t$ sub-HoloSs. In this case, however, the integrand function of \eqref{eq:genGBarTXn} depends on all possible pairs $(m,n)$, i.e., $p^n(\mathbf{r}_{Tx}', \mathbf{r}_{Rx})$ is replaced by $p^{(m,n)}(\mathbf{r}_{Tx}', \mathbf{r}_{Rx})$. Even applying the  approximations in Lemma 15, the addends inside the exponential term of \eqref{eq:nonParaxialSA_app} depend on terms of the kind $D_u^{(m,n)}u_i$, $D_u^{(m',n)}u_i'$, $D_v^{(m,n)}v_i$, $D_v^{(m',n)}v_i'$ with $D_{(\cdot)^{(\cdot)}}$ constant factors. Since $D_u^{(m,n)} \ne D_u^{(m',n)}$ and  $D_v^{(m,n)} \ne D_v^{(m',n)}$ if $m \ne m'$, it is not possible to express the self-adjoint operator in \eqref{eq:nonParaxialSA_app} in terms of differences $u_i - u_{i}'$ and $v_i - v_{i}'$. Thus, Lemma 8 is not applicable. The analysis of non-convolutional operators is beyond the scope of this paper, and is left to future work \cite{Landau1975}}.

\vspace{-0.15cm}
\section*{Appendix K -- Proof of Proposition 3} 
Consider ${{\bar G}_{Tx}}\left( {{{\bf{r}}_{Tx}},{\bf{r}}_{Tx}'} \right)$ in \eqref{eq:GBarTXn}, setting $\tau_{12} = \tau_{21} = 0$ and omitting the superscript $n$. The eigenproblem in \eqref{eq:eigenproblemSimp} becomes
%
\begin{align} \label{Eq:KernelPSWF}
{{{\mu _m}} }{{\bar \phi }_m}\left( {{u_i},{v_i}} \right)& = {\left| {{{\bar g}_{i,o}}\left( {{{\bf{c}}_{Tx}};{{\bf{c}}_{Rx}}} \right)} \right|^2} \\ & \hspace{-0.75cm}\times \int_{ - {U_{Tx}}}^{{U_{Tx}}} {\int_{-{V_{Tx}}}^{{V_{Tx}}} {{A_{Rx}}f\left( {u_i',v_i'} \right){{\bar \phi }_m}\left( {{u_i'},{v_i'}} \right)du_i'dv_i'} } \nonumber 
\end{align} 
where $f\left( {u_i',v_i'} \right) = {\rm{sinc}}\left( {{U_o}{\tau _{11}}\left( {{u_i} - u_i'} \right)} \right){\rm{sinc}}\left( {{V_o}{\tau _{22}}\left( {{v_i} - v_i'} \right)} \right)$.

Define ${ {{{\bar \mu }_{m,i,o}}} } = A_{Rx}^{ - 1}{\left| {{{\bar g}_{i,o}}\left( {{{\bf{c}}_{Tx}};{{\bf{c}}_{Rx}}} \right)} \right|^{-2}}{ {{\mu _m}}}$, ${{{{\bar \mu }_{m,i,o}}} } = {{{{\bar \mu }_{m,i,o,u}}} }{ {{{\bar \mu }_{m,i,o,v}}} }$, and assume to look for eigenfunctions that can be written as ${{\bar \phi }_m}\left( {{u_i},{v_i}} \right) = {{\bar \phi }_{m,u}}\left( {{u_i}} \right){{\bar \phi }_{m,v}}\left( {{v_i}} \right)$. Then, the eigenproblem in \eqref{Eq:KernelPSWF} boils down to solving the eigenproblems
\begin{align}
& {{\tilde \mu }_{m,i,o,u}}{{\bar \phi }_{m,u}}\left( {{u_i}} \right) = \int_{ - {U_{Tx}}}^{{U_{Tx}}} {\frac{{\sin \left( {\pi {U_o}{\tau _{11}}\left( {{u_i} - u_i'} \right)} \right)}}{{\pi \left( {{u_i} - u_i'} \right)}}{{\bar \phi }_{m,u}}\left( {{u_i'}} \right)du_i'} \nonumber \\ & {{\tilde \mu }_{m,i,o,v}}{{\bar \phi }_{m,v}}\left( {{v_i}} \right) = \int_{ - {V_{Tx}}}^{{V_{Tx}}} {\frac{{\sin \left( {\pi {V_o}{\tau _{22}}\left( {{v_i} - v_i'} \right)} \right)}}{{\pi \left( {{v_i} - v_i'} \right)}}{{\bar \phi }_{m,v}}\left( {{v_i'}} \right)dv_i'} \nonumber 
\end{align} 
where ${{\tilde \mu }_{m,i,o,u}} = {U_o}{\tau _{11}}{{\bar \mu }_{m,i,o,u}}$ and ${{\tilde \mu }_{m,i,o,v}} = {V_o}{\tau _{22}}{{\bar \mu }_{m,i,o,v}}$. The proof follows from \cite[Eq. (11)]{Slepian1961}.

\vspace{-0.15cm}
\section*{Appendix L -- Proof of Proposition 4}
\textcolor{black}{Let us insert the eigenfunctions given in \eqref{eq:NP_eigenfunctions}, for $k=1,2,\ldots, N_r$, and the kernel given in \eqref{eq:nonParaxialSA_app} into the eigenproblem in \eqref{eq:eigenproblemSimpNP}. By letting $\mathcal{S}_{Tx} = r \mathcal{S}_{Tx}'$, we obtain}
\begin{align}
\label{eq:eigenfunctionLarge2}
        \mu_m &{\bar{\phi}}_m^{k}(\mathbf{r}_{Tx}) = \sum\nolimits_{n=1}^{N_r}
     |{{\bar g}_{i,o}^{n}}|^2  \int\nolimits_{r \mathcal{S}_{Tx}'} {\bar G}_{Tx} ^{n} \left( {{{\bf{r}}_{Tx}},{\bf{r}}_{Tx}'} \right) \nonumber \\
     & \times
     \exp \biggl\{- j (\Delta k_u^{n} - \Delta k_u^{k}) (u_i - u_i')\biggl\}\nonumber \\
     & \times \exp \biggl\{- j  (\Delta k_v^{n} - \Delta k_v^{k})(v_i - v_i') \biggl\}
        {\bar{\phi}}_m^{k}(\mathbf{r}_{Tx}')\, d\mathbf{r}_{Tx}'
\end{align}

\textcolor{black}{In the asymptotic regime $r \rightarrow \infty$, the integral in \eqref{eq:eigenfunctionLarge2} becomes (approximately) a convolution. By virtue of Lemma 17, therefore, the functions ${\bar G}_{Tx} ^{n} \left( {{{\bf{r}}_{Tx}},{\bf{r}}_{Tx}'} \right)$ behave as filters in the wavenumber domain, which reject the eigenfunctions ${\bar{\phi}}_m^{k}(\mathbf{r}_{Tx})$ for any $n \ne k$. Accordingly, we obtain}
\begin{align}
\label{eq:eigenfunctionLarge3}
        \mu_m {\bar{\phi}}_m^{(k)}(\mathbf{r}_{Tx}) \approx &
     |{{\bar g}_{i,o}^{(k)}}|^2 \int_{{{\mathcal{S}}_{Tx}}}   
       {\bar G}_{Tx} ^{(k)} \left( {{{\bf{r}}_{Tx}},{\bf{r}}_{Tx}'} \right) {\bar{\phi}}_m^{(k)}(\mathbf{r}_{Tx}')\, d\mathbf{r}_{Tx}' \nonumber
\end{align}
\textcolor{black}{hence proving that \eqref{eq:NP_eigenfunctions} are eigenfunctions of \eqref{eq:eigenproblemSimpNP} with ${\bar{\phi}}_m^{k}(\mathbf{r}_{Tx})$ being eigenfunctions of \eqref{eq:eigenproblemSimp}}.


\bibliographystyle{IEEEtran}
\bibliography{sample}


\balance

\end{document}